\documentclass[twocolumn]{aastex631}
\usepackage{makecell}
\usepackage{multirow} 
\usepackage{longtable}
\usepackage[utf8]{inputenc}
\usepackage{newunicodechar}
\usepackage{newunicodechar}
\usepackage{lineno}
\usepackage{soul}

\setstcolor{red} % Define a cor do risco para vermelho
\newunicodechar{ʼ}{'}
\newunicodechar{́}{\'}
\shorttitle{Multiepoch stellar occultation by (50000) Quaoar}
\shortauthors{Margoti et al.}
\graphicspath{{./}{figures/}}
\begin{document}
%\linenumbers
\title{Size, shape, density, and atmospheric limit of (50000) Quaoar revealed from 14 years of stellar occultation}

\author[0000-0002-2103-4408]{Giuliano Margoti} % giulianomargoti@gmail.com
\affiliation{Observatório Nacional/MCTI, R. General José Cristino 77, Rio de Janeiro, RJ 20.921-400, Brasil}
\affiliation{Federal University of Technology - Paraná (PPGFA/UTFPR-Curitiba), Av. Sete de Setembro, 3165, CEP 80230-901 - Curitiba - PR - Brazil}
\affiliation{Laboratório Interinstitucional de e-Astronomia - LIneA, Av. Pastor Martin Luther King Jr 126, CEP: 20765-000 Rio de Janeiro, RJ, Brasil}

\author[0000-0003-2311-2438]{Felipe Braga-Ribas} % fribas@utfpr.edu.br
\affiliation{Federal University of Technology - Paraná (PPGFA/UTFPR-Curitiba), Av. Sete de Setembro, 3165, CEP 80230-901 - Curitiba - PR - Brazil}
\affiliation{Observatório Nacional/MCTI, R. General José Cristino 77, Rio de Janeiro, RJ 20.921-400, Brasil}
\affiliation{Laboratório Interinstitucional de e-Astronomia - LIneA, Av. Pastor Martin Luther King Jr 126, CEP: 20765-000 Rio de Janeiro, RJ, Brasil}
\affiliation{LIRA, CNRS UMR8254, Observatoire de Paris, Universit\'e PSL, Sorbonne Universit\'e, Universit\'e Paris Cit\'e, CY Cergy Paris Universit\'e, 92190 Meudon, France}

\author[0000-0002-8690-2413]{José Luis Ortiz} % ortiz@iaa.es
\affiliation{Instituto de Astrofı́sica de Andalucı́a (IAA-CSIC), Glorieta de la Astronomı́a s/n, 18008 Granada, Spain}

\author[0000-0003-1995-0842]{Bruno Sicardy} % bruno.sicardy@obspm.fr
\affiliation{Laboratoire Temps Espace (LTE), Observatoire de Paris, Universit\'e PSL, CNRS UMR 8255, Sorbonne Universit\'e, LNE, 61 Av. de l'Observatoire, F75014 Paris, France}

\author[0000-0002-2193-8204]{Josselin Desmars}
\affiliation{Laboratoire Temps Espace (LTE), Observatoire de Paris, Universit\'e PSL, CNRS UMR 8255, Sorbonne Universit\'e, LNE, 61 Av. de l'Observatoire, F75014 Paris, France}
\affiliation{Institut Polytechnique des Sciences Avancées IPSA, Ivry-sur-Seine 94200, France}

\author[0000-0003-0088-1808]{Morgado, B. E.} 
\affiliation{Universidade Federal do Rio de Janeiro - Observatório do Valongo, Ladeira do Pedro Antonio 43, Rio de Janeiro, RJ 20.080-090, Brasil}

\author[0009-0001-5754-4688]{Eros de Oliveira Gradovski} % erosgradovski@gmail.com
\affiliation{Observatório Nacional/MCTI, R. General José Cristino 77, Rio de Janeiro, RJ 20.921-400, Brasil}
\affiliation{Federal University of Technology - Paraná (PPGFA/UTFPR-Curitiba), Av. Sete de Setembro, 3165, CEP 80230-901 - Curitiba - PR - Brazil}
\affiliation{Laboratório Interinstitucional de e-Astronomia - LIneA, Av. Pastor Martin Luther King Jr 126, CEP: 20765-000 Rio de Janeiro, RJ, Brasil}

\author[0000-0003-1000-8113]{Chrystian Luciano Pereira}
\affiliation{Observatório Nacional/MCTI, R. General José Cristino 77, Rio de Janeiro, RJ 20.921-400, Brasil}
\affiliation{Laboratório Interinstitucional de e-Astronomia - LIneA, Av. Pastor Martin Luther King Jr 126, CEP: 20765-000 Rio de Janeiro, RJ, Brasil}

\author[0000-0002-1123-983X]{Pablo Santos-Sanz} % psantos@iaa.es
\affiliation{Instituto de Astrofı́sica de Andalucı́a (IAA-CSIC), Glorieta de la Astronomı́a s/n, 18008 Granada, Spain}

\author[0000-0002-3362-2127]{Altair Ramos Gomes-Júnior} % altairgomesjr@gmail.com
\affiliation{Universidade Federal de Uberlândia (UFU), Uberlândia, MG, Brasil}
\affiliation{Laboratório Interinstitucional de e-Astronomia - LIneA, Av. Pastor Martin Luther King Jr 126, CEP: 20765-000 Rio de Janeiro, RJ, Brasil}

% ...

%%%%%%%%%%%%%%%%%%%%%%%%%%%%%%%%%%%%%%%%%%%

%% Note that the \and command from previous versions of AASTeX is now
%% depreciated in this version as it is no longer necessary. AASTeX 
%% automatically takes care of all commas and "and"s between authors names.

%% AASTeX 6.31 has the new \collaboration and \nocollaboration commands to
%% provide the collaboration status of a group of authors. These commands 
%% can be used either before or after the list of corresponding authors. The
%% argument for \collaboration is the collaboration identifier. Authors are
%% encouraged to surround collaboration identifiers with ()s. The 
%% \nocollaboration command takes no argument and exists to indicate that
%% the nearby authors are not part of surrounding collaborations.

%% Mark off the abstract in the ``abstract'' environment. 
\begin{abstract}

We present results from 28 stellar occultations by the large Trans-Neptunian Object (50000)~Quaoar registered between 2018 and 2025. By performing a joint analysis of this occultation data-set, along with other 9 published events, we were able to fit an oblate ellipsoid shape, with equatorial semi-axes, $a$ and $b$ of $566.1^{+2.5}_{-2.2}$~km, and a polar semi-axis, $c$, of $511.2^{+3.6}_{-3.7}$~km. It provides an equivalent volumetric diameter of 1094.4~$\pm$~4.6~km and polar oblateness of 0.097~$\pm$~0.011. Considering an absolute magnitude of $H=2.79\pm0.35$, we derive a geometric albedo of $p_{\rm V}=0.125\pm0.038$. 
%\st{We have derived a new upper limit to the presence of a CH{$_4$} atmosphere of 1.5~nano~bar, and we also provide} 
{\color{black}We have derived new upper limits to the surface pressure of a CH$_4$ atmosphere of 0.15 nbar (1$\sigma$) and 0.65 nbar (3$\sigma$). We also provide} a table with the 36 new astrometric positions for Quaoar. Using the new system mass derived from Weywot's orbit around Quaoar, we calculated a density of 1.760~$\pm$~0.109~g/cm$^3$. Moreover, from the derived size and rotation period (8.8394~$\pm$~0.0002 hours \citep{Ortiz_2003}), we calculate that, if Quaoar is in Maclaurin hydrostatic equilibrium state, it would have a density of 1.859~$\pm$~0.200~g/cm$^3$. This result, within the error bars, is compatible with the value we found. Therefore, this work shows that Quaoar can be a Maclaurin object, being eligible as a dwarf planet. 

\end{abstract}

%% Keywords should appear after the \end{abstract} command. 
%% The AAS Journals now uses Unified Astronomy Thesaurus concepts:
%% https://astrothesaurus.org
%% You will be asked to selected these concepts during the submission process
%% but this old "keyword" functionality is maintained in case authors want
%% to include these concepts in their preprints.

%% From the front matter, we move on to the body of the paper.
%% Sections are demarcated by \section and \subsection, respectively.
%% Observe the use of the LaTeX \label
%% command after the \subsection to give a symbolic KEY to the
%% subsection for cross-referencing in a \ref command.
%% You can use LaTeX's \ref and \label commands to keep track of
%% cross-references to sections, equations, tables, and figures.
%% That way, if you change the order of any elements, LaTeX will
%% automatically renumber them.
%%
%% We recommend that authors also use the natbib \citep
%% and \citet commands to identify citations.  The citations are
%% tied to the reference list via symbolic KEYs. The KEY corresponds
%% to the KEY in the \bibitem in the reference list below. 

\section{\textbf{Introduction}}\label{sec:intro}

Since the first stellar occultation by (50000) Quaoar observed in 2011 \citep{Person_2011, Ribas_2013}, knowledge about this object has steadily increased. In total, Quaoar's shadow has been detected during 36 distinct stellar occultation events, though only restricted analysis of 13 of these events have been published so far (see Table \ref{tab:ev list}). 
%\citep{Person_2011, Ribas_2013, Arimatsu_2019, Morgado_2023, Pereira_2023, Ribas_2025}.
Furthermore, two rings outside the Roche Limit were found around Quaoar \citep{Morgado_2023,Pereira_2023}.

A stellar occultation occurs when a solar system object passes in front of a background star, temporarily blocking its light and casting a shadow at a given location. Each observational detection of the object's shadow at different points on the ground provides a positive chord (i.e., a unidimensional measurement of its profile), while non-detections are called negative chords and can constrain the object's size and position (when associated with positive detections) \citep{Sicardy_2024}.

%\st{In total, 107 positive chords were detected across Quaoar} 
{\color{black}In this work, we present 83 new positive chords for Quaoar, out of a total dataset of 107 positive chords}, not counting additional observations made with instruments that simultaneously acquired data in multiple filters. 
This extensive set of occultations over diverse epochs allows for a comprehensive analysis of the global properties of Quaoar, such as those made for Chariklo \citep{Leiva_2017, Morgado_2021}, 2002~KX$_{14}$ \citep{Rizos_2025}, 2003~AZ$_{84}$ \citep{dias_2017} and others. 

Quaoar is one of the largest Trans‑Neptunian Objects (TNOs) known to date. Because it is not in mean motion resonance with Neptune, Quaoar is classified as a classical Kuiper belt object, 
%\st{(cubewano),} 
and its orbital inclination ($>$8°) places it among the dynamically hot population \citep{Lykawka_2007}. 
%Early size estimates show a large discrepancy for Quaoar's diameter and have big error bars, for example, data derived from direct imaging find diameter of 1,260~$\pm$~190~km \citep{Brown_2004}, while thermal measurements initially suggested 853~$\pm$~199~km \citep{Stansberry_2008}. Subsequent combined analyzes produced values of 890~$\pm$~70~km \citep{Fraser_2010} and later 1,074~$\pm$~138~km \citep{Fornasier_2013}. 
The first multi‑chord stellar occultation by Quaoar reported by \citet{Ribas_2013} %marked a major advance, reducing the size uncertainty to the kilometer scale. \citet{Ribas_2013} 
derived an equivalent diameter of 1,110~$\pm$~5~km, assuming a Maclaurin spheroid shape with equatorial radius 572.5~$\pm$~20.5~km and polar oblateness of 0.0917~$\pm$~0.0222. In 2022, a ten‑chord occultation reported by \citet{Pereira_2023} produced a sky-plane ellipse fit with semi‑major axis 579.5~$\pm$~4.0~km and apparent flattening 0.12~$\pm$~0.01, consistent with \citet{Ribas_2013} projection. Finally, \citet{Kiss_2024} combined \citet{Ribas_2013} occultation data with K2 photometry and Herschel PACS thermal light curves to obtain an equivalent diameter of 1,090~$\pm$~40~km.

%Quaoar has a known satellite, Weywot, whose orbital parameters were recently improved. 
Weywot, Quaoar's known satellite, was discovered in 2007 \citep{Brown_2007}, and its first orbit determinations were given by \cite{Vachier_2012} and \cite{Fraser_2013}. Using four stellar occultations by Weywot, 
%Quaoar's known satellite,
\cite{Ribas_2025} derived precise relative positions, which were used to improve its orbital parameters and derive the system mass (1.208~$\pm$~0.063~$\times$10$^{21}$~kg.).
%, thanks to the use of stellar occultation-derived relative positions \citep{Ribas_2025}.
Weywot orbits Quaoar at a distance of 13,309~$\pm$~231~km, with an eccentricity of 0.018~$\pm$~0.016 and an orbital inclination relative to the Ecliptic of 38.0$^\circ$~$\pm$~1.4$^\circ$, yielding an orbital period of 12.431013~$\pm$~0.00021~days. \citet{Ribas_2025} have also determined Weywot’s orbital pole as $\alpha_W$:~266.9$^\circ$~$\pm$~1.5$^\circ$ and $\delta_W$:~51.9$^\circ$~$\pm$~1.4$^\circ$.
%and system mass, 1.208~$\pm$~0.063~$\times$10$^{18}$~kg. 
%These orbital parameters constitute a component of the field to refine our understanding of Quaoar. Knowing both the satellite's orbit and the primary's shape enables us to determine, for example, whether Quaoar is in tidal hydrostatic equilibrium or to derive its density when combined with the shape model.

Additionally, as mentioned, Quaoar possesses two known rings \citep{Morgado_2023, Pereira_2023}. More recently, a new opaque structure has been detected at a distance of ~5{,}700~km from the main body center, indicating that more material is present around it \citep{Nolthenius_2025}.  The outer ring (QR1) orbits Quaoar at a distance of 4,057~$\pm$~5~km, with its pole orientation given by $\alpha_{QR}$: 259.82~$\pm$~0.23$^\circ$ and $\delta_{QR}$: 53.45~$\pm$~0.3$^\circ$ \citep{Pereira_2023}, and the inner (QR2) at 2,520~$\pm$~20~km, with the same pole. There is a small discrepancy between the orientation of Quaoar's ring system \citep{Pereira_2023} and Weywot's orbit \citep{Ribas_2025}, causing an inclination between the two orbital planes of $4.8 \pm 1.4^{\circ}$. 

{\color{black} There is no current evidence indicating that Weywot significantly interferes with the stability of Quaoar's rings. Therefore, its orbital inclination is consistent with current ring models \citep{Rodriguez_2023}. Nevertheless, studies by the authors of this paper are currently underway to verify if Weywot and it's orbital characteristics are related to Quaoar's 3D shape, and whether it could influence the object's figure, potentially leading to a specific triaxial configuration \citep{margoti2024determinaccao}.}
Ring systems are generally expected to orbit along the equatorial plane of their central bodies; therefore, Quaoar and its rings are assumed to share the same pole orientation. We used this assumption to derive Quaoar's three-dimensional shape in this work.  

For objects the size of Quaoar, it is expected that they presented an atmosphere at some point in their formation history \citep{Sicardy_2023}. Stellar occultation light curves have been used 
%\st{to determine the current atmospheric limits for Quaoar} 
{\color{black}to determine the current 1$\sigma $ atmospheric limits for Quaoar}, which are: 21~nbar \citep{Ribas_2013}, 6~nbar \citep{Arimatsu_2019}, 85~nbar \citep{morgado2022}, and {\color{black}0.}2~nbar \citep{proudfoot2025}.

Specifically, using the measured chords, we investigated the dimensions of Quaoar assuming it to be an oblate spheroid, assessed whether it could be in Maclaurin hydrostatic equilibrium, and compared our results with those obtained by \citet{Ribas_2013}. For the latter purpose, a rotation period is also necessary. \citet{Ortiz_2003} proposed two rotation periods for Quaoar: 8.8394~$\pm$~0.0002~hours if Quaoar's rotational light curve is single-peaked (implying an oblate shape), and 17.6788~$\pm$~0.0004~hours if it is double-peaked (indicating a triaxial shape). \citet{Kiss_2024} also proposed similar rotation periods: 8.8364~$\pm$~0.0031~hours for an oblate shape, and 17.6728~$\pm$~0.0062~hours for a triaxial shape. The main issue with adopting an oblate shape is that the entire amplitude of Quaoar's rotational light curve would then have to be explained solely by albedo variegation, which has amplitudes of 0.133~$\pm$~0.028~mag according to \citet{Ortiz_2003}, and 0.154~$\pm$~0.040~mag according to \citet{Kiss_2024}.

This paper presents the results of the analysis of new stellar occultations by the large Trans-Neptunian Object (50000) Quaoar, registered between 2018 and 2025. Following this introduction, Section \ref{sec:Predictions and campaigns} details the predictions and observational campaigns that provided the data for this work. Section \ref{sec:Analysis} describes the analysis of the data, including the occultation light curves, the fitting of an oblate ellipsoid model to determine Quaoar's shape, and we establish a limit on Quaoar's atmosphere. %based on the Gemini Z' light curve. 
Finally, in Section \ref{sec: Conclusion}, we present our conclusions, discussing the implications of our findings for Quaoar's density and its potential to be classified as a dwarf planet. The Appendix provides additional details regarding the stellar information (Appendix \ref{sec:Star informations}), observer information (Appendix \ref{sec:Observers informations}), figures of the light curves (Appendix \ref{sec:Light curves}) and new astrometric position for Quaoar (Appendix \ref{sec:New astrometric position}).

\section{\textbf{Predictions and campaigns}}\label{sec:Predictions and campaigns}

Predicting a stellar occultation requires the sky position of the occulting body over time (its ephemeris) and the occulted star’s apparent position (as seen by a given observer). Both should be known to within a few tens of milliarcseconds (\textit{mas}) \citep{Sicardy_2024}, on the order of the object’s own angular size in the sky (around 50~mas in the case of Quaoar), for having a precise prediction \citep{Ribas_2013}.

In this work, we present 28 new stellar occultations by Quaoar that occurred between 2018 and 2025. These events were predicted within the Lucky-Star Collaboration\footnote{Lucky-Star Collaboration: \url{https://lesia.obspm.fr/lucky-star/team.php}}. The Lucky-Star Collaboration aims to provide reliable stellar occultation predictions to enhance our understanding of solar system objects. To achieve this, the ephemerides of a select list of TNOs, Centaurs, and Trojans are frequently updated using \textit{NIMA} integrator \citep{Desmars_2015} with astrometric data from recurring observations, obtained from our own observations and analysis \citep{Assafin2023b}; information is publicly available on the Lucky-Star website\footnote{Website: \url{https://lesia.obspm.fr/lucky-star/nima.php}}. Moreover, occultation events themselves produce astrometric positions with uncertainties on the order of fractions of \textit{mas}, which are also used to refine the ephemerides \citep{Rommel_2020}. All events analyzed in this work were predicted using stars from the GAIA DR2 and DR3 catalogs \citep{Gaia_2018, Gaia_2021}. Details on the occultation stars' information, along with their propagated positions to the event date, are shown in the Appendix \ref{sec:Star informations}, Table~\ref{tab:stars}. %The position of GAIA's star has uncertainties at the sub‑mas level, which has enabled a much larger number of occultation predictions; today, the dominant source of prediction error is the uncertainty in the object’s ephemeris \citep{Ribas_2019}.

The occultation data were acquired using a wide range of telescopes, from those with large apertures, such as Gran~Telescopio~Canarias~(GTC,~10.4~m), the Gemini~North~(8.0 m), and the Southern~Astrophysical~Research~Telescope~(SOAR,~4.1~m), 
%to apertures as small as 20–30 centimeters, being the vast majority from citizen scientists around the world 
to apertures smaller than 60 centimeters. The majority of observations were carried out by small observatories and citizen scientists around the world (e.g., IOTA members\footnote{International Occultation Timing Association}).
%apertures as small as 20–30 centimeters to professional telescopes with apertures as large as 10 meters. Observations here reported were observed with the Gran Telescópio Canarias (GTC, 10.4~m) \citep{Morgado_2023}, the Genini North (8.0 m), and the Southern Astrophysical Research Telescope (SOAR, 4.1~m).
%, three at Pico dos Dias Observatory (OPD, 1.6~m), three with the Transiting Planets and Planetesimals Small Telescope (Tappist S, 0.6~m), and two with Trappist North (0.6~m). Moreover, **theless, the vast majority come from citizen scientists around the world (e.g., IOTA members\footnote{International Occultation Timing Association}), most of whom used telescopes with apertures between 10 and 50 centimeters. 
In addition to the diverse range of telescopes, a wide array of instruments was also used, such as the HiPERCAM \citep{Dhillon_2021} at GTC and SPARC4 \citep{Bernardes_2025} at Observatório~Pico~dos~Dias~(OPD), which enabled simultaneous image acquisition in multiple channels, to CMOS and video detectors such as the QHY174M-GPS and Watec 910BD models, on mobile telescopes.

%All stellar occultation events, with locations, telescope, and camera details, in which Quaoar's shadow was detected, are summarized in Table~\ref{tab:ev list}.  

All stellar occultation events for which Quaoar's shadow was detected are summarized in Table~\ref{tab:ev list}. Details on the locations, telescope, camera, exposure time, and cycle are provided in the Appendix~\ref{sec:Observers informations}, Tables~\ref{tab:obs_data_pt1}~and~\ref{tab:obs_data_pt2}.
%\st{, with Table}~\ref{tab:obs_data_pt1}\st{ specifically including the timing data for the positive occultation chords}
A subset of events the events in Table~\ref{tab:ev list} overlapped with this work has been published previously, but those studies focused exclusively on Quaoar’s ring or Weywot detections rather than the primary body itself. 
%The Appendix \ref{sec:Observers informations} presents the positive chords of this work in Table~\ref{tab:observacoes pos}, and negative chords, technical failure, or weather-related issues are included in Table~\ref{tab:observacoes neg}. 
%Finaly, the plot of each new light curve is shown in Appendix \ref{sec:Light curves}, Figures~\ref{fig:2018,1019,2020,2022}~to~\ref{fig:2024,2025}.

Since 2022, the Lucky Star's Occultation Portal \citep{Kilic_2022} has been used to store the occultation data, including events by Quaoar. This has greatly streamlined the organization of observer information, data backups, distribution, and long-term studies like this work. Therefore, from 2022 on, the reports and data related to Quaoar occultations were collected using the Occultation Portal\footnote{Occultation Portal: \url{https://opop.obspm.fr/}}.
%, as previously, only the campaign PI maintained the data on a personal computer, with backups made on institutional computers. 

\begin{deluxetable}{lcc}
\startlongtable
\tablecaption{Summary of Stellar Occultation Events by Quaoar. The columns are as follows: Event (the date of the occultation); P: Positive (the number of positive chords detected), N: Negative (the number of negative chords), and O: Other (observations that did not yield conclusive data due to technical issues, poor weather, or inconclusive photometry); and reference. The symbols are used to indicate specific types of detections: an star ($\star$) represents a QR1 detection; two stars ($\star\star$) indicate QR1 and QR2 detections; a bullet point ($\bullet$) symbolizes a positive chord observed by a space telescope; a dagger ($\dagger$) marks the detection of both Quaoar and Weywot; and a double dagger $\ddagger$ marks the detection of a possible double star.\label{tab:ev list}}
\tablewidth{0pt}
\tablehead{
\colhead{Event} & \colhead{P / N / O} & \colhead{Reference} 
}
\startdata
2011-02-11                      & 2 / 0 / 0  & \makecell{\citet{Person_2011}\\This Work} \\
2011-05-04                      & 6 / 9 / 1  & \citet{Ribas_2013} \\
2012-02-17                      & 4 / 0 / 0  & \citet{Ribas_2013} \\
2012-10-15                      & 1 / 0 / 3  & \citet{Ribas_2013} \\
2013-07-09                      & 1 / 0 / 0  & \citet{Rommel_2020} \\
2018-07-26                      & 4 / 0 / 0  & This Work \\
2018-09-02$\star\ddagger$               & 1 / 0 / 0  & \makecell{\citet{Morgado_2023}\\This Work} \\
2019-03-27                      & 1 / 0 / 0  & This Work  \\
2019-04-28                      & 2 / 0 / 0  & This Work  \\
2019-05-28                      & 4 / 0 / 10 & This Work  \\
2019-06-05                      & 5 / 0 / 0  & \makecell{\citet{Morgado_2023}\\This Work} \\
2019-06-28                      & 1 / 0 / 3  & \citet{Arimatsu_2019}  \\
2019-08-04$\dagger$             & 1 / 0 / 0  & \makecell{\citet{Ribas_2025}\\This Work}  \\
2019-09-26                      & 4 / 0 / 0  & This Work  \\
2019-10-16                      & 3 / 0 / 0  & This Work  \\
2020-06-11$\star \bullet$       & 1 / 1 / 0  & \citet{morgado2022}\\
2020-06-16                      & 4 / 0 / 0  & This Work  \\
2020-07-01                      & 1 / 1 / 0  & This Work  \\
2022-06-24                      & 1 / 0 / 0  & This Work  \\
2022-08-09$\star \star$         & 10/ 1 / 19 & \citet{Pereira_2023}  \\
2023-05-13                      & 4 / 1 / 10 & This Work  \\
2023-05-20$\star$               & 1 / 0 / 0  & This Work  \\
2023-05-24$\star$               & 1 / 4 / 0  & This Work  \\
2023-05-26$\dagger$             & 4 / 4 / 2  & \makecell{\citet{Ribas_2025}\\This Work}  \\
2023-07-15                      & 7 / 0 / 13 & This Work  \\
2023-08-01                      & 4 / 0 / 1  & This Work  \\
2023-08-24                      & 5 / 0 / 5  & This Work  \\
2024-04-10$\star \star$         & 5 / 0 / 3  & This Work  \\
2024-05-29                      & 1 / 0 / 0  & This Work  \\
2024-07-04$\star$               & 7 / 1 / 3  & This Work  \\
2024-08-28$\star \star \bullet$ & 1 / 0 / 0  & \citet{proudfoot2025} \\
2025-06-12$\star$               & 3 / 2 / 1  & This Work  \\
2025-06-20                      & 6 / 0 / 7  & This Work  \\
2025-07-04$\star$               & 1 / 1 / 4  & This Work  \\
2025-08-28$\ddagger$            & 2 / 1 / 1  & This Work  \\
2025-08-31                      & 2 / 0 / 1  & This Work  \\
\hline
  \enddata
\end{deluxetable}

\section{\textbf{Analysis}}\label{sec:Analysis}

This section is dedicated to explaining the methodology utilized to analyze the data obtained from the stellar occultations. Specifically, Section \ref{sec:occs} details the procedures for obtaining the occultation light curves, Section \ref{sec:fit} covers the fitting of the oblate model, and Section \ref{sec:atm} describes our atmosphere fitting analysis.

\subsection{Occultation light curves}\label{sec:occs}

The variety file formats was as diverse as the variety of telescopes and cameras. The most common format was \texttt{FITS} \citep{Hanisch_2001}, although video sequences in formats such as \texttt{AVI}, \texttt{SER}\footnote{Documentation: \url{https://siril.readthedocs.io/en/latest/file-formats/SER.html}}, \texttt{AAV}\footnote{Documentation: \url{http://www.hristopavlov.net/OccuRec/AavFormat.html}}, \texttt{ADV} \citep{Pavlov_2020} and \texttt{H5}\footnote{Documentation: \url{https://support.hdfgroup.org/documentation/hdf5/latest/}} were also present. When images were supplied in a format other than \texttt{FITS}, they were first converted to \texttt{FITS} using custom Python software based on the \texttt{numpy} \citep{numpy_2020} and \texttt{astropy} \citep{Astropy_2022} libraries.

When calibration images were provided, bias (or darks) and flats, the science images were calibrated accordingly. In some cases, additional calibration was necessary by dividing each pixel by the mean of its corresponding row or column. This step was required because cameras such as the Watec and QHY174M-GPS, and colour sensors, often exhibit line-by-line (or column-by-column) noise that varies from frame to frame, and normalizing each pixel by the average of its row (or column) effectively suppressed this systematic effect.

For video files like \texttt{AVI}, acquired from analog cameras, images are recorded with a fixed minimum exposure time of 0.040~s, which corresponds to the exposure time of an individual extracted frame in Phase Alternating Line (PAL) mode. Consequently, the science images had to be stacked to achieve the actual exposure time, which is a multiple of this minimum value. In addition, video acquisitions are typically accompanied by timestamps written in the image, which requires extracting these timestamps for each frame. 

Furthermore, cameras like the Watec and Raptor require specific time corrections \citep{Barry_2015}. For the Watec cameras, the corrections are indicated in publicly avaliable tables\footnote{Website: \url{http://www.dangl.at/ausruest/vid\_tim/vid\_tim1.htm\#wat\_910bd}}. The Raptor Merlin camera records, on its header, the computer time at 1.5 exposure intervals after the actual image's mid-acquisition time. All these corrections were already applied in the event instants shown in Table \ref{tab:obs_data_pt2}.

{\color{black} The vast majority of the instruments used in the Quaoar observation campaigns were equipped with GPS antennas for time calibration. In the case of Raptor cameras, the acquisition start time is retrieved from the control computer and recorded in the FITS headers, while QHY cameras retrieve timing directly from an internal GPS. Observers who did not have access to GPS-based acquisition utilized the Network Time Protocol (NTP) most suitable for their observatory. Although this protocol is less reliable than a dedicated GPS antenna, such chords were allowed to undergo time shifts to align them with the GPS-timed chords. Since a regular shape for Quaoar is a physical prior, the alignment of these chords is a natural consequence of the modeling process. In our campaigns, only one observatory utilized neither a GPS antenna nor an NTP protocol. This specific chord, indicated in Table \ref{tab:obs_data_pt2}, exhibited a significant initial offset; therefore, a random offset was applied to center it with the remaining chords.}

{\color{black}Combining occultation data from dozens of distinct observational setups naturally introduces non-trivial systematic timing uncertainties and cross-calibration issues. While internal random errors are derived directly from light curve pipeline fittings, the absolute time anchoring depends heavily on the synchronization method utilized, such as dedicated GPS, various NTP protocols, or internal computer clocks. To mitigate these potential cross-calibration offsets, chords backed by absolute GPS timing were used as anchors for each specific event. Chords obtained via less reliable protocols (e.g., NTP) or those presenting known systematic instrumentation delays were allowed to undergo time shifts. This alignment procedure is physically justified by adopting a regular shape for Quaoar as a physical prior for individual single-epoch projections. Consequently, while individual instrumentation combinations present systematic footprints, our cross-calibration framework absorbs these misaligned chords before feeding the dataset into the global shape optimization.}

{\color{black} This multi-epoch dataset is strongly heteroskedastic, combining observations from telescopes ranging in aperture from 10.4 m down to 30 mm. To prevent individual, highly precise chords from dominating the global solution, and to account for residual cross-calibration timing offsets across distinct equipment, our optimization relies on the systematic uncertainty floor $\sigma_{model} = 2\text{ km}$ defined in Equation \ref{eq:chi2}. This parameter acts as an empirical weighting factor across the network: by enforcing a reduced $\chi^2_{pdf}$ close to unity, the MCMC exploration naturally inflates the posterior probability contours. Consequently, the derived parameter uncertainties (Table \ref{tab:results}) effectively incorporate both the random noise of individual chords and the underlying instrumentation cross-calibration scatter.}

Differential aperture photometry of each dataset was primarily obtained with the Platform for Reduction of Astronomical Images Automatically \texttt{PRAIA} \citep{Assafin_2023}. For datasets requiring extra care due to issues like undersampling of the target star (i.e., SNR less than 1.5), we employed our custom Python code. This code automatically detects the field and uses a genetic algorithm to determine the optimal centroid and aperture for each star in the photometry, particularly for the occultation target. It is based on \texttt{numpy} and \texttt{astropy}, and also utilizes \texttt{astrometry.net}, \texttt{scikit-learn} \citep{Scikitlearn_2011}, and \texttt{deap} \citep{deap_2012}. 

All photometric data for the positive chords obtained in this work are organized by year in the Appendix~\ref{sec:Light curves},~Figures~\ref{fig:2018,1019,2020,2022}~to~\ref{fig:2024,2025}. The occultation timings for each of these light curves were determined using our custom reduction pipeline based on the Python library \texttt{SORA} \citep{Gomes_2022}; the timings for each chord are listed in Table \ref{tab:obs_data_pt2}. Finally, the projection of the chords onto the plane of the sky was also performed with \texttt{SORA} using Quaoar ephemerides \texttt{NIMAv19}\footnote{The ephemeris is available at the following link: \url{https://lesia.obspm.fr/lucky-star/obj.php?p=1177}} and planetary ephemeris \texttt{DE442}, the chords are shown in Figure \ref{fig:oblato}.

\subsection{Oblate fitting and uncertainty determination}\label{sec:fit}

Considering the predicted diameter for Quaoar and its composition of rock and ice \citep{Ribas_2013}, it is reasonable to assume that it is in hydrostatic equilibrium, and the simplest equilibrium figure is that of an oblate spheroid \citep{Tancredi_2008}.

Since 2011, Quaoar has been the target of 36 successful stellar occultations. %Ten of these events have already been published (\citet{Person_2011, Ribas_2013, Arimatsu_2019, Rommel_2020, morgado2022}, and \citet{Pereira_2023}), and their timings have been incorporated into our final dataset. 
Using these occultations, we aimed to employ an oblate model to represent all the observed chords.

The fitting of the oblate model to the observational data was performed through a bayesian statistical inference approach, using the Markov Chain Monte Carlo (MCMC) method. The goal was not only to find the best-fit set of parameters but also to map the posterior probability distribution to characterize the uncertainties and correlations among all the model's variables.

The geometric model of the oblate ellipsoid has four global parameters: the equatorial semi-axes $a=b$, the rotational polar-axis $c$, and the orientation, described by the right ascension ($\alpha_Q$) and declination ($\delta_Q$) of its rotation axis. In addition, the model was fitted to the 36 data sets from the different occultations, requiring 70 additional parameters related to the projected center coordinates of the ellipsoid ($f_i$, $g_i$, for i=1,...,36). This resulted in a total of 76 parameters to be determined.

The quality of the fit for each occultation set was quantified by a chi-squared ($\chi^2$) test. The final $\chi^2$ was obtained by summing the individual $\chi^2$ values from each occultation, according to the equation:
\begin{equation}\label{eq:chi2}
\chi^2 = \sum_{i=1}^{N} \frac{(r_i - r'_i)^2}{\sigma_i^2 + \sigma_{\text{model}}^2},    
\end{equation}
where $r_i$ is the radial distance of the $i$th observed chord extremity, $r'_i$ is the radial distance of the ellipsoid shadow limb, $\sigma_i$ is the observed radial uncertainty at each extremity, and $\sigma_{\text{model}}$ represents the uncertainty in our ellipsoidal model relative to Quaoar’s 3D shape (for example, due to topographic features \citep{Rommel_2025}). We iteratively estimated $\sigma_{\text{model}}$ as 2 km, ensuring that $\chi^2$ per degree of freedom remains close to 1.

{\color{black}The uncertainties quoted for our final solutions, presented in Table \ref{tab:results}, provide a robust assessment of our model constrained to an oblate object. Combining 14 years of multi-epoch data across distinct geometries and variety of equipment requires accounting for localized topographic features or systemic discrepancies, especially in the chord timings, which are critical to shape determination. This population of possible true variations is encapsulated in our framework via the introduction of the $\sigma_{model} = 2\text{ km}$ parameter in Equation \ref{eq:chi2}. Rather than enforcing an uncalibrated weight on high-dispersion or over-precise chords, this baseline systematic error acts as an absolute uncertainty floor applied in the direction of every single chord extremity. The physical magnitude of this floor ($2\text{ km}$) was systematically cross-evaluated against expected global topography bounds derived from \citet{Johnson_1973} using the results from \citet{Ribas_2013}. By ensuring our final MCMC joint probability exploration yields a minimum $\chi^2_{pdf} = 0.998$ close to unity, the calculated marginal distributions inherently incorporate both the statistical observational noise and the underlying systematic profile dispersion, providing representative true uncertainty values for the oblate model.}

For the analysis, we first assumed that Quaoar’s pole orientation is aligned with that of its rings ($\alpha_Q = \alpha_{QR} = 259.82 \pm 0.23^{\circ}$ and $\delta_Q = \delta_{QR} = 53.45 \pm 0.30^{\circ}$ \cite{Pereira_2023}), which helped constrain the range of possible solutions. By modeling the object as an oblate spheroid with no pole precession, we ensure the pole's orientation remains constant. The only variation in the projection of its limb, therefore, is caused by the changing aspect angle, which is unrelated to the object's rotational phase. This allows us to disregard the object's rotation period and initial phase.
%, removing two variables from the model. 

To optimize the fitting process and deal with the dependent nature of the local parameters ($f_i$, $g_i$), a profile likelihood analysis method was employed. For each set of global parameters ($a$, $c$, $\alpha_{QR}$ and $\delta_{QR}$) sampled by the MCMC, the optimal values of the projected centers ($f_i$, $g_i$) were determined by a minimization for each of the 36 dates. This procedure ensures that the $\chi^2$ evaluation for each point in the global parameter space represents the best possible fit, incorporating the uncertainties of the local parameters into the uncertainty analysis of the global parameters.

The MCMC algorithm was used to explore the 76-dimensional space and generate a long chain of samples (points), whose density is proportional to the posterior probability function.

The uncertainties for the parameters were obtained from this posterior probability distribution. The marginal probability distributions for each parameter were constructed from histograms of the sample chain. The best-fit value for each variable was considered the median of its respective distribution, and the 1$\sigma$ uncertainties (68.3\% confidence interval) were determined by the 16th and 84th percentiles of the distribution. 

\begin{figure}[htb!]
    \centering
    \includegraphics{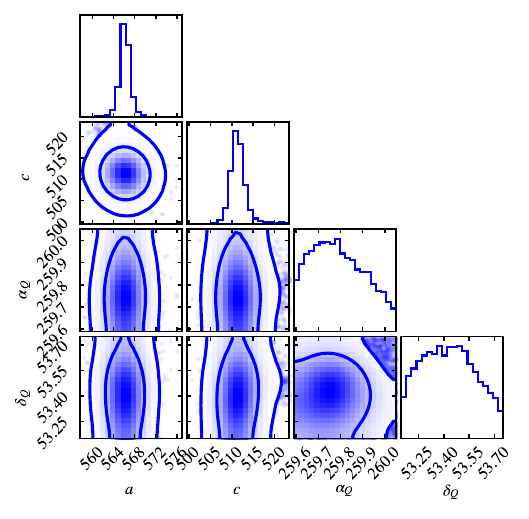}
    \caption{Corner plot showing the posterior probability distributions and correlations for the oblate ellipsoid model parameters. The samples used to generate this plot were obtained from an MCMC simulation and filtered to a $\chi^2$ below $\chi^2_{min}+100$. The histograms on the diagonal show the marginal probability distributions for each parameter. The off-diagonal plots show the joint probability distributions for each pair of parameters, revealing their correlations, and the level curves show the 1-$\sigma$ and 3-$\sigma$ regions. The fit is based on all the occultations.}
    \label{fig:corner plot}
\end{figure}

The correlations among the parameters and the shape of the marginal and joint probability distributions can be visualized in the corner plot, figure \ref{fig:corner plot}, which displays all the histograms and two-dimensional projections of the posterior probability distribution.

\begin{figure*}[htb!]
    \centering
    \includegraphics{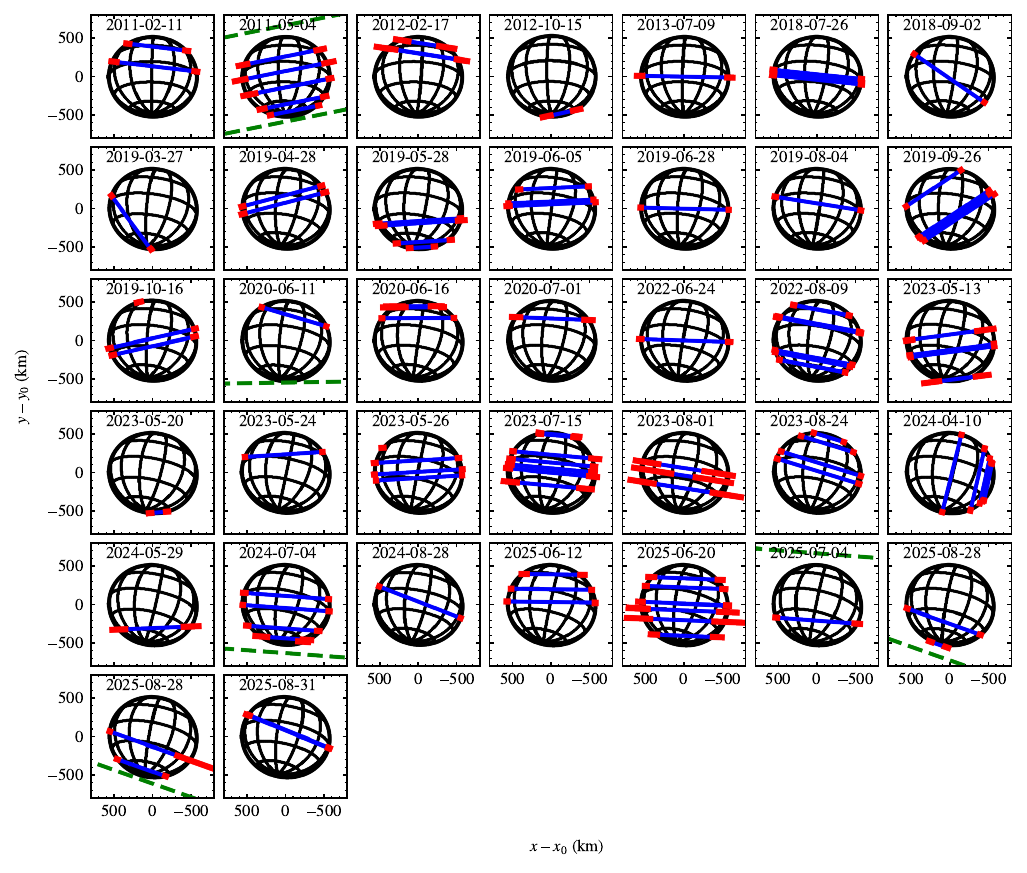}
    \caption{Best fit of the oblate ellipsoidal model: $a = b = 566.1$~km, $c=511.2$~km, {\color{black}$\alpha_{Q} = 259.7${\color{black}${^{\circ}}$}} and $\delta_{Q} = 53.4$${^{\circ}}$. This model does not account for the rotation period or initial phase. The blue lines are the observed positive chords with their associated uncertainties, shown as red segments. The green dashed lines are the negative chords used to constrain the model. The 2025-08-28 occultation shows duplicated frames due to a dual detection of Quaoar, suggesting a possible duplicity in the target star. The fit where the star was too far from zero ephemeris was excluded from both the ellipsoid fitting and the determination of Quaoar's astrometric position.}
    \label{fig:oblato}
\end{figure*}

This resulted the semi-axis and pole-position, with the respective uncertainties, which best represent the 36 observed events, being: $a=b=566.1^{+2.5}_{-2.2}$~km, $c=511.2^{+3.6}_{-3.7}$~km, 
%\st{$\alpha_{Q} = 257.7^{+0.2}_{-0.2}$~${^{\circ}}$}
{\color{black} $\alpha_{Q} = 259.7^{+0.2}_{-0.2}$~${^{\circ}}$} and $\delta_{Q} = 53.4^{+0.2}_{-0.2}$~${^{\circ}}$. The best fit ($\chi^2_{min}$) was used to present the model along with the detected chords in Figure \ref{fig:oblato}. {\color{black} We note that the posterior distribution for the pole orientation mirrors the adopted prior. Because the occultation chords alone do not independently constrain the rotational axis, the coordinates $\alpha_Q$ and $\delta_Q$ are treated throughout this work as fixed model inputs based on \citet{Pereira_2023}, rather than newly derived measurements.}

%\st{Therefore, using the ($f_i$, $g_i$) values derived from each occultation, we were able to determine Quaoar's astrometric position by calculating the difference between the occultation center and the ephemeris center ($f_i=0$, $g_i=0$). This difference provides the offset correction, which is then applied to the star's known position. The resulting astrometric positions were corrected for Solar, Jupiter and Saturn gravitational deflection and are presented in Appendix C, Table 4}

{\color{black}The remaining fitted parameters correspond to the projected center of the ellipsoidal model on the tangent plane associated with each occultation event. Since the origin of each tangent plane is defined by the ephemeris position of Quaoar, the fitted center provides the offset relative to the ephemeris at the corresponding occultation epoch. This offset was further corrected for gravitational light deflection due to the Sun, Jupiter, and Saturn, treating the correction as a rigid translation of the fitted occultation geometry. The resulting offset was then applied to the ephemeris position of Quaoar, obtaining the astrometric positions, presented in Appendix \ref{sec:New astrometric position}, Table \ref{tab:astrometric position}}

It was observed that, on certain dates, the data did not constrain the model, resulting in an individual $\chi^2$ close to zero. In these cases, the MCMC algorithm sampled the center parameters ($f_i$, $g_i$) uniformly, indicating that the data were not sensitive to them. The conclusion is that their values are consistent with the prior range used in the simulation. 

{\color{black}The ability of an individual event to constrain the global shape model depends on its chord coverage and the resulting balance in degrees of freedom. For each event, adjusting the center coordinates introduces two local free parameters ($f_i, g_i$), meaning that a successful constraint requires the observational geometric inputs to outweigh these added variables.} 

{\color{black}Multi-chord events with limited or asymmetric distribution may fail to restrict the model. Conversely, some few-chord events can provide significant constraints. For instance, the two-chord event of 2019-10-16, which consisted of a central chord and a northern chord, effectively restricted the global solution because its four limb points easily countered the two local center variables.} 

{\color{black}For single-chord events, providing a constraint is generally more challenging since the two extremity points match the two local degrees of freedom. However, central single chords can constrain the model if their measured lengths exceed the projected semi-minor axis $b^{\prime}$, establishing a strict minimum size for the body. This is the case for the single-chord events of 2013-07-09, 2018-09-02, 2019-06-28, and 2019-08-04. On the other hand, single chords crossing far to the north or south of the body do not impose this size boundary; the center optimization can freely shift them, driving their individual $\chi^2$ close to zero without bounding the global parameters (e.g., the events of 2019-03-27, 2020-06-16, 2020-07-01, 2023-05-20, and 2023-05-24).} 

{\color{black}A unique case is the single-chord event of 2020-06-11. Although the positive chord is short, it is closely bounded by a negative chord. If Quaoar's limb were larger than a specific threshold, it would intersect the negative chord's path, invalidating the model. Therefore, the combination of this tight negative constraint makes the 2020-06-11 event highly effective in bounding the shape parameters.}

Using only the events that constrain the model, we are left with a total of 23 occultations {\color{black}(2011-05-04, 2013-07-09, 2018-07-26, 2018-09-02, 2019-04-28, 2019-05-28, 2019-06-05, 2019-06-28, 2019-08-04, 2019-09-26, 2019-10-16, 2020-06-11, 2022-06-24, 2022-08-09, 2023-05-13, 2023-05-26, 2023-07-15, 2023-08-24, 2024-04-10, 2024-07-04, 2024-08-28, 2025-06-12, 2025-06-20)}, totalizing 90 chords, which means 180 points on Quaoar's limb. Our minimum $\chi^2$, considering a sigma model of 2~km, was 179.6 for our MCMC fit. Therefore, the minimum $\chi^2$ per degree of freedom for this fit was 0.998.

\subsection{Atmosphere}\label{sec:atm}

We use the occultation light curves obtained during the August 9, 2022 occultation at Gemini North (z' band), which is our curve with the lowest dispersion and highest spatial resolution, to derive an upper limit for a global atmosphere around Quaoar. Using a direct ray tracing code, we generate synthetic light curves, following the protocol presented in \cite{dias2015}. We assume a CH$_4$ atmosphere and we use the same thermal profile $T(r)$ as \cite{Ribas_2013}, where $r$ is the distance to Quaoar's center. The $T(r)$ profile starts with a surface temperature of $T(r_{\rm surf})=42$~K and a thermal gradient of  $dT/dz(r_{\rm surf})=5.7$~K~km$^{-1}$, which ramps up to an isothermal branch at 102~K at an altitude of about $z=10$~km, see \cite{morgado2022}. We fit synthetic light curves with various surface pressure $p_{\rm surf}$ to derive a $\chi^2(p_{\rm surf})$ function. 

We have applied this procedure to the three data sets obtained at Gemini (r' and z' bands) and at CFHT (Ks band). Concerning Gemini, the light curve obtained in the z' band has about twice the SNR obtained in the r' band. As a consequence, the latter does not bring significant improvement when compared to the z' band. On the other hand, the CFHT light curve has an SNR comparable to that of the Gemini z' band. However, diffraction effects limit the detection of an atmosphere in the immediate vicinity of the surface. Fresnel fringes are present over a typical distance of a few times $\lambda_{\rm F}= \sqrt{\lambda \Delta/2}$, where $\lambda$ is the wavelength of observation, $\Delta=42.0$~AU is the Quaoar geocentric distance in August 2022, and  $\lambda_{\rm F}$ is the Fresnel scale. For the r', z' and Ks bands (0.620, 0.947, 2.15 $\mu$m, respectively), we have $\lambda_{\rm F}=1.4, 1.7$ and 2.6~km, respectively.

\begin{figure*}
\centerline{%
\includegraphics[totalheight=70mm, trim=0 4.5cm 0 4.5cm, clip]{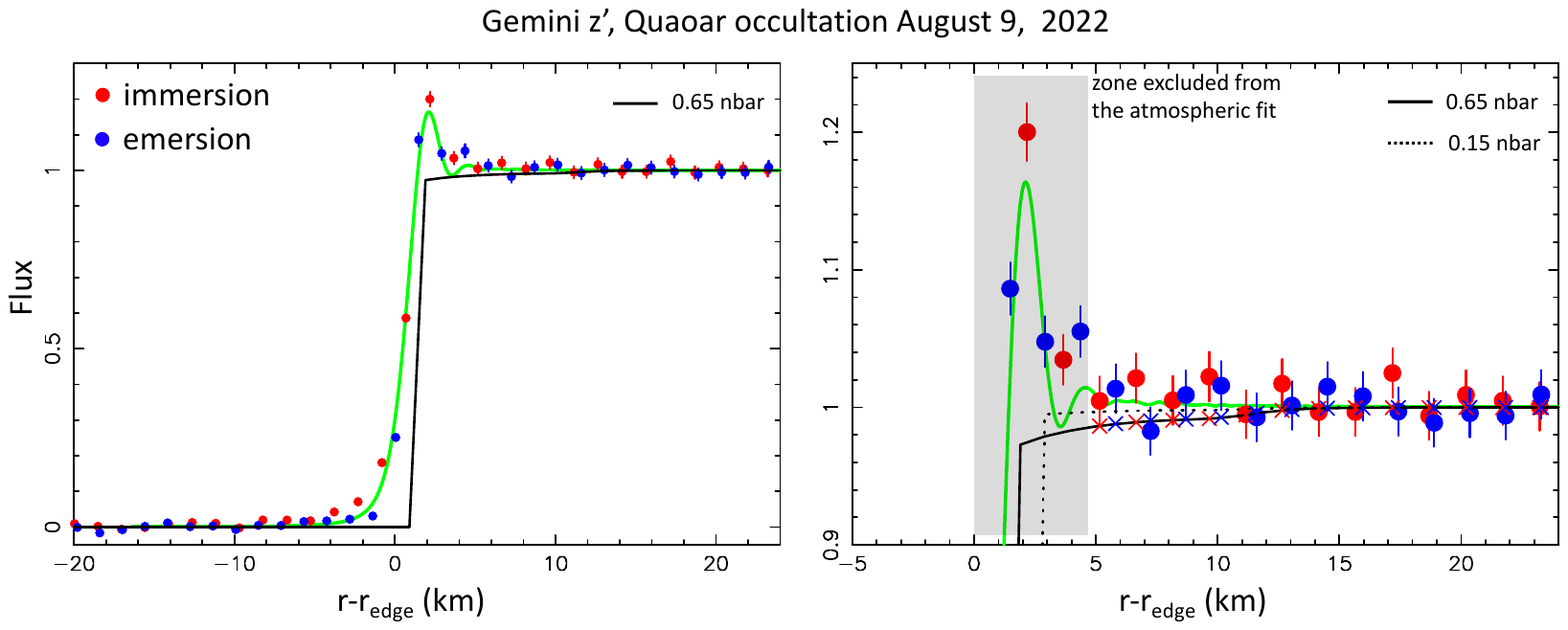}
}%
\caption{%
\textit{Left panel}: the immersion (red) and emersion (blue) data points obtained at the the Gemini telescope (z' band) are plotted against their perpendicular distance to the local edge, $r-r_{\rm edge}$, where $r$ is the radial distance to Quaoar's center, known with a precision of 5.1~km, projected in the sky plane and $r_{\rm edge}$ is the fitted values of each edge radius. 
The green curve is the best-fitting Fresnel diffraction pattern accounting for the wavelength ($\lambda=0.947$~$\mu$m), bandwidth ($\Delta \lambda=0.236$~$\mu$m), stellar diameter (1.33~km), and finite integration time (1.5~km when projected in the sky plane). Although the diffraction model basically captures the presence of the main Fresnel fringe, some disagreement occurs , possibly due to unaccounted properties of the stellar disk profile or local topographic features. {\color{black}This means that in practice the data points within 5 km from the edge are not used in the fit. Also, the ray tracing program that generates the synthetic curves assumes purely geometric optics, i.e. ignores diffraction effects. The black curve is a synthetic light curve accounting for a CH$_4$ atmosphere with surface pressure $p_{\rm surf} = 0.65$~nbar, our 3$\sigma$ upper limit for such atmosphere. \textit{Right panel}: An enlarged version of the left panel, where the dotted line is a synthetic light curve for a CH$_4$ atmosphere with surface pressure $p_{\rm surf} = 0.15$~nbar, our 1$\sigma$ upper limit for an atmosphere. The crosses indicate the points of the model that are used to calculate the $\chi^2$ value of the fits.}}
\label{fig_fit_gemini_z}
\end{figure*}

Atmospheres with sub-nanobar surface pressure have detectable effects on the light curve below altitudes of $\sim$10 km (Fig.~\ref{fig_fit_gemini_z}). Thus, diffraction competes with atmospheric effects in this region. In principle, it is possible to disentangle the effects of diffraction and an atmosphere \citep{French1976,sicardy2025}. In practice, however, we found that small changes in the modeling of the stellar disk luminosity profile or of possible local topographic features prevent the derivation of a Fresnel fringe pattern that is accurate enough to be disentangled from the small effects caused by a sub-nanobar atmosphere.

The CFHT light curve was obtained in the Ks band (2.15~$\mu$m), yielding the largest value of $\lambda_{\rm F}$ among the three light curves considered in this work. Consequently, most of the useful part of the CFHT light curve is dominated by diffraction effects and cannot be used for constraining the presence of an atmosphere. Consequently, only the Gemini z' light curve is used in this work to derive an atmospheric upper limit. 

The Fig.~\ref{fig_fit_gemini_z} presents our results. The immersion and emersion data points have been compared with models using various surface pressures. We found that the minimum value $\chi^2_{\rm min}$ of $\chi^2$ is obtained for $p_{\rm surf}=0$~nbar, meaning that no atmosphere is detected. Applying the $\chi^2_{\rm min}+1$ and $\chi^2_{\rm min}+9$ criteria, we obtain upper limits of $p_{\rm surf}=\text{{\color{black}0.15}}$~nbar (1$\sigma$) and  $p_{\rm surf}={\color{black}0.65}$~nbar (3$\sigma$).
This result improves the 1$\sigma$ upper limits obtained from other stellar occultations by Quaoar. 

{\color{black}The same exercise assuming a CO atmosphere provides upper limits of $p_{\rm surf} = 0.2$~nbar (1$\sigma$) and $p_{\rm surf} = 0.85$~nbar (3$\sigma$). Assuming a N$_2$, we obtain respective upper limits of $p_{\rm surf} = 0.2$~nbar and $p_{\rm surf} = 0.95$~nbar.}
%21~nbar \citep{Ribas_2013},
%6~nbar  \citep{Arimatsu_2019},
%85~nbar \citep{morgado2022} and
%2~nbar \citep{proudfoot2025}.

\section{\textbf{Conclusions}}\label{sec: Conclusion}

We present here a comprehensive analysis of 14 years of observations of stellar occultations by the large TNO (50000) Quaoar, obtained on 36 different occultation campaigns, which resulted in 107 chords. We implemented a method, using Markov Chain Monte Carlo (MCMC), to fit 76 parameters, assuming an oblate spheroid that shares its rotational pole with its rings, and all parameters obtained for Quaoar using this model are summarized in Table \ref{tab:results}.

\begin{table}[h!]
    \caption{Summary results. The J2 and J4 were obtained using \citep{Rossi_1999} equations. {\color{black}All the values in this table are in 1-$\sigma$ confidence level.}}
    \centering
    \begin{tabular}{lc}
        \hline \hline
        Variable & Result\\
        \hline
        Equatorial semi-axis (a=b)     & $566.1^{+2.5}_{-2.2}$~km           \\
        Polar semi-axis (c)            & $511.2^{+3.6}_{-3.7}$~km           \\
        Oblateness               & $0.097 \pm 0.011$                  \\
        J2                             & \ $0.03960 \pm 0.00048$   \\
        J4                             & -$0.00222  \pm 0.00005$    \\
        D$_{\text{e}}$ volumetric      & 1,094.4~$\pm$~4.6~km               \\
        D$_{\text{e}}$ area            & 1,095.4~$\pm$~4.6~km               \\
        Density                        & 1.760~$\pm$~0.109~g/cm$^3$         \\
        Density Maclaurin              & 1.859~$\pm$~0.200~g/cm$^3$         \\
        {\color{black} CH$_4$} upper limit $p_{\rm surf}$                & {\color{black}0.15}~nbar                \\
        {\color{black}CO upper limit $p_{\rm surf}$}                   &  {\color{black}0.20~nbar}                \\
        {\color{black}N$_2$ upper limit $p_{\rm surf}$}                &  {\color{black}0.20~nbar}                \\
        \hline
    \end{tabular}
    \label{tab:results}
\end{table}

The oblate model we obtained has equatorial semi-axes of $566.1^{+2.5}_{-2.2}$~km, polar semi-axis of $511.2^{+3.6}_{-3.7}$~km, polar oblateness of $0.097 \pm 0.011$
%, and pole orientation of $\alpha_p = 17.32^{+0.01}_{-0.01}$~h and $\delta_p = 53.4^{+0.2}_{-0.2}$~${^{\circ}}$ 
that produces an equivalent surface diameter of 1,095.4~$\pm$~4.6~km and an equivalent volumetric diameter of 1,094.4~$\pm$~4.6~km. This can be compared to previous values reported by \citet{Ribas_2013} of 1,110~$\pm$~5~km in equivalent volumetric diameter and $0.090^{+0.027}_{-0.018}$ in oblateness. Using the obtained volume and the mass from \citet{Ribas_2025}, we derive a density of 1.760~$\pm$~0.109~g/cm$^3$. Alternatively, using the rotation period determined by \citet{Ortiz_2003}, the density calculated for a body in Maclaurin hydrostatic equilibrium with the here obtained dimensions is 1.859~$\pm$~0.200~g/cm$^3$. For more details on the Maclaurin density calculations, see \citet{Ribas_2013}, \citet{Ortiz_2012}, or \citet{Rommel_2025}. The density proposed by \citet{Ribas_2013} was 1.99~$\pm$~0.46~g/cm$^3$, still consistent with the density found in our study, but based on old
%\st{, and thus worse,} 
value for the system mass: {\color{black}1.65~$\pm$~0.16~$\times$10$^{21}$~kg. \citet{Vachier_2012}}.

We note that the density obtained from mass and volume and the one from the Maclaurin equilibrium figure overlap within uncertainties. This suggests that Quaoar may indeed be in hydrostatic equilibrium, thus strengthening the case for its classification as a dwarf planet. 

Quaoar has an oblate shape and is in Maclaurin hydrostatic equilibrium implies a single-peaked rotation period of 8.8394~$\pm$~0.0002 hours. Consequently, the entire amplitude of the rotational light curve must be attributed solely to albedo variegation. Furthermore, this model places the QR1 ring closer to the 1/6 spin-orbit resonance instead of the 1/3 and QR2 at 5/14 instead 5/7. These characteristics weaken the assumption that Quaoar is oblate and in hydrostatic equilibrium. This data set can also be used to explore a putative triaxial shape. Considering that the rotation light curve amplitude is caused by its shape or a combination of albedo variegation and shape, to account for the double peak solution and the 17.6788h rotation period.

%In a follow-up study, we will use this dataset, along with a re-analysis of the \cite{Ortiz_2003} rotational light curve and data from surveys like Gaia DR3 \citep{Gaia_2021} to constrain Quaoar's triaxial shape.

A value for Quaoar's absolute magnitude, $H_V=2.79\pm0.35$, was derived by \cite{Kiss_2024}, eliminating the contributions from Weywot and the rings and using 2003 data originally published by \cite{Rabinowitz_2007}. 
%An absolute magnitude of $H_V = 2.79 \pm 0.35$, eliminating Weywot's and the rings' contributions, was derived by \cite{Kiss_2024}, using 2003 data published by \cite{Rabinowitz_2007}. 
To derive Quaoar's geometric albedo, we used the model's projected semi-axis a'= 566.3~$\pm$~3.1~km and b'= 518.8~$\pm$~4.2~km at the epoch of the absolute magnitude observation. 
With that, we obtain a geometric albedo of 0.125~$\pm$~0.038 using the equation in \cite{Ribas_2013}. This result shows a small discrepancy when compared to the geometric albedo of 0.109~$\pm$~0.007 proposed by \cite{Ribas_2013}, when no projection and higher absolute magnitude values were used.

Quaoar was one of the objects that potentially retained sufficient volatiles to sustain a global atmosphere \cite{Young_2022}. 
%\st{Therefore, the possible absence of an atmosphere in Quaoar’s system, smaller than 1.5~nbar}
{\color{black}Therefore, the possible absence of a CH$_4$ atmosphere in Quaoar’s system, smaller than 0.15~nbar}, aids in constraining the mechanisms responsible for maintaining volatiles in the outer solar system. {\color{black} We also duplicated our ray-tracing analysis for putative $N_2$ and $CO$ atmospheres. The derived $1\sigma$ upper limits for these gases (both equal to 0.2) are higher than those for $CH_4$. Considering these limits alongside the high volatility of these species, we conclude that the presence of global $N_2$ or $CO$ atmospheres around Quaoar is highly improbable.}

Finally, although the data set is fully explained by the oblate model presented here, we want to mention that other models, like a triaxial {\color{black}ellipsoid}, may also be compatible. Therefore, Quaoar's definitive tridimensional shape remains to be defined.

\section*{Acknowledgments}
%\begin{acknowledgments}
%\textit{Acknowledgements:}
This work is based on observations collected by a large network of institutions and individual observers whose collaboration and support were essential. We are grateful for the efforts in data acquisition, campaigns, and analysis. This work would not have been possible without the support of this community. We dedicate this work to Tony George, deceased.
The following observers have also contributed to data acquisition used in this work: 
A. Aia, 
M. Bachini, T. Bridges, P. Barroy, A. Barry, K. L. Bath, Z. Bora, G. Brabant, D. Briggs, 
S. Calavia, E. Campbell, G. Canaud, M. Conjat, M. Craciun, 
P. Delincak, I. Dinev, J. Dostal, S. Dostal, C. Drebber, E. Ducrot, 
B. Emptage, R. Evans, 
K. Hill, G. Holtkamp, R. Horvat, G. Hudson,  
E. Karampotsiou, P. Kilmartin, A. Klotz, M. Korec, P. Krabbendam, 
E. Lacruz, B. Lambert, B. Loader, J. Loucks, 
T. Mollier, 
E. Ogiza, J. Oliveira, A. Olsen, 
R. Paine, B. Paton, I. Pérez-Garcia, J. Polák, J. Pollak, S. Pyrzas,
C. Ratinaud, A. Ribera, S. Robinson, M. Rottenborn, L. Rousselot, 
M. Serrau, M. Serau, H. Simon, M. F. Skrutskie, P. Sogorb, 
J. Talbot, J. P. Tengk, S. Todd, D. Tomko,  
M. Unwin, 
D. Vernet, J. M. Vienney, 
W. Witte, 
and S. Yusuf.
We gratefully acknowledge financial support from the following institutions and grants: This study was financed in part by the \textbf{Coordenação de Aperfeiçoamento de Pessoal de Nível Superior – Brasil (CAPES)} – Finance Code 001, and the \textbf{National Institute of Science and Technology of the e-Universe project (INCT do e-Universo)}, CNPq grant 465376/2014-2. FBR acknowledges \textbf{CNPq grant 316604/2023-2} and the financial support of the NAPI “Fenômenos Extremos do Universo” of \textbf{Fundação Araucária}. R. V-M thanks grant \textbf{CNPq 307368/2021-1}. MA acknowledges \textbf{CNPq 427700/2018-3, 310683/2017-3, and 473002/2013-2}, and \textbf{FAPERJ E-26/210.705/2024}. JIBC acknowledges \textbf{CNPq grants 305917/2019-6 and 306691/2022-1}, and \textbf{FAPERJ 201.681/2019}. T.F.L.L. Pinheiro acknowledges support from \textbf{CNPq - Proc.313994/2025-0}. CLP thanks the \textbf{FAPERJ/DSC-10 E-26/204.141/2022, FAPERJ/PDR-10 E-26/200.107/2025, and FAPERJ 200.108/2025}. GR acknowledges \textbf{FAPESP 2024/20150-1}. ARGJ thanks \textbf{FAPEMIG grant APQ-02987-24}. BEM thanks \textbf{CAPES Grant 23079.212658/2024-30}. {\color{black}BS acknowledges support by the French ANR project Roche, number ANR-23-CE49-0012. ER  would like to thanks to Fundação de Amparo à Pesquisa do Estado do Rio de Janeiro (FAPERJ) for their support through a fellowship (E-26/204.602/2021). The authors are grateful to the IMPACTON team, in special to R. Souza, A. Santiago and A. Santos for the technical support in the observations at the Observatório Astronômico do Sertão de Itaparica (OASI). BEM thanks CAPES grant 23079.212658/2024-30, CNPq/Universal grant 408543/2025-6 and FAPERJ grant E-26/204.205/2025.}. Support was also provided by the \textbf{Fundação de Amparo à Pesquisa e Inovação do Espírito Santo (FAPES)}, No. 2021-X4ZNX. Further support from \textbf{CNPq} (project 307400/2025-5), the \textbf{State Secretariat of Science, Technology, and Higher Education of Paraná (SETI–Fundo Paraná, grant 031/2024)} is acknowledged.
We acknowledge funding from the \textbf{European Research Council (ERC)} under the European Community's H2020 (2014-2020/ERC Grant Agreement no. 669416, “LUCKY STAR”), the \textbf{ERC AdG SUBSTELLAR} (GA 101054354), and ERC grant agreement n$^\circ$ 803193/BEBOP. \textbf{HiPERCAM was funded by the European Research Council under the European Union’s Seventh Framework Programme (FP/2007-2013) under ERC-2013-ADG Grant Agreement no. 340040 (HiPERCAM), with additional funding for operations and enhancements provided by the UK Science and Technology Facilities Council (STFC)}. J.L.O., Y.K., P.S.-S., N.M., M.V.-L., J.M.G.L and R.D. acknowledge financial support from the Severo Ochoa grant \textbf{CEX2021-001131-S} (MCIN/AEI/10.13039/501100011033) and Spanish projects \textbf{PID2020-112789GB-I00} (AEI) and \textbf{Proyecto de Excelencia de la Junta de Andalucía PY20-01309}. P.S.-S. and Y.K. acknowledge financial support from the Spanish I+D+i project \textbf{PID2022-139555NB-I00}. T.S-R. acknowledges funding from \textbf{PID2021-125883NB-C21}. Further Spanish support includes grants \textbf{PID2021-122842OB-C21} and \textbf{CEX2019-000918-M}. J. M. Gómez-Limón acknowledges funding from the university training programme \textbf{FPU2022/00492}.
Operation of the \textbf{University of Haifa's H80 (800.0) telescope at the Wise Observatory} is partly supported by ISF grant 3200/20. Observations were also collected at the Wise Observatory with the C28 (711.0) Jay Baum Rich telescope. This article includes observations made in the \textbf{Two-meter Twin Telescope (TTT)} at the Teide Observatory of the \textbf{IAC}, operated by Light Bridges, and supported by Indefeasible Computer Rights (ICR) through PEI "PLANETIX25." The photometric observations from the \textbf{University of Athens Observatory (UOAO)} utilized robotic instruments \citep{Gazeas2016}. The authors acknowledge the use of Sonja Itting-Enke’s C14 telescope and the facilities at the Cuno Hoffmeister Memorial Observatory (CHMO). \textbf{IST40}, one of the observational facilities of the Istanbul University Observatory, was funded by the Scientific Research Projects Coordination Unit of Istanbul University (BAP-3685 and FBG-2017-23943). We also wish to thank the \textbf{Adıyaman University Astrophysics Application and Research Center (Türkiye)} for their support and collaboration. The \textbf{Joan Oró Telescope (TJO)} at the Montsec Observatory (OdM) is acknowledged, as are observations made with the \textbf{Tx40 telescope at the Observatorio Astrofísico de Javalambre (OAJ)}. Partially based on observations collected at the \textbf{La Silla European Southern Observatory}. The 1.2-m Kryoneri telescope is operated by the Institute for Astronomy, Astrophysics, Space Applications and Remote Sensing of the National Observatory of Athens.
Individual acknowledgments include: Funding for KB was provided by the European Union \textbf{ERC AdG SUBSTELLAR, GA 101054354)},\textbf{E. S. thanks the Adiyaman University Astrophysics Application and Research Center for their support in the acquisition of data with the ADYU60 telescope.} Work of KH supported by the project \textbf{RVO:67985815}. R.S. acknowledges funding from the \textbf{K-138962 grant} (NKFIH, Hungary). A. Takey and E.G. Elhosseiny acknowledge financial support from the \textbf{Egyptian Science, Technology \& Innovation Funding Authority (STDF)} under grant number 48102. A. Liakos acknowledges financial support by the \textbf{European Space Agency} under the NELIOTA program. The contributions of R. Scott Fisher were funded by the \textbf{Heising-Simons Foundation through grant 2023-4845}. D.I. acknowledges funding by the \textbf{University of Belgrade - Faculty of Mathematics} through grant from the \textbf{Ministry of Science, Technological Development and Innovation of the Republic of Serbia} (contract No. \textbf{451-03-136/2025-03/200104}). O.V. acknowledges support by the \textbf{Astronomical station Vidojevica}, funding from the \textbf{Ministry of Science, Technological Development and Innovation of the Republic of Serbia} (contract No. \textbf{451-03-136/2025-03/200002}), by the \textbf{EC} through project \textbf{BELISSIMA} (call \textbf{FP7-REGPOT-2010-5}, No. \textbf{265772}), and \textbf{Chinese Academy of Sciences (CAS) President's International Fellowship Initiative (PIFI)} (grant No. \textbf{2024VMB0006}). Q.C.T. and S.R. wish to thank \textbf{Roy Kilgard} for his assistance at the 24-inch telescope at Van Vleck Observatory. We thank the \textbf{Friends of MIRA} for their support.
This publication uses data products from the observational networks and research programs.
This publication uses data products from the \textbf{ESO La Silla Observatory}. We gratefully acknowledge support from the \textbf{SPECULOOS} network and the \textbf{TRAPPIST} project \citep{Jehin_2011}. The \textbf{TRAPPIST} project is funded by the \textbf{Belgian F.R.S.-FNRS} under grant \textbf{PDR T.0120.21}. \textbf{E.J.} and \textbf{M.G.} are directors of research at the \textbf{FNRS}.
The ULiege's contribution to SPECULOOS has received funding from the \textbf{European Research Council (ERC)} under the European Union's Seventh Framework Programme (FP/2007-2013) (grant Agreement n$^\circ$ 336480/SPECULOOS), from the \textbf{Balzan Prize and Francqui Foundations}, from the \textbf{Belgian Scientific Research Foundation (F.R.S.-FNRS; grant n$^\circ$ T.0109.20)}, from the University of Liege, and from the \textbf{ARC grant} financed by the Wallonia-Brussels Federation. J.d.W. and MIT gratefully acknowledge financial support from the \textbf{Heising-Simons Foundation}, Dr. and Mrs. Colin Masson and Dr. Peter A. Gilman for \textbf{Artemis} (the first telescope of the SPECULOOS-North Observatory situated in Tenerife, Spain). M.G. is F.R.S-FNRS Research Director. E.J. and M.G. are Directors of Research at the Belgian F.R.S.-FNRS.
SPECULOOS and related projects are further supported by the \textbf{Swiss National Science Foundation} (PP00P2-163967, PP00P2-190080 and the \textbf{National Centre for Competence in Research PlanetS}). Support was also received from the \textbf{Simons Foundation (PI Queloz, grant number 327127)}, and the \textbf{European Research Council (ERC)} under the European Union's Horizon 2020 research and innovation programme (grant agreement n$^\circ$ 803193/BEBOP), from the \textbf{MERAC foundation}, and from the \textbf{Science and Technology Facilities Council (STFC; grant n$^\circ$ ST/S00193X/1)}.

%\end{acknowledgments}

\appendix

\section{\textbf{Star information}}\label{sec:Star informations}

This appendix provides detailed astrometric and photometric information for the stellar sources used in the analysis of the Quaoar occultations. Table \ref{tab:stars} lists the parameters for each candidate star involved in a successful occultation event. These parameters include the star's coordinates and magnitude from the Gaia DR3 catalog, the estimated stellar angular diameter, and the predicted event velocity relevant to the occultation geometry.

\begin{longtable*}{ccccccc}
\caption{Stellar occultation candidates. The columns are: Event (date of the occultation), Designation (GaiaDR3 Source ID), R.A. (ICRS star Right Ascension at epoch), Decl. (ICRS star Declination at epoch), G (star's magnitude in the Gaia G-band), star angular diameter for a main sequence star caulculated using \cite{vanbelle_1999} equations and predicted event velocity.\label{tab:stars}} \\
\hline
\multicolumn{1}{l}{\textbf{Event}} &
\multicolumn{1}{l}{\textbf{Designation}} &
\multicolumn{1}{l}{\textbf{\shortstack[l]{R.A. \\ h m s $\pm$ mas}}} &
\multicolumn{1}{l}{\textbf{\shortstack[l]{Decl. \\ d m s $\pm$ mas}}} &
\multicolumn{1}{l}{\textbf{\shortstack[l]{G \\ mag}}} &
\multicolumn{1}{l}{\textbf{\shortstack[l]{star d. \\ mas}}} &
\multicolumn{1}{l}{\textbf{\shortstack[l]{ev. vel. \\ km/s}}} \\ 
\hline
\endfirsthead
\multicolumn{7}{c}{\tablename\ \thetable{} -- Continuation} \\
\hline
\multicolumn{1}{l}{\textbf{Event}} &
\multicolumn{1}{l}{\textbf{Designation}} &
\multicolumn{1}{l}{\textbf{\shortstack[l]{R.A. \\ h m s $\pm$ mas}}} &
\multicolumn{1}{l}{\textbf{\shortstack[l]{Decl. \\ d m s $\pm$ mas}}} &
\multicolumn{1}{l}{\textbf{\shortstack[l]{G \\ mag}}} &
\multicolumn{1}{l}{\textbf{\shortstack[l]{star d. \\ mas}}} & 
\multicolumn{1}{l}{\textbf{\shortstack[l]{ev. vel. \\ km/s}}} \\ 
\hline
\endhead     
2011-02-11 & 4136779291133721856 & 17 28 47.6205987 +/- 0.133 & -15 41 59.088361 +/- 0.182 & 15.6 & 0.038 & 20.3  \\
2011-05-04 & 4136810730296078208 & 17 28 50.8012677 +/- 0.149 & -15 27 42.725259 +/- 0.191 & 15.6 & 0.019 & -18.3 \\
2012-02-17 & 4125098801168495104 & 17 34 21.8456254 +/- 0.088 & -15 42 10.493594 +/- 0.110 & 14.8 & 0.039 & 18.5  \\
2012-10-15 & 4136784376375275264 & 17 28 10.1277047 +/- 0.187 & -15 36 23.267173 +/- 0.261 & 16.9 & 0.008 & 20.0  \\
2013-07-09 & 4125147729438520576 & 17 34 40.4637157 +/- 0.051 & -15 23 37.568216 +/- 0.062 & 14.4 & 0.025 & -22.5 \\
2018-07-26 & 4145438426258652032 & 18 00 39.4071069 +/- 0.099 & -15 21 58.417719 +/- 0.112 & 16.0 & 0.023 & -20.0 \\
2018-09-02 & 4145444851530756224 & 17 59 02.0588533 +/- 1.224 & -15 27 30.206675 +/- 1.228 & 12.7 & 0.122 & -6.3  \\
2019-03-27 & 4145808549349443456 & 18 12 50.7872483 +/- 0.700 & -15 27 52.149623 +/- 0.629 & 19.0 & 0.020 & 5.5   \\
2019-04-28 & 4145998451292755840 & 18 12 23.2150735 +/- 0.956 & -15 22 07.193200 +/- 0.957 & 18.7 & 0.010 & -12.3 \\
2019-05-28 & 4146010893851176704 & 18 10 38.9330116 +/- 0.233 & -15 18 37.726232 +/- 0.296 & 16.8 & 0.023 & -21.9 \\
2019-06-05 & 4146058516454329344 & 18 10 07.3111037 +/- 0.356 & -15 18 12.353542 +/- 0.438 & 17.4 & 0.035 & -23.2 \\
2019-06-28 & 4145978492632029696 & 18 08 15.5287147 +/- 0.116 & -15 17 58.148575 +/- 0.135 & 15.7 & 0.106 & -24.6 \\
2019-08-04 & 4146716230551709696 & 18 05 34.7866006 +/- 0.130 & -15 21 02.046296 +/- 0.143 & 15.8 & 0.026 & -17.8 \\
2019-09-26 & 4145210002691776000 & 18 04 27.3879636 +/- 0.128 & -15 30 01.808631 +/- 0.152 & 15.5 & 0.010 & 7.3   \\
2019-10-16 & 4146695060624747264 & 18 05 07.9990422 +/- 0.153 & -15 33 24.821037 +/- 0.177 & 16.2 & 0.010 & 16.1  \\
2020-06-11 & 4146202445050212352 & 18 15 03.0557899 +/- 0.064 & -15 15 04.946217 +/- 0.064 & 12.7 & 0.034 & -24.0 \\
2020-06-16 & 4145839022104776192 & 18 14 40.8010099 +/- 0.210 & -15 15 02.309810 +/- 0.276 & 16.1 & 0.011 & -24.4 \\
2020-07-01 & 4145841908322772736 & 18 13 25.4865633 +/- 0.392 & -15 15 23.885899 +/- 0.474 & 17.4 & 0.010 & -24.5 \\
2022-06-24 & 4098181274706042624 & 18 25 04.0901845 +/- 0.464 & -15 07 59.491317 +/- 0.539 & 16.9 & 0.020 & -24.7 \\
2022-08-09 & 4098214367441486592 & 18 21 42.8696462 +/- 0.240 & -15 12 45.963919 +/- 0.219 & 15.3 & 0.094 & -17.6 \\
2023-05-13 & 4103283867659271424 & 18 33 31.7127652 +/- 0.271 & -15 04 52.193652 +/- 0.261 & 15.2 & 0.012 & -15.8 \\
2023-05-20 & 4103284421808448768 & 18 33 09.7151861 +/- 0.397 & -15 04 16.183796 +/- 0.435 & 16.5 & 0.022 & -18.1 \\
2023-05-24 & 4103286105383315840 & 18 32 52.8648095 +/- 0.246 & -15 03 56.396132 +/- 0.317 & 15.3 & 0.036 & -19.4 \\
2023-05-26 & 4103286891316477056 & 18 32 44.9288941 +/- 0.195 & -15 03 48.978547 +/- 0.199 & 14.8 & 0.038 & -20.0 \\
2023-07-15 & 4103961617817078400 & 18 28 56.6548653 +/- 0.832 & -15 05 12.815200 +/- 1.063 & 17.8 & 0.010 & -23.7 \\
2023-08-01 & 4103986872241883008 & 18 27 39.0746912 +/- 1.737 & -15 07 33.231370 +/- 1.599 & 18.7 & 0.010 & -20.2 \\
2023-08-24 & 4103983505031369088 & 18 26 27.7596490 +/- 1.123 & -15 11 20.147339 +/- 0.354 & 16.9 & 0.014 & -12.9 \\
2024-04-10 & 4103400420265295488 & 18 39 51.9082536 +/- 0.142 & -15 04 33.742587 +/- 0.179 & 13.8 & 0.026 & -4.1  \\
2024-05-29 & 4103441029097671680 & 18 37 58.1148584 +/- 2.364 & -14 58 37.782025 +/- 1.577 & 18.5 & 0.010 & -20.6 \\
2024-07-04 & 4103264802318239360 & 18 35 10.7002403 +/- 0.694 & -14 59 34.936931 +/- 0.726 & 17.2 & 0.008 & -24.6 \\
2024-08-28 & 4103291293710742656 & 18 31 41.2280262 +/- 1.171 & -15 07 52.300912 +/- 1.222 & 18.0 & 0.010 & -11.8 \\
2025-06-12 & 4103487277307240704 & 18 42 29.3682255 +/- 0.407 & -14 53 02.547910 +/- 0.561 & 16.1 & 0.033 & -23.1 \\
2025-06-20 & 4103489407614726016 & 18 41 49.2834017 +/- 1.502 & -14 53 18.937250 +/- 1.322 & 18.2 & 0.011 & -24.2 \\
2025-07-04 & 4103410659417963520 & 18 40 41.6345537 +/- 0.745 & -14 54 18.517783 +/- 0.832 & 13.3 & 0.492 & -24.7 \\
2025-08-28 & 4103437150790068352 & 18 37 10.4131641 +/- 0.998 & -15 02 54.481853 +/- 1.293 & 16.3 & 0.010 & -12.2 \\
2025-08-31 & 4103436390532307456 & 18 37 03.1467727 +/- 0.703 & -15 03 35.634275 +/- 0.738 & 17.1 & 0.008 & -10.9 \\
\hline
\end{longtable*}

\section{\textbf{Light curves}}\label{sec:Light curves}

\begin{figure}
    \centering
    \includegraphics[trim=0cm 0.5cm 0cm 0cm,clip]{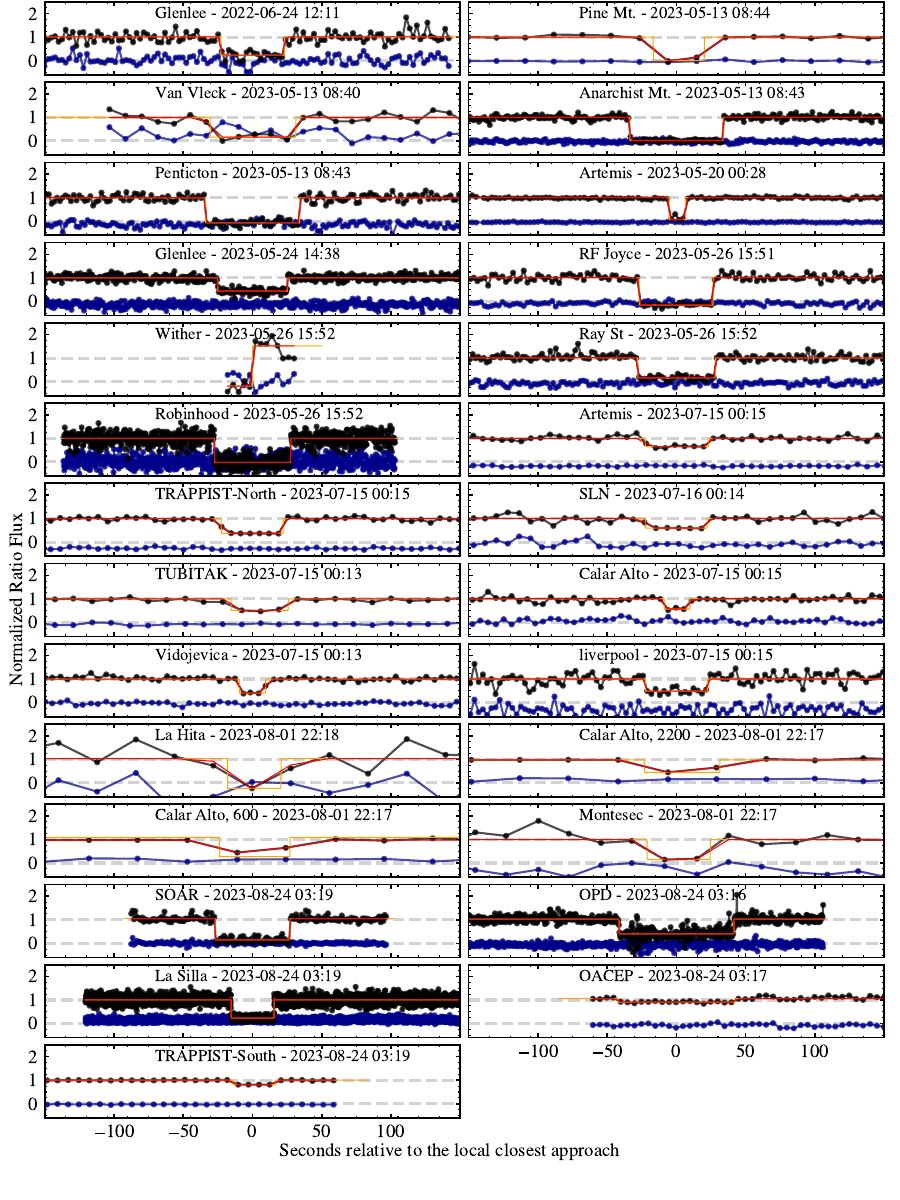}
    \caption{Light curves of Quaoar's occultations that occurred between 2022 and 2023. Black line points represent the normalized target/calibrator flux ratio, while blue line points represent the flux from a star-free region of the sky near the target (ghost). The solid yellow line is the geometric light curve, and the solid red line is the light curve model.}
    \label{fig:2023}
\end{figure}

This appendix presents the set of positive light curves obtained from all stellar occultation events utilized in this work. Light curves that have been previously published are not included in this work. Figures \ref{fig:2018,1019,2020,2022}, \ref{fig:2023}, and \ref{fig:2024,2025} display the normalized flux as a function of time, ordered chronologically. Each light curve includes the normalized flux ratio (black points), the background flux from the star-free region (blue points), the theoretical geometric light curve (solid yellow line), and the final fitted model (solid red line).

\begin{figure}[htp!]
    \centering
    \includegraphics{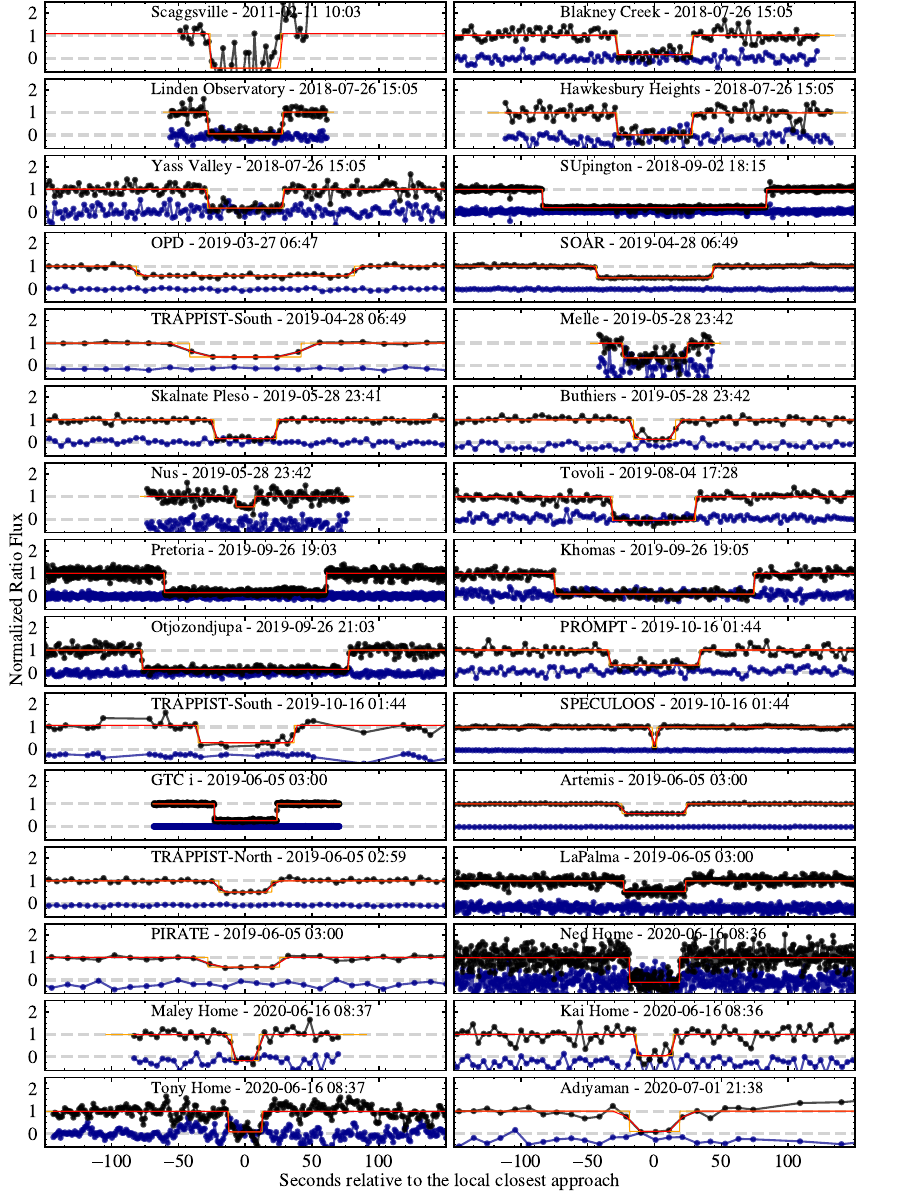}
    \caption{Light curves of Quaoar's occultations that occurred between 2011 and 2020. Black line points represent the normalized target/calibrator flux ratio, while blue line points represent the flux from a star-free region of the sky near the target (ghost). The solid yellow line is the geometric light curve, and the solid red line is the light curve model.}
    \label{fig:2018,1019,2020,2022}
\end{figure}

\begin{figure}
    \centering
    \includegraphics[trim=0cm 1.75cm 0cm 0cm,clip]{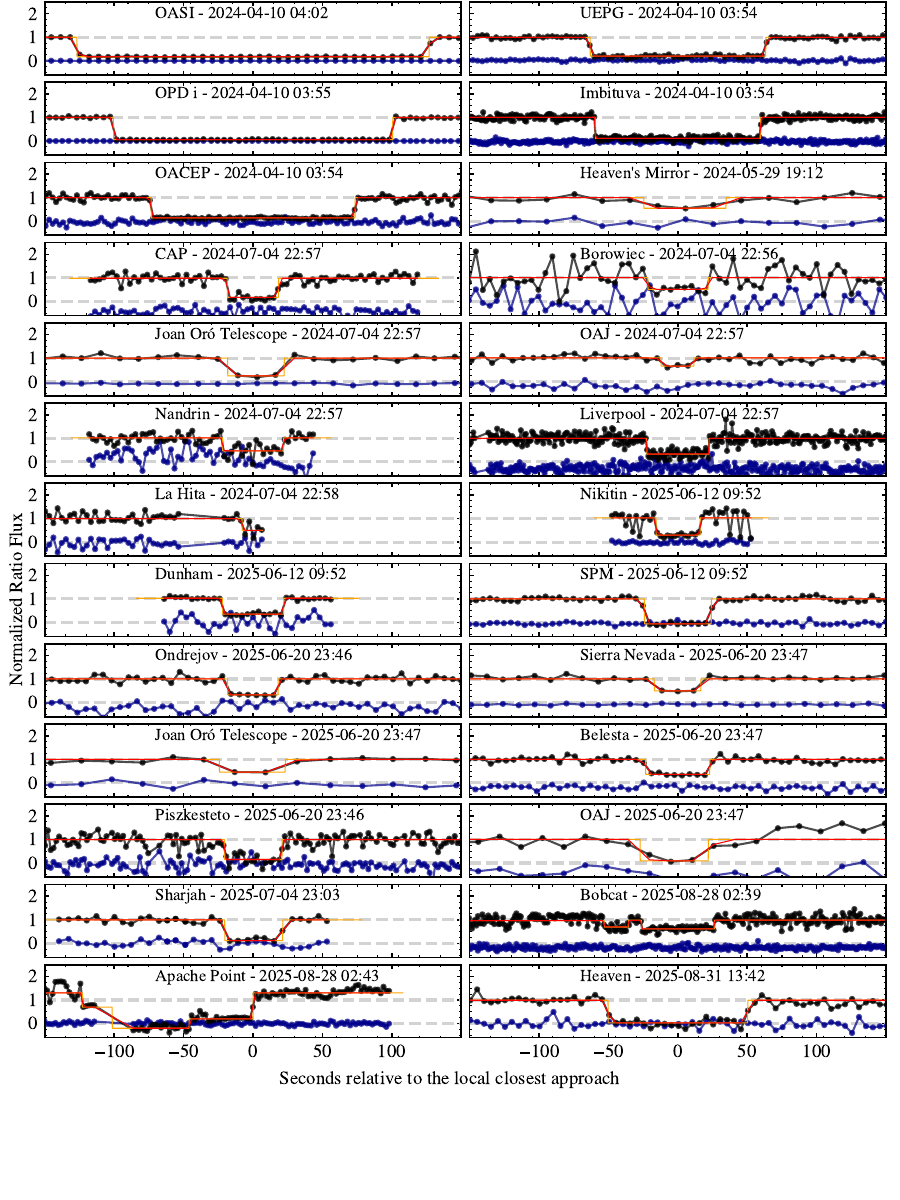}
    \caption{Light curves of Quaoar's occultations that occurred between 2024 and 2025. Black line points represent the normalized target/calibrator flux ratio, while blue line points represent the flux from a star-free region of the sky near the target (ghost). The solid yellow line is the geometric light curve, and the solid red line is the light curve model.}
    \label{fig:2024,2025}
\end{figure}

\clearpage

\section{\textbf{New astrometric position}}\label{sec:New astrometric position}

%\st{This appendix presents the determined astrometric positions of Quaoar on all stellar occultation dates. Table 4 lists the coordinates at the time of closest approach for each occultation event. These positions were derived from the ellipsoid center  and star position and corrected for gravitational deflection effects caused by the Sun, Jupiter, and Saturn.}

{\color{black}This appendix presents the astrometric positions of Quaoar derived from the stellar occultations analysed in this work. For each event, the fitted center of the ellipsoidal model provides the offset of Quaoar relative to its ephemeris position (NIMAv19) on the corresponding tangent plane. The occultation geometry was constructed using the Gaia DR3 astrometric position of the occulted star.

The fitted offsets were corrected for gravitational light deflection due to the Sun, Jupiter, and Saturn using the post-Newtonian formulation of \citet{Klioner_2003}. For each deflecting body, we computed the differential deflection between the stellar ray and the finite-distance direction associated with the ephemeris position of Quaoar. The individual contributions were combined vectorially and applied as a rigid translation of the fitted occultation geometry on the tangent plane, following the occultation-specific formulation of \citet{Poiani_2026}.

The corrected offsets were then applied to the ephemeris position of Quaoar to obtain the final astrometric coordinates at the time of closest approach for each occultation event. These positions are listed in Table \ref{tab:astrometric position}.}

\begin{longtable}{ccccccc}
\caption{This table contains Quaoar's astrometric positions at the time of closest approach for each stellar occultation. These positions have been corrected for the gravitational deflection caused by the Sun, Jupiter and Saturn.\label{tab:astrometric position}} \\
\hline
\multicolumn{1}{l}{\textbf{Event}} &
\multicolumn{1}{l}{\textbf{\shortstack[l]{{\color{black}f }\\ {\color{black}(km)}}}} &
\multicolumn{1}{l}{\textbf{\shortstack[l]{{\color{black}g }\\ {\color{black}(km)}}}} &
\multicolumn{1}{l}{\textbf{\shortstack[l]{{\color{black}df} \\{\color{black} (km)}}}} &
\multicolumn{1}{l}{\textbf{\shortstack[l]{{\color{black}dg} \\{\color{black} (km)}}}}&
\multicolumn{1}{l}{\textbf{\shortstack[l]{R.A. \\ h m s $\pm$ mas}}} &
\multicolumn{1}{l}{\textbf{\shortstack[l]{Decl. \\ d m s $\pm$ mas}}} \\
\hline
\endfirsthead
\multicolumn{3}{c}{\tablename\ \thetable{} -- Continuation} \\
\hline
\multicolumn{1}{l}{\textbf{Event}} &
\multicolumn{1}{l}{\textbf{\shortstack[l]{{\color{black}f }\\ {\color{black}(km)}}}} &
\multicolumn{1}{l}{\textbf{\shortstack[l]{{\color{black}g }\\ {\color{black}(km)}}}} &
\multicolumn{1}{l}{\textbf{\shortstack[l]{{\color{black}df} \\{\color{black} (km)}}}} &
\multicolumn{1}{l}{\textbf{\shortstack[l]{{\color{black}dg} \\{\color{black} (km)}}}}&
\multicolumn{1}{l}{\textbf{\shortstack[l]{R.A. \\ h m s $\pm$ mas}}} &
\multicolumn{1}{l}{\textbf{\shortstack[l]{Decl. \\ d m s $\pm$ mas}}} \\
\hline
\endhead
2011-02-11 10:05:49.940 & {\color{black}-10.2 $\pm$ 68.8  }  & {\color{black}-18.0 $\pm$ 82.0 }   & {\color{black}-5.0 }& {\color{black} 0.7} & 17 28 47.6192086 $\pm$ 3.164 & -15 41 58.929638 $\pm$ 3.025 \\
2011-05-04 02:38:30.680 & {\color{black}-4.3 $\pm$ 10.4   }  & {\color{black}22.0 $\pm$ 24.7  }   & {\color{black}-1.0 }& {\color{black}-0.1} & 17 28 50.7999863 $\pm$ 0.369 & -15 27 42.812496 $\pm$ 0.827 \\
2012-02-17 04:30:44.960 & {\color{black}37.4 $\pm$ 62.6   }  & {\color{black}11.7 $\pm$ 88.3  }   & {\color{black}-4.6 }& {\color{black} 0.5} & 17 34 21.8438798 $\pm$ 1.985 & -15 42 10.328825 $\pm$ 2.798 \\
2012-10-15 00:41:18.820 & {\color{black}0.0 $\pm$ 100.0   }  & {\color{black}0.0 $\pm$ 100.0  }   & {\color{black} 5.0 }& {\color{black} 0.1} & 17 28 10.1273220 $\pm$ 3.171 & -15 36 23.292881 $\pm$ 3.176 \\
2013-07-09 02:40:03.760 & {\color{black}0.0 $\pm$ 100.0   }  & {\color{black}0.0 $\pm$ 100.0  }   & {\color{black} 0.6 }& {\color{black}-0.2} & 17 34 40.4636051 $\pm$ 3.274 & -15 23 37.482333 $\pm$ 3.274 \\
2018-07-26 15:08:21.300 & {\color{black}25.0 $\pm$ 15.7   }  & {\color{black}-3.6 $\pm$ 90.2  }   & {\color{black} 0.9 }& {\color{black}-0.2} & 18 00 39.4079045 $\pm$ 0.524 & -15 21 58.514347 $\pm$ 2.959 \\
2018-09-02 18:16:51.600 & {\color{black}-478.2 $\pm$ 33.2 }  & {\color{black}-508.4 $\pm$ 44.0}   & {\color{black} 2.1 }& {\color{black}-0.1} & 17 59 02.0596130 $\pm$ 1.630 & -15 27 30.258388 $\pm$ 1.881 \\
2019-03-27 06:56:33.480 & {\color{black}-6.3 $\pm$ 24.2   }  & {\color{black}58.8 $\pm$ 22.9  }   & {\color{black}-2.8 }& {\color{black}-0.1} & 18 12 50.7835708 $\pm$ 1.049 & -15 27 52.104642 $\pm$ 0.969 \\
2019-04-28 06:47:02.040 & {\color{black}44.9 $\pm$ 55.1   }  & {\color{black}0.0 $\pm$ 100.0  }   & {\color{black}-1.5 }& {\color{black}-0.2} & 18 12 23.2137245 $\pm$ 2.034 & -15 22 07.269225 $\pm$ 3.398 \\
2019-05-28 23:40:49.480 & {\color{black}-40.7 $\pm$ 40.0  }  & {\color{black}-5.0 $\pm$ 32.2  }   & {\color{black}-0.6 }& {\color{black}-0.2} & 18 10 38.9338179 $\pm$ 1.335 & -15 18 37.534872 $\pm$ 1.099 \\
2019-06-05 03:00:57.040 & {\color{black}-21.6 $\pm$ 4.6   }  & {\color{black}59.9 $\pm$ 25.8  }   & {\color{black}-0.5 }& {\color{black}-0.2} & 18 10 07.3114764 $\pm$ 0.386 & -15 18 12.210859 $\pm$ 0.954 \\
2019-06-28 12:35:40.720 & {\color{black}-3.0 $\pm$ 4.4    }  & {\color{black}-8.9 $\pm$ 62.1  }   & {\color{black} 0.1 }& {\color{black}-0.2} & 18 08 15.5284488 $\pm$ 0.186 & -15 17 57.989302 $\pm$ 2.052 \\
2019-08-04 17:24:51.320 & {\color{black}5.7 $\pm$ 32.1    }  & {\color{black}0.0 $\pm$ 100.0  }   & {\color{black} 1.1 }& {\color{black}-0.2} & 18 05 34.7868586 $\pm$ 1.061 & -15 21 02.068371 $\pm$ 3.280 \\
2019-09-26 18:55:33.920 & {\color{black}-6.3 $\pm$ 4.0    }  & {\color{black}32.1 $\pm$ 4.7   }   & {\color{black} 3.0 }& {\color{black} 0.1} & 18 04 27.3892080 $\pm$ 0.181 & -15 30 01.779453 $\pm$ 0.214 \\
2019-10-16 01:39:00.780 & {\color{black}15.9 $\pm$ 37.5   }  & {\color{black}25.8 $\pm$ 13.9  }   & {\color{black} 4.2 }& {\color{black} 0.2} & 18 05 07.9984539 $\pm$ 1.206 & -15 33 24.860443 $\pm$ 0.478 \\
2020-06-11 16:32:13.560 & {\color{black}13.4 $\pm$ 6.2    }  & {\color{black}14.5 $\pm$ 9.1   }   & {\color{black}-0.3 }& {\color{black}-0.2} & 18 15 03.0558375 $\pm$ 0.215 & -15 15 04.934584 $\pm$ 0.306 \\
2020-06-16 08:38:20.160 & {\color{black}-33.2 $\pm$ 22.4  }  & {\color{black}9.0 $\pm$ 30.5   }   & {\color{black}-0.2 }& {\color{black}-0.2} & 18 14 40.8009527 $\pm$ 0.766 & -15 15 02.166913 $\pm$ 1.042 \\
2020-07-01 21:38:55.520 & {\color{black}0.0 $\pm$ 100.0   }  & {\color{black}0.0 $\pm$ 100.0  }   & {\color{black} 0.2 }& {\color{black}-0.2} & 18 13 25.4861259 $\pm$ 3.320 & -15 15 23.732643 $\pm$ 3.331 \\
2022-06-24 12:09:29.520 & {\color{black}24.8 $\pm$ 55.6   }  & {\color{black}0.0 $\pm$ 100.0  }   & {\color{black}-0.1 }& {\color{black}-0.2} & 18 25 04.0903196 $\pm$ 1.894 & -15 07 59.527209 $\pm$ 3.347 \\
2022-08-09 06:34:02.600 & {\color{black}-21.0 $\pm$ 2.6   }  & {\color{black}-5.1 $\pm$ 4.4   }   & {\color{black} 1.1 }& {\color{black}-0.1} & 18 21 42.8677879 $\pm$ 0.255 & -15 12 45.829889 $\pm$ 0.263 \\
2023-05-13 08:40:33.160 & {\color{black}39.7 $\pm$ 8.3    }  & {\color{black}37.3 $\pm$ 28.7  }   & {\color{black}-1.2 }& {\color{black}-0.2} & 18 33 31.7144466 $\pm$ 0.385 & -15 04 52.022706 $\pm$ 0.979 \\
2023-05-20 00:24:21.460 & {\color{black}0.0 $\pm$ 100.0   }  & {\color{black}-3.6 $\pm$ 63.7  }   & {\color{black}-1.0 }& {\color{black}-0.2} & 18 33 09.7160077 $\pm$ 3.311 & -15 04 16.055219 $\pm$ 2.139 \\
2023-05-24 14:36:05.100 & {\color{black}17.1 $\pm$ 10.2   }  & {\color{black}18.3 $\pm$ 20.5  }   & {\color{black}-0.9 }& {\color{black}-0.2} & 18 32 52.8646139 $\pm$ 0.417 & -15 03 56.444422 $\pm$ 0.746 \\
2023-05-26 15:53:07.440 & {\color{black}34.4 $\pm$ 6.0    }  & {\color{black}3.1 $\pm$ 35.2   }   & {\color{black}-0.9 }& {\color{black}-0.2} & 18 32 44.9285463 $\pm$ 0.277 & -15 03 49.081984 $\pm$ 1.176 \\
2023-07-15 00:16:17.840 & {\color{black}81.7 $\pm$ 18.3   }  & {\color{black}-31.2 $\pm$ 35.9 }   & {\color{black} 0.4 }& {\color{black}-0.2} & 18 28 56.6541066 $\pm$ 1.029 & -15 05 12.672837 $\pm$ 1.592 \\
2023-08-01 22:18:44.900 & {\color{black}0.0 $\pm$ 100.0   }  & {\color{black}0.0 $\pm$ 100.0  }   & {\color{black} 0.8 }& {\color{black}-0.1} & 18 27 39.0728406 $\pm$ 3.724 & -15 07 33.063778 $\pm$ 3.662 \\
2023-08-24 03:22:10.460 & {\color{black}12.0 $\pm$ 4.6    }  & {\color{black}-12.8 $\pm$ 6.0  }   & {\color{black} 1.5 }& {\color{black}-0.1} & 18 26 27.7618401 $\pm$ 1.134 & -15 11 20.245411 $\pm$ 0.405 \\
2024-04-10 03:58:50.740 & {\color{black}-52.2 $\pm$ 5.0   }  & {\color{black}58.4 $\pm$ 3.2   }   & {\color{black}-2.4 }& {\color{black}-0.2} & 18 39 51.8964551 $\pm$ 0.215 & -15 04 33.772079 $\pm$ 0.206 \\
2024-05-29 19:15:55.320 & {\color{black}0.0 $\pm$ 100.0   }  & {\color{black}0.0 $\pm$ 100.0  }   & {\color{black}-0.8 }& {\color{black}-0.2} & 18 37 58.1146688 $\pm$ 4.057 & -14 58 37.852399 $\pm$ 3.655 \\
2024-07-04 22:57:30.880 & {\color{black}1.5 $\pm$ 14.3    }  & {\color{black}-8.9 $\pm$ 69.8  }   & {\color{black} 0.1 }& {\color{black}-0.2} & 18 35 10.6992779 $\pm$ 0.839 & -14 59 34.749734 $\pm$ 2.420 \\
2024-08-28 10:38:30.040 & {\color{black}62.4 $\pm$ 32.5   }  & {\color{black}-29.9 $\pm$ 70.1 }   & {\color{black} 1.6 }& {\color{black}-0.1} & 18 31 41.2281212 $\pm$ 1.582 & -15 07 52.300062 $\pm$ 2.600 \\
2025-06-12 09:53:32.740 & {\color{black}-6.3 $\pm$ 13.4   }  & {\color{black}-12.0 $\pm$ 41.8 }   & {\color{black}-0.5 }& {\color{black}-0.2} & 18 42 29.3680578 $\pm$ 0.601 & -14 53 02.399948 $\pm$ 1.491 \\
2025-06-20 23:46:41.500 & {\color{black}85.7 $\pm$ 14.3   }  & {\color{black}52.5 $\pm$ 47.5  }   & {\color{black}-0.3 }& {\color{black}-0.2} & 18 41 49.2831008 $\pm$ 1.575 & -14 53 18.759540 $\pm$ 2.054 \\
2025-07-04 23:06:15.360 & {\color{black}-63.4 $\pm$ 36.6  }  & {\color{black}0.0 $\pm$ 100.0  }   & {\color{black} 0.1 }& {\color{black}-0.2} & 18 40 41.6337573 $\pm$ 1.423 & -14 54 18.400168 $\pm$ 3.414 \\
2025-08-28 02:44:10.100 & {\color{black}-55.9 $\pm$ 38.9  }  & {\color{black}-35.3 $\pm$ 30.2 }   & {\color{black} 1.5 }& {\color{black}-0.05} & 18 37 10.4092945 $\pm$ 1.618 & -15 02 54.335238 $\pm$ 1.629 \\
2025-08-31 13:46:22.600 & {\color{black}75.2 $\pm$ 24.8   }  & {\color{black}40.9 $\pm$ 59.1  }   & {\color{black} 1.7 }& {\color{black}-0.03} & 18 37 03.1506750 $\pm$ 1.074 & -15 03 35.760080 $\pm$ 2.070 \\
\hline
\end{longtable}

\clearpage

\section{\textbf{Observational information}}\label{sec:Observers informations}

This appendix provides a summary of the observational campaigns, including successful occultation detections, as well as negative and unsuccessful observations. Due to the multi-site nature of this work, {\color{black}the following tables} serve to document the instrumentation and the geographical location of each observer. 
%\st{Table 5 lists the positive occultation observations, presenting the ingress and egress times, and the resulting chord length. We note that previously published positive events are not included here unless new timing information was derived. Table 6 summarizes the negative and unsuccessful campaigns, detailing the observation status (e.g., 'Overcast', 'Subsampled Star', or 'Technical failure') for the non-detection events. It should also be noted that observation campaigns consisting exclusively of non-detection chords are not presented in this work. Both tables provide the telescope aperture, camera model, exposure/cycle time, and the geographic coordinates for each observing site.} 
{\color{black}To accommodate the extensive dataset, the observation logs have been combined and divided into two continuous tables linked by a unique observation ID. Table \ref{tab:obs_data_pt1} details the setup and status of each event (e.g., Positive, Negative, 'Overcast', or 'Technical failure'), along with the observing site, telescope aperture, and camera model. Table \ref{tab:obs_data_pt2} provides the corresponding timing and coordinate parameters, including exposure and cycle times, ingress and egress times, and the resulting chord lengths. We note that previously published positive events are not included here unless new timing information was derived, and observation campaigns consisting exclusively of non-detection chords are not presented in this work.}

\begin{longrotatetable}        
\begin{deluxetable}{ccccccccc}
\tablewidth{0pt}
\tabletypesize{\scriptsize}
\tablecaption{Observational data for all Quaoar events: Part 1 (Setup and Status). \label{tab:obs_data_pt1}}
\tablehead{
\colhead{id} & \colhead{Date} & \colhead{Site} & \colhead{Country} & \colhead{Observer} & \colhead{Ap. (mm)} & \colhead{Camera} & \colhead{Time} & \colhead{Status}
}
\startdata
001 & 2011-02-11 &  & U.S.A. & A. Scheck & 203.0 & MallinCam & GPS & Positive \\
002 & 2018-07-26 &  & Australia & W. Hanna & 508.0 & QHY174M & GPS & Positive \\
003 & 2018-07-26 &  & Australia & D. Herald & 40.0 & Watec 910BD & GPS & Positive \\
004 & 2018-07-26 &  & Australia & D. Gault & 30.0 & Watec 910BD & GPS & Positive \\
005 & 2018-07-26 &  & Australia & T. Barry & 750.0 & Point Grey 28S5M & NTP & Positive \\
006 & 2018-09-02 &  & South Africa & M. Kretlow &  & Raptor 247 & GPS & Positive \\
007 & 2019-03-27 & OPD & Brasil & F. L. Rommel & 1600.0 & IXon 4269 & GPS & Positive \\
008 & 2019-04-28 & SOAR & Chile & J. I. B. Camargo & 4100 & Raptor 247 & GPS & Positive \\
009 & 2019-04-28 & TRAPPIST-South & Chile & E. Jehin & 600 & FLI ProLine & GPS & Positive \\
010 & 2019-05-28 & Ciprian &  & C. Vîntdevară &  &  &  & Subsampled Star \\
011 & 2019-05-28 & Saint-Caprais/Rabastens &  & E. Frappa & 940 & Watec 910HX & GPS & Overcast \\
012 & 2019-05-28 & Nick Haigh - UK &  & N. J. Haigh & 300 & asssi1600 &  & Overcast \\
013 & 2019-05-28 & Pallini &  & V. Tsamis & 400 & ATIK 16 HR &  & Technical failure  \\
014 & 2019-05-28 & C2PU &  & J.P. Rivet & 1040 & iXon Ultra 888 & GPS & Overcast \\
015 & 2019-05-28 & Adiyaman Observatory &  & E. Sonbas & 600 & Andor Ikon-m 934 & GPS & Negative \\
016 & 2019-05-28 & Kryoneri &  & A. Liakos & 1200 & Apogee Aspen CG47 &  & Negative \\
017 & 2019-05-28 & Puimichel &  & J. Lecacheux & 105 & Raptor 247 & GPS & Negative \\
018 & 2019-05-28 & Vlad Dumitrescu &  & V. Dumitrescu &  & ZWO ASI1600 &  & Subsampled Star \\
019 & 2019-05-28 & UOAO &  & K. Gazeas & 400 & SBIG STF-3200W &  & Negative \\
020 & 2019-05-28 &  & Italy & V. Aosta &  & Raptor ??? & GPS & Positive \\
021 & 2019-05-28 & Skalnate Pleso & Slovakia & R. Komzik & 1300 & FLI ProLine & NTP & Positive \\
022 & 2019-05-28 &  & Germany & O. Eberhard & 0.6 m & Watec 910 HX & GPS & Positive \\
023 & 2019-05-28 &  & France & A. Leroy & 600 & QHY174M & GPS & Positive \\
024 & 2019-06-05 & GTC & Spain & J. L. Ortiz & 10.4 m & HiPERCAM & GPS & Positive \\
025 & 2019-06-05 & Sanchez &  & C. Sánchez &  &  &  & Technical failure  \\
026 & 2019-06-05 & TRAPPIST-North & Morocco & E. Jehin & 0.6 m & Andor IKONL & GPS & Positive \\
027 & 2019-06-05 & PIRATE Mark III & Spain & U. Kolb & 0.425 m & FLI PL 16803 & GPS & Positive \\
028 & 2019-06-05 & Liverpool & Spain & P. S. Sanz & 2.0 m & RISE & GPS & Positive \\
029 & 2019-06-05 & Artemis & Spain & E. Jehin & 1.0 m & Andor iKon-L & GPS & Positive \\
030 & 2019-08-04 & Tovoli & Namibia & K. Poschinger & 350 & ZWO ASI 174 MM & GPS & Positive \\
031 & 2019-09-26 & Wolfgang Hakos &  & W. Beisker &  &  &  & Technical failure \\
032 & 2019-09-26 & Les Makes &  & J. P. Teng &  &  &  & Overcast \\
033 & 2019-09-26 &  & South Africa & C. Foster & 0.355 & ZWO ASI290MM & NTP & Positive \\
034 & 2019-09-26 &  & Namibia & W. Beisker & 0.35 m & Raptor 247 & GPS & Positive \\
035 & 2019-09-26 &  & Namibia & M. Backes & 0.35 m & Raptor 247 & GPS & Positive \\
036 & 2019-09-26 &  & Namibia & M. Kretlow &  & Raptor 247 & GPS & Positive \\
037 & 2019-10-16 & SPECULOOS & Chile & E. Jehin & 1000 & Andor IKONL & GPS & Positive \\
038 & 2019-10-16 & PROMPT5 & Chile & J. Pollock & 400 & AltaU-47 & NTP & Positive \\
039 & 2019-10-16 & TRAPPIST-South & Chile & E. Jehin & 600 & FLI ProLine & GPS & Positive \\
040 & 2020-06-16 &  & U.S.A & T. George & 300 & WAT-910HX & GPS & Positive \\
041 & 2020-06-16 &  & U.S.A & P. Maley &  & WAT-910HX & GPS & Positive \\
042 & 2020-06-16 &  & U.S.A & K. Getrost & 254 & QHY174M & GPS & Positive \\
043 & 2020-06-16 &  & U.S.A & N. Smith &  & QHY174M & GPS & Positive \\
044 & 2020-07-01 & AUO & Turkey & W. Ogloza & 600 & Andor Tech & GPS & Positive \\
045 & 2022-06-24 &  & Australia & S. Kerr & 304 & Watec 910BD & GPS & Positive \\
046 & 2023-05-13 & MIRA &  & K. Bender & 914 & Watec 910hx/rc & GPS & Negative \\
047 & 2023-05-13 & Olsen Urbana - Illinois &  & A. Olsen & 508 & QHY174M & GPS & Overcast \\
048 & 2023-05-13 & Harvest Moon &  & S. Messner & 450 & WAT-910HX-RC & GPS & Overcast \\
049 & 2023-05-13 & Naylor &  & R. Kamin & 355 & QHY174M & GPS & Overcast \\
050 & 2023-05-13 & Swift Home &  & T. Swift & 200 & Watek910HX/RC & GPS & Technical failure \\
051 & 2023-05-13 & HAA-Occultation-Mobil &  & J. J. Kavelaars & 305 & QHY174M & GPS & Technical failure \\
052 & 2023-05-13 & Mendel Observatory &  & C. Duston & 457 & Moravian-C4 & GPS & Technical failure \\
053 & 2023-05-13 & MIRA &  & R. Nolthenius & 356 & Watec 910hx/rc & GPS & Negative \\
054 & 2023-05-13 &  & U.S.A & Q. C. Tian & 610 & FLI PL4240 & PC Time & Positive \\
055 & 2023-05-13 &  & U.S.A & S. Fisher & 356 & Andor F9000 & PC Time & Positive \\
056 & 2023-05-13 &  & Canada & B. Gowe & 400 & QHY174M & GPS & Positive \\
057 & 2023-05-13 &  & Canada & P. Ceravolo & 355 & QHY174M & GPS & Positive \\
058 & 2023-05-20 & Artemis & Spain & A. Burdanov & 1000 & Andor iKon-L & GPS & Positive \\
059 & 2023-05-24 & Mt John &  & A. Gilmore & 610 & Watec 120N+ & GPS & Negative \\
060 & 2023-05-24 & Murrumbateman &  & D. Herald & 400 & Watec 910BD & GPS & Negative \\
061 & 2023-05-24 & Linden &  & T. Barry & 750 & Point Grey 28S5M &  & Negative \\
062 & 2023-05-24 & Heaven's Mirror &  & W. Hanna & 508 & QHY174M & GPS & Negative \\
063 & 2023-05-24 &  & Australia & S. Kerr & 304 & Watec 910BD & GPS & Positive \\
064 & 2023-05-26 &  & New Zealand & A. Pennell & 350 & Watec 910BD & GPS & Positive \\
065 & 2023-05-26 & Greenhill Observatory &  & K. Hill & 1270 & Merlin 247 & GPS & Overcast \\
066 & 2023-05-26 & Kuriwa Observatory &  & D. Gault & 300 & Watec WT-910BD & GPS & Negative \\
067 & 2023-05-26 & Linden Observatory &  & T. Barry & 750 & Point Grey 28S5M &  & Negative \\
068 & 2023-05-26 & Heaven's Mirror Observatory &  & W. Hanna & 508 & QHY174M & GPS & Negative \\
069 & 2023-05-26 & Murrumbateman &  & D. Herald & 400 & Watec 910BD & GPS & Negative \\
070 & 2023-05-26 & Flynn &  & J. Newman & 35 & Wat-910BD & GPS & Technical failure \\
071 & 2023-05-26 &  & New Zealand & G. McKay & 250 & QHY174M & GPS & Positive \\
072 & 2023-05-26 &  & New Zealand & R. Paine & 235 & QHY174M & GPS & Positive \\
073 & 2023-05-26 &  & New Zealand & R. Glassey & 405 & ASI 178MM & NTP & Positive \\
074 & 2023-07-15 & Albox &  & J. L. Maestre & 406 & Atik314L+ &  & Subsampled Star \\
075 & 2023-07-15 & TUBITAK (T100) &  & Y. Kilic & 1000 & SI 1100 &  & Overcast \\
076 & 2023-07-15 & Spain - OAJ &  & R. Iglesias-Marzoa & 400 & ProLine &  & Subsampled Star \\
077 & 2023-07-15 & TTT2 &  & M. R. Alarcon & 800 & QHY411M & GPS & Subsampled Star \\
078 & 2023-07-15 & TTT1 &  & M. R. Alarcon & 800 & QHY411M &  & Subsampled Star \\
079 & 2023-07-15 & TÜBİTAK (T60) & Türkiye & Y.kilic & 600 & Andor iKon-L & GPS & Positive \\
080 & 2023-07-15 & SLN & Italy & A. Frasca & 910 & Moravian-C4 & GPS & Positive \\
081 & 2023-07-15 & Artemis & Spain & A. Burdanov & 1000 & Andor iKon-L & GPS & Positive \\
082 & 2023-07-15 & Calar Alto & Spain & J. L. Ortiz & 1230 & ASI-6200MM & GPS & Positive \\
083 & 2023-07-15 & Liverpool & Spain & N. Morales & 2000 & RISE & GPS & Positive \\
084 & 2023-07-15 & La Hita &  & N. Morales & 770 & SBIG-16803 &  & Subsampled Star \\
085 & 2023-07-15 & Sierra Nevada  &  & J. L. Ortiz & 1500 & Andor Ikon & GPS & No Data \\
086 & 2023-07-15 & La Sagra &  & N. Morales & 356 & QHY174M & GPS & No Data \\
087 & 2023-07-15 & Presa de Rules &  & A. Román & 635 & ASI 174 MM Mono &  & Negative \\
088 & 2023-07-15 & MOSS &  & C. Rinner & 500 & ZWO6200 &  & No Data \\
089 & 2023-07-15 & A.S. Vidojevica & Serbia & D. llic & 1400 & Andor iKon-L & GPS & Positive \\
090 & 2023-07-15 & TRAPPIST-North & Morocco & K. Barkaoui & 600 & Andor IKONL & GPS & Positive \\
091 & 2023-07-15 & BOOTES-2 &  & E. J. Fernandez & 600 & Andor Ixon & GPS & Subsampled Star \\
092 & 2023-08-01 & Tübingen University &  & R. Sfair & 800 & SBIG STL-1001E &  & Technical failure \\
093 & 2023-08-01 & La Palma-Liverpool &  & N. Morales & 2000 & RISE &  & Subsampled Star \\
094 & 2023-08-01 & Joan Oró Telescope & Spain & T. Santana-Ros & 800 & CCD42-40 & NTP & Positive \\
095 & 2023-08-01 & La Hita & Spain & N. Morales & 770 & SBIG-16803 & GPS & Positive \\
096 & 2023-08-01 & Calar Alto & Spain & N. Morales & 600 & ASI-6200MM & GPS & Positive \\
097 & 2023-08-01 & Calar Alto & Spain & N. Morales & 2200 & Cafos &  & Positive \\
098 & 2023-08-24 & GOA &  & M. Malacarne & 304 & ZWOASI1600MMPro &  & Subsampled Star \\
099 & 2023-08-24 & OACEP &  & F. Braga-Ribas & 114 & IMX224 & GPS & Subsampled Star \\
100 & 2023-08-24 & SONEAR3 &  & C. Jacques & 450 & QHY600 &  & Subsampled Star \\
101 & 2023-08-24 & Spaceobs - ASH2 &  & N. Morales & 407 & SBIG STL11000 &  & Negative \\
102 & 2023-08-24 & UEPG &  & M. Emilio & 406 & SBIG &  & Technical failure \\
103 & 2023-08-24 & SOAR & Chile & J. I. B. Camargo & 4100 & Raptor 247 & GPS & Positive \\
104 & 2023-08-24 & OACEP & Brasil & F. Braga-Ribas & 254 & Raptor 247 & GPS & Positive \\
105 & 2023-08-24 & TRAPPIST-South & Chile & E. Jehin & 600 & FLI ProLine & GPS & Positive \\
106 & 2023-08-24 & Danish & Chile & C. Snodgrass & 1540 & Andor - Ixon & GPS & Positive \\
107 & 2023-08-24 & OPD & Brazil & C. L. Pereira & 1600 & iXon 4335 & GPS & Positive \\
108 & 2024-04-10 & OPD & Brasil & M. Emilio & 1600 & SPARC4 & GPS & Positive \\
109 & 2024-04-10 & OACEP & Brasil & F. B. Ribas & 114 & IMX224 & NTP & Positive \\
110 & 2024-04-10 & OASI & Brasil & E. Rondon & 1000 & Apo-U47 & NTP & Positive \\
111 & 2024-04-10 & IAG USP &  & M. F. Neto & 406.4 & Apogee Alta U9000 &  & Overcast \\
112 & 2024-04-10 &  & Brasil & C. L. Pereira & 305 & Raptor 247 & GPS & Positive \\
113 & 2024-04-10 & SONEAR3 &  & C. Jacques & 450.0 & QHY600 & NTP & Overcast \\
114 & 2024-04-10 & UEPG & Brasil & M. L. Castanheira & 304.8 & ASI294MC & GPS & Positive \\
115 & 2024-04-10 & UNESP - FEG &  & R. Sfair & 406.0 & Raptor 247 & GPS & Overcast \\
116 & 2024-05-29 &  & Australia & W. Hanna & 508 & QHY174M & GPS & Positive \\
117 & 2024-07-04 & CAP & Spain & A. Selva & 406 & ZWO ASI & NTP & Positive \\
118 & 2024-07-04 & OAJ & Spain & R. Iglesias-Marzoa & 400 & ProLine PL4720 & NTP & Positive \\
119 & 2024-07-04 &  & Schweiz & A. Schweizer & 850 & DVTI+CAM 432 & GPS & Positive \\
120 & 2024-07-04 & Borowiec & Poland & A. Marciniak & 400 & Andor Zyla & GPS & Positive \\
121 & 2024-07-04 & La Hita & Spain & N. Morales & 770 & QHY174M & GPS & Positive \\
122 & 2024-07-04 & Joan Oró Telescope & Spain & T. Santana-Ros & 800 & CCD42-40 & NTP & Positive \\
123 & 2024-07-04 &  & Belgium & O. Schreurs & 406 & Watec 910HX & GPS & Positive \\
124 & 2024-07-04 & Chalin &  & M. K. Kaminska & 700.0 & FLI Kepler 4040 &  & Subsampled Star \\
125 & 2024-07-04 & Sierra Nevada &  & J. L. Ortiz & 900.0 & ANDOR & GPS & Negative \\
126 & 2024-07-04 & La Hita &  & N. Morales & 770.0 & QHY174M & GPS & Subsampled Star \\
127 & 2024-07-04 & BOTORRITA &  & O. Canales & 500.0 & QHY174 & GPS & Technical failure \\
128 & 2025-06-12 & Gemini North &  & C. L. Pereira & 81000.0 & 'Alopeke & GPS & Negative \\
129 & 2025-06-12 & CFHT &  & D. Souami & 3580.0 & WIRCam & GPS & Negative \\
130 & 2025-06-12 & MIRA &  & K. Bender & 203.0 & Watec WAT-910HX & GPS & Subsampled Star \\
131 & 2025-06-12 & BOOTES-2 & Mexico - SPM & E. J. Fernandez & 600 & Ixon & GPS & Positive \\
132 & 2025-06-12 &  & U.S.A & V. Nikitin & 279 & QHY174 & GPS & Positive \\
133 & 2025-06-12 &  & U.S.A & J. Dunham & 356 & QHY174M & GPS & Positive \\
134 & 2025-06-20 & Calar Alto &  & J. L. Ortiz & 1200.0 & QHY600 & GPS & Subsampled Star \\
135 & 2025-06-20 & TUG &  & S. Eryilmaz & 1000.0 & SI 1100 & GPS & Subsampled Star \\
136 & 2025-06-20 & Joan Oró  & Spain & T. Santana-Ros & 800 & CCD42-40 & NTP & Positive \\
137 & 2025-06-20 & İSTEK  &  & M. Acar & 300.0 & QHY 5 III 178 M &  & Overcast \\
138 & 2025-06-20 & İST40 &  & S. Fişek & 400.0 & Moravian G2 8300 & GPS & Overcast \\
139 & 2025-06-20 & Ondokuz Mayıs &  & S. Kalkan & 370.0 & SBIG &  & Overcast \\
140 & 2025-06-20 &  & Czechia & K. Hornoch & 650 & G2CCD-3200 &  & Positive \\
141 & 2025-06-20 & Piszkesteto & Hungary & A. Pal & 1000 & Andor iXon-888 & GPS & Positive \\
142 & 2025-06-20 & Sierra Nevada & Spain & F. J. Aceituno & 1500 & ANDOR & GPS & Positive \\
143 & 2025-06-20 & Teplice &  & Z. Moravec & 600.0 & FLI Kepler FL4040 &  & Subsampled Star \\
144 & 2025-06-20 & OAJ & Spain & R. Iglesias-Marzoa & 400 & ProLine PL4720 & NTP & Positive \\
145 & 2025-06-20 & Bélesta & France & P. André & 820 & C3-PRO-61000 & NTP & Positive \\
146 & 2025-06-20 & TURKSAT &  & M. N. Bagiran & 500.0 & Kepler KL4040 &  & Subsampled Star \\
147 & 2025-07-04 & Sharjah &  Emirates & Mohammad F. Talafha & 431 & SBIG & NTP & Positive \\
148 & 2025-07-04 & Al Khatim &  & M. Odeh & 356.0 & ZWO ASI2600MM &  & Overcast \\
149 & 2025-07-04 & Wise &  & S. Kaspi & 457.2 & QSI683 &  & Negative \\
150 & 2025-07-04 & Wise &  & S. Kaspi & 800.0 & QHY461 &  & Negative \\
151 & 2025-07-04 & Wise &  & S. Kaspi & 711.0 & FLI-PL16801 &  & Negative \\
152 & 2025-07-04 & KAO-Egypt &  & A. Takey & 1880.0 & E2V 42-40 2k & NTP & Overcast \\
153 & 2025-08-28 & Westport &  & S. Diamond & 356.0 & QHY174M & GPS & Negative \\
154 & 2025-08-28 & JJM Observatory &  & K. D. Green & 430.0 & QHY174M & GPS & Negative \\
155 & 2025-08-28 &  Mobile &  & V. Nikitin & 279 & QHY174M & GPS & Overcast \\
156 & 2025-08-28 &  & U.S.A & S. Conard & 356 & Watec 910 & GPS & Positive \\
157 & 2025-08-28 &  & U.S.A & S. Conard & 356 & Watec 910 & GPS & Positive \\
158 & 2025-08-28 & Apache & U.S.A & M. Skrutskie & 356 & ASI432 & GPS & Positive \\
159 & 2025-08-28 & Apache & U.S.A & M. Skrutskie & 356 & ASI432 & GPS & Positive \\
160 & 2025-08-31 & LCO &  & J. M. Gómez-Limón Gallardo & 350.0 & QHY600 &  & Technical failure \\
161 & 2025-08-31 &  & Australia & D. Herald & 508 & Watec 910BD & GPS & Positive \\
162 & 2025-08-31 &  & Australia & W. Hanna & 508 & QHY174M & GPS & Positive \\
\enddata
\end{deluxetable}
\end{longrotatetable}

\begin{longrotatetable}
\begin{deluxetable}{ccccccccccccc}
\tablewidth{0pt}
\tabletypesize{\scriptsize}
\tablecaption{Observational data for all Quaoar events: Part 2 (Timing and Coordinates). \label{tab:obs_data_pt2}}
\tablehead{
\colhead{id} & \colhead{Exp (s)} & \colhead{Cycle (s)} & \colhead{Offset (s)} & \colhead{Latitude} & \colhead{Longitude} & \colhead{Height (m)} & \colhead{Ingress} & \colhead{E. (s)} & \colhead{Egress} & \colhead{E. (s)} & \colhead{Chord (km)} & \colhead{Chord e. (km)}
}
\startdata
001 & 2.135 & 2.135 &  & 39 08 58.73 & -76 53 13.42 & 120.0 & 10:02:55.9 & 1.7 & 10:03:48.7 & 1.0 & 1059.1 & 53.5 \\
002 & 2.0 & 2.0 &  & -34 51 51.16 & 148 58 35.19 & 536.0 & 15:04:51.5 & 0.4 & 15:05:48.1 & 0.5 & 1143.7 & 18.7 \\
003 & 2.56 & 2.56 & -1.3 & -34 57 31.6 & 148 59 54.80 & 594.0 & 15:04:50.8 & 0.3 & 15:05:47.2 & 0.5 & 1139.4 & 17.5 \\
004 & 2.56 & 2.56 & -1.3 & -33 39 51.90 & 150 38 27.90 & 286.0 & 15:04:43.6 & 0.2 & 15:05:40.1 & 0.6 & 1141.2 & 16.8 \\
005 & 1.0 & 1.0 &  & -33 42 27.30 & 150 29 43.50 & 574.0 & 15:04:44.4 & 0.2 & 15:05:40.4 & 0.2 & 1130.9 & 9.0 \\
006 &  &  &  & -28 03 35.15 & 21 01 31.63 & 873.0 & 18:14:26.2 & 0.1 & 18:17:13.6 & 0.1 & 1107.5 & 0.7 \\
007 & 6.0 & 6.0 &  & -22 32 07.75 & -45 34 57.54 & 1810.0 & 06:46:27.0 & 0.7 & 06:49:10.6 & 0.8 & 855.2 & 7.8 \\
008 & 3.2 & 3.2 & -4.8 & -30 14 16.89 & -70 44 21.12 & 2693.0 & 06:48:53.3 & 1.6 & 06:50:20.0 & 1.6 & 1092.7 & 39.2 \\
009 & 15.0 & 15.98 &  & -29 15 16.56 & -70 44 21.84 & 2315.0 & 06:48:58.9 & 1.0 & 06:50:22.5 & 1.0 & 1054.1 & 24.9 \\
010 & 30.0 & 31.0 &  & 44 07 58.213 & 26 13 07.378 & 90.0 &  &  &  &  &  &  \\
011 & 1.28 & 1.28 &  & 43 52 26.6 & 01 43 07.5 & 190.0 &  &  &  &  &  &  \\
012 & 10.0 & 12.0 &  & 50 55 19.5 & -01 22 30.1 & 10.0 &  &  &  &  &  &  \\
013 & 5.0 & 5.0 &  & 37 59 52 & 23 53 35 & 165.0 &  &  &  &  &  &  \\
014 & 0.3 & 0.3 &  & 43 45 13.2000 & 06 55 22.700 & 1270.0 &  &  &  &  &  &  \\
015 & 2.0 & 2.4 &  & 37 45 06.000 & 38 13 32.000 & 700.0 &  &  &  &  &  &  \\
016 & 6.0 & 6.5 &  & 37 58 19.000 & 22 37 07.000 & 890.0 &  &  &  &  &  &  \\
017 & 1.6 & 1.6 & -2.4 & 43 58 48.000 & 06 01 15.000 & 730.0 &  &  &  &  &  &  \\
018 & 7.0 & 7.11 &  & 46 13 54.000 & 27 40 10.000 & 70.0 &  &  &  &  &  &  \\
019 & 10.0 & 12.5 &  & 37 58 06.800 & 23 47 00.100 & 250.0 &  &  &  &  &  &  \\
020 & 1.0 & 1.0 &  & 45 47 22.00 & 07 28 42.00 & 1670.0 & 23:42:17.8 & 0.4 & 23:42:32.1 & 0.5 & 316.7 & 18.6 \\
021 & 1.0 & 4.65 &  & 20 14 01.0 & 49 11 22.0 & 1786.0 & 23:41:16.9 & 2.0 & 23:42:03.7 & 2.0 & 1038.1 & 87.3 \\
022 & 1.28 & 1.28 & -0.66 & 52 15 09.66 & 08 19 24.16 & 220.0 & 23:41:50.2 & 0.5 & 23:42:38.5 & 1.1 & 1071.0 & 33.3 \\
023 & 5.0 & 5.0 &  & 48 17 30.72 & 02 26 16.84 & 90.0 & 23:42:19.6 & 0.6 & 23:42:51.1 & 0.6 & 697.8 & 27.5 \\
024 & 0.1302282 & 0.1348 &  & 28 45 21.60 & -17 53 30.20 & 2270.0 & 03:00:08.45 & 0.03 & 03:00:55.32 & 0.02 & 1107.1 & 0.9 \\
025 &  &  &  &  &  &  &  &  &  &  &  &  \\
026 & 5.0 & 6.0 &  & 31 12 22.32 & -07 51 59.04 & 2777.0 & 02:59:32.5 & 0.3 & 03:00:12.2 & 0.7 & 900.8 & 19.3 \\
027 & 8.0 & 11.65 &  & 28 17 57.30 & -16 30 36.70 & 2390.0 & 03:00:00.1 & 1.4 & 03:00:53.9 & 2.9 & 1271.1 & 101.6 \\
028 & 0.6 & 0.63 &  & 28 45 45.05 & -17 52 45.12 & 2387.0 & 03:00:08.9 & 0.2 & 03:00:55.2 & 0.2 & 1094.1 & 9.1 \\
029 & 2.5 & 3.78 &  & 28 18 02.00 & -16 30 41.00 & 2440.0 & 03:00:01.5 & 0.1 & 03:00:49.6 & 0.8 & 1135.7 & 21.4 \\
030 & 2.0 & 2.4961 &  & -23 27 39.63 & 18 00 56.54 & 1340.0 & 17:27:54.4 & 0.3 & 17:28:56.2 & 0.5 & 1120.7 & 14.7 \\
031 &  &  &  &  &  &  &  &  &  &  &  &  \\
032 &  &  &  &  &  &  &  &  &  &  &  &  \\
033 & 1.3 & 1.2139 &  & -25 43 55.00 & 28 13 47.00 & 1300.0 & 19:04:39.9 & 0.1 & 19:07:09.1 & 0.2 & 1051.7 & 1.9 \\
034 & 1.0 & 1.0 & -1.5 & -23 14 11.0 & 16 21 42.0 & 1834.0 & 19:02:28.6 & 0.2 & 19:05:03.2 & 0.2 & 1082.7 & 2.8 \\
035 & 1.0 & 1.0 & -1.5 & -22 41 54.50 & 17 06 31.70 & 1910.0 & 19:02:28.9 & 0.2 & 19:05:03.5 & 0.3 & 1082.3 & 2.9 \\
036 & 0.3 & 0.3 & -0.45 & -19 05 15.50 & 18 36 01.70 & 1200.0 & 19:02:36.86 & 0.05 & 19:04:38.05 & 0.05 & 848.0 & 0.6 \\
037 & 1.5 & 2.44 &  & -24 36 58.00 & -70 23 26.00 & 2440.0 & 01:44:48.8 & 0.2 & 01:44:52.0 & 0.1 & 50.3 & 4.7 \\
038 & 2.0 & 3.02 &  & -30 10 03.52 & -70 48 18.91 & 2200.0 & 01:44:10.5 & 1.1 & 01:45:18.4 & 1.1 & 1082.1 & 34.7 \\
039 & 3.0 & 4.12 &  & -29 15 16.56 & -70 44 21.84 & 2315.0 & 01:44:11.0 & 1.1 & 01:45:21.9 & 0.8 & 1130.5 & 29.3 \\
040 & 1.068 & 1.068 &  & 33 49 00.10 & -111 52 07.30 & 843.0 & 08:37:37.4 & 2.0 & 08:38:00.8 & 2.0 & 581.3 & 99.5 \\
041 & 4.2291 & 4.235 &  & 33 48 42.90 & -111 57 07.90 & 654.0 & 08:37:37.2 & 2.7 & 08:37:57.2 & 2.6 & 496.8 & 132.8 \\
042 & 4.0 & 4.0 &  & 36 25 37.90 & -87 41 43.80 & 90.0 & 08:36:13.3 & 0.7 & 08:36:38.3 & 5.8 & 617.5 & 160.9 \\
043 & 0.5 & 0.5 &  & 34 53 34.32 & -85 28 15.88 & 218.0 & 08:35:58.3 & 0.1 & 08:36:35.7 & 0.2 & 924.1 & 9.3 \\
044 & 10.0 & 10.0 &  & 37 45 06.00 & 38 13 32.00 & 695.0 & 21:37:52.3 & 2.2 & 21:38:30.0 & 2.2 & 935.8 & 108.9 \\
045 & 2.56 & 2.56 & -1.3 & -23 16 10.06 & 150 30 01.62 & 53.0 & 12:11:10.0 & 0.6 & 12:11:56.0 & 0.6 & 1150.1 & 31.7 \\
046 & 0.133 & 0.133 &  & 36 18 20.8800 & -121 33 59.7600 & 1603.0 &  &  &  &  &  &  \\
047 &  &  &  & 40 05 12.4008 & -88 11 46.2984 & 224.0 &  &  &  &  &  &  \\
048 &  &  &  & 44 29 57.5232 & -93 07 45.0480 & 289.46 &  &  &  &  &  &  \\
049 &  &  &  & 40 08 55.6008 & -76 53 43.4796 & 181.6 &  &  &  &  &  &  \\
050 &  &  &  & 38 33 07.8012 & -121 47 08.1600 & 18.0 &  &  &  &  &  &  \\
051 &  &  &  & 48 29 22.1364 & -123 20 15.0433 & 34.36 &  &  &  &  &  &  \\
052 & 10.0 & 12.0 &  & 42 40 9.57 & -71 07 19.056 & 71.64 &  &  &  &  &  &  \\
053 & 0.25 & 0.25 & -0.15 & 36 18 20.8800 & -121 33 59.7600 & 1603.0 &  &  &  &  &  &  \\
054 & 10.0 & 11.69 &  & 41 33 18.62 & -72 39 33.12 & 68.0 & 08:40:01.0 & 6.6 & 08:41:01.9 & 2.5 & 985.6 & 148.6 \\
055 & 15.0 & 20.59 &  & 43 47 30.59 & -120 56 28.99 & 1907.0 & 08:43:54.0 & 5.2 & 08:44:35.5 & 5.9 & 653.3 & 178.8 \\
056 & 2.0 & 2.0 &  & 49 32 01.67 & -119 33 26.50 & 455.0 & 08:43:21.4 & 0.1 & 08:44:29.4 & 0.2 & 1093.1 & 5.8 \\
057 & 1.3 & 1.3 &  & 49 00 32.18 & -119 21 47.30 & 512.0 & 08:43:22.2 & 0.1 & 08:44:29.6 & 0.1 & 1083.8 & 3.2 \\
058 & 2.0 & 2.94 &  & 28 18 02.00 & -16 30 41.00 & 2440.0 & 00:28:32.2 & 0.6 & 00:28:44.1 & 0.9 & 218.0 & 27.5 \\
059 & 0.64 & 0.64 & -0.34 & -43 59 12.0120 & 170 27 54.0000 & 1030.0 &  &  &  &  &  &  \\
060 & 0.64 & 0.64 & -0.34 & -34 57 31.3164 & 148 59 56.004 & 595.91 &  &  &  &  &  &  \\
061 & 0.266 &  &  & -33 42 26.6868 & 150 29 43.4652 & 590.0 &  &  &  &  &  &  \\
062 & 0.5 & 0.5 &  & -34 51 50.8900 & 148 58 35.0602 & 536.0 &  &  &  &  &  &  \\
063 & 0.64 & 0.64 & -0.34 & -23 16 10.06 & 150 30 01.62 & 53.0 & 14:37:52.2 & 0.1 & 14:38:42.5 & 0.1 & 996.7 & 4.0 \\
064 & 0.32 & 0.32 & -0.18 & -45 52 21.37 & 170 29 29.86 & 103.0 & 15:51:38.6 & 0.1 & 15:52:33.5 & 0.1 & 1116.6 & 1.9 \\
065 &  &  &  & -42 25 51.8160 & 147 43 10.5600 & 646.0 &  &  &  &  &  &  \\
066 & 0.64 & 0.64 & -0.34 & -33 39 52.4844 & 150 38 27.9600 & 273.17 &  &  &  &  &  &  \\
067 & 0.266 &  &  & -33 42 26.6868 & 150 29 43.4652 & 590.0 &  &  &  &  &  &  \\
068 & 0.25 & 0.25 &  & -34 51 50.8900 & 148 58 35.0602 & 536.0 &  &  &  &  &  &  \\
069 & 0.32 & 0.32 & -0.18 & -34 57 31.3164 & 148 59 56.004 & 595.91 &  &  &  &  &  &  \\
070 & 0.64 & 0.64 &  & -35 11 55.3400 & 149 02 57.5300 & 657.0 &  &  &  &  &  &  \\
071 & 4.0 & 4.0 &  & -41 32 08.66 & 173 57 25.31 & 14.0 &  &  & 15:52:11.6 & 0.7 &  &  \\
072 & 2.0 & 2.0 &  & -45 02 44.80 & 169 12 36.70 & 200.0 & 15:51:42.6 & 0.2 & 15:52:38.5 & 0.2 & 1137.3 & 9.7 \\
073 & 2.0 & 2.53 &  & -43 29 55.35 & 172 20 59.03 & 103.0 & 15:51:31.5 & 0.5 & 15:52:24.8 & 0.5 & 1083.7 & 20.8 \\
074 & 15.0 &  &  & 37 24 20.0304 & -2 09 06.4800 & 491.0 &  &  &  &  &  &  \\
075 &  &  &  & 36 49 17.4000 & 30 20 07.9846 & 2538.72 &  &  &  &  &  &  \\
076 & 13.0 &  &  & 40 02 30.4800 & -1 00 58.6800 & 1957.0 &  &  &  &  &  &  \\
077 & 4.53 &  &  & 28 17 55.3633 & -16 30 34.1358 & 2359.11 &  &  &  &  &  &  \\
078 & 4.53 &  &  & 28 17 55.3604 & -16 30 34.4481 & 2359.12 &  &  &  &  &  &  \\
079 & 8.0 & 13.2 &  & 36 49 29.00 & 30 20 08.28 & 2455.0 & 00:13:03.1 & 3.4 & 00:13:43.7 & 3.5 & 973.7 & 166.9 \\
080 & 8.0 & 8.22 &  & 37 41 05.00 & 14 58 04.01 & 1735.0 & 00:13:42.3 & 2.2 & 00:14:26.7 & 0.4 & 1067.2 & 62.9 \\
081 & 6.0 & 6.93 &  & 28 18 02.00 & -16 30 41.00 & 2440.0 & 00:15:31.4 & 2.2 & 00:16:17.1 & 1.1 & 1102.8 & 80.1 \\
082 & 4.0 & 5.61 &  & 37 13 24.71 & -02 32 44.88 & 2173.0 & 00:14:50.1 & 1.4 & 00:15:10.0 & 0.7 & 480.0 & 48.3 \\
083 & 3.0 & 3.03 &  & 28 45 45.05 & -17 52 45.12 & 2387.0 & 00:15:37.0 & 1.0 & 00:16:21.8 & 1.1 & 1080.2 & 50.0 \\
084 & 7.0 &  &  & 39 34 04.8000 & -3 10 59.8800 & 770.0 &  &  &  &  &  &  \\
085 &  &  &  & 37 03 50.8896 & -3 23 04.9200 & 2930.53 &  &  &  &  &  &  \\
086 &  &  &  & 37 58 51.6000 & -2 33 50.4000 & 1530.0 &  &  &  &  &  &  \\
087 & 4.319 &  &  & 36 51 07.6603 & -3 30 12.5857 & 155.21 &  &  &  &  &  &  \\
088 &  &  &  & 31 12 23.6052 & -7 51 58.9248 & 2750.0 &  &  &  &  &  &  \\
089 & 4.0 & 5.0 &  & 43 08 24.6 & 21 33 21.4 & 1150.0 & 00:13:43.9 & 1.4 & 00:14:02.8 & 0.3 & 452.9 & 40.9 \\
090 & 6.0 & 6.86 &  & 31 12 22.32 & -07 51 59.04 & 2777.0 & 00:14:57.1 & 0.6 & 00:15:41.7 & 0.9 & 1077.8 & 37.6 \\
091 & 10.0 &  &  & 36 45 33.2676 & -4 02 27.4920 & 70.0 &  &  &  &  &  &  \\
092 & 1.0 &  &  & 48 32 03.0156 & 9 04 12.0036 & 400.0 &  &  &  &  &  &  \\
093 & 5.0 &  &  & 28 45 45.0576 & -17 52 45.1200 & 2387.63 &  &  &  &  &  &  \\
094 & 20.0 & 23.2 &  & 42 03 05.96 & 00 43 46.74 & 1564.0 & 22:17:08.2 & 9.1 & 22:17:53.8 & 7.2 & 936.4 & 334.7 \\
095 & 25.0 & 27.9 &  & 39 34 04.80 & -03 10 59.88 & 770.0 & 22:17:26.8 & 11.3 & 22:18:05.4 & 11.9 & 793.2 & 476.8 \\
096 & 30.0 & 30.8 &  & 37 13 24.71 & -02 32 44.88 & 2173.0 & 22:17:18.3 & 5.6 & 22:18:08.8 & 7.4 & 1038.4 & 267.3 \\
097 & 20.0 & 35.0 &  & 37 13 24.71 & -02 32 44.88 & 2173.0 & 22:17:14.0 & 8.2 & 22:18:08.4 & 1.3 & 1118.6 & 195.3 \\
098 & 15.0 &  &  & -20 18 01.99998 & -40 19 02.0000 & 24.0 &  &  &  &  &  &  \\
099 & 4.0 &  &  & -25 20 57.3942 & -49 21 55.4240 & 1055.01 &  &  &  &  &  &  \\
100 & 2.0 &  &  & -19 49 27.2635 & -43 41 24.0250 & 1498.97 &  &  &  &  &  &  \\
101 & 8.0 &  &  & -22 57 10.8000 & -68 10 44.4000 & 2400.0 &  &  &  &  &  &  \\
102 &  &  &  & -25 05 22.1251 & -50 05 56.3925 & 909.79 &  &  &  &  &  &  \\
103 & 1.0 & 1.0 & -1.5 & -30 14 16.89 & -70 44 21.12 & 2693.0 & 03:18:39.4 & 0.1 & 03:19:33.3 & 0.1 & 711.7 & 3.1 \\
104 & 5.0 & 5.0 & -7.5 & -25 20 57.39 & -49 21 55.42 & 1055.0 & 03:16:21.2 & 0.5 & 03:17:45.7 & 0.5 & 1106.5 & 12.4 \\
105 & 5.0 & 7.64 &  & -29 15 16.56 & -70 44 21.84 & 2315.0 & 03:18:47.3 & 1.6 & 03:19:18.1 & 2.3 & 406.1 & 50.6 \\
106 & 0.1 & 0.101 &  & -29 15 31.76 & -70 44 01.46 & 2345.0 & 03:18:46.82 & 0.01 & 03:19:17.85 & 0.02 & 409.1 & 0.4 \\
107 & 0.4 & 0.43 &  & -22 32 07.75 & -45 34 57.54 & 1810.0 & 03:15:53.5 & 0.1 & 03:17:16.6 & 0.3 & 1084.8 & 4.4 \\
108 & 2.0 & 4.64 &  & -22 32 07.75 & -45 34 57.54 & 1810.0 & 03:53:58.8 & 1.4 & 03:57:20.6 & 1.5 & 815.9 & 11.4 \\
109 & 2.5 & 2.5 &  & -25 20 57.39 & -49 21 55.42 & 1055.0 & 03:52:54.2 & 0.2 & 03:55:21.0 & 0.3 & 592.5 & 2.1 \\
110 & 5.0 & 6.8 &  & -08 47 32.10 & -38 41 18.70 & 390.0 & 04:00:06.4 & 0.1 & 04:04:20.1 & 0.1 & 1030.1 & 0.5 \\
111 &  &  &  & -23 00 09 & -46 57 55 & 870.0 &  &  &  &  &  &  \\
112 & 1.0 & 1.0 & -1.5 & -25 13 39.65 & -50 36 41.71 & 968.0 & 03:53:07.1 & 0.1 & 03:55:06.3 & 0.1 & 480.5 & 0.8 \\
113 &  &  &  & -19 49 27.2635 & -43 41 24.0250 & 1498.97 &  &  &  &  &  &  \\
114 & 3.0 & 3.35 &  & -25 05 22.15 & -50 05 56.40 & 909.0 & 03:53:09.3 & 0.2 & 03:55:15.8 & 0.1 & 510.3 & 1.2 \\
115 &  &  &  & -22 48 06.0012 & -45 11 25.5984 & 540.0 &  &  &  &  &  &  \\
116 & 20.0 & 20.0 &  & -34 51 50.89 & 148 58 35.06 & 536.0 & 19:12:28.9 & 4.6 & 19:13:13.3 & 4.0 & 928.3 & 179.8 \\
117 & 3.0 & 3.1 &  & 42 13 42.95 & 1 44 16.37 & 1185.0 & 22:57:28.4 & 0.6 & 22:58:05.2 & 0.6 & 919.6 & 31.3 \\
118 & 7.0 & 7.4 &  & 40 02 30.48 & -01 00 58.68 & 1957.0 & 22:57:44.6 & 2.2 & 22:58:08.2 & 3.1 & 589.7 & 131.8 \\
119 & 0.9 & 0.9 & 1.35 & 47 31 10.48 & 8 34 14.32 & 546.0 & 22:56:59.3 & 0.3 & 22:57:44.2 & 0.1 & 1119.0 & 10.7 \\
120 & 5.0 & 5.0 &  & 52 16 36.87 & 17 04 30.75 & 83.0 & 22:56:35.1 & 2.4 & 22:57:18.5 & 1.7 & 1082.4 & 101.6 \\
121 & 3.0 & 3.0 &  & 39 34 04.80 & -03 10 59.88 & 770.0 & 22:57:54.2 & 2.6 &  &  &  &  \\
122 & 10.0 & 13.7 &  & 42 03 05.96 & 00 43 46.74 & 1564.0 & 22:57:28.8 & 3.5 & 22:58:09.5 & 2.7 & 1016.5 & 155.5 \\
123 & 2.56 & 2.56 & -1.3 & 50 31 24.93 & 05 26 29.36 & 261.0 & 22:57:08.0 & 0.2 & 22:57:51.8 & 0.7 & 1092.6 & 23.4 \\
124 & 1.0 &  &  & 52 36 14.0270 & 16 02 33.4403 & 65.0 &  &  &  &  &  &  \\
125 & 5.0 &  &  & 37 03 50.8896 & -3 23 04.9200 & 2930.53 &  &  &  &  &  &  \\
126 &  &  &  & 39 34 04.8000 & -3 10 59.8800 & 770.0 &  &  &  &  &  &  \\
127 &  &  &  & 41 29 50.5500 & -1 01 15.1212 & 403.0 &  &  &  &  &  &  \\
128 & 0.1 &  &  & 19 49 25.7016 & -155 28 08.6160 & 4213.0 &  &  &  &  &  &  \\
129 & 0.1 &  &  & 19 49 31.0080 & -155 28 07.9320 & 4206.11 &  &  &  &  &  &  \\
130 & 0.266 & 0.266 & -0.15 & 36 18 21.0069 & -121 34 00.3195 & 1520.42 &  &  &  &  &  &  \\
131 & 5.0 & 5.0 & -7.5 & 31 02 36.26 & -115 27 47.88 & 2824.0 & 09:52:13.2 & 0.2 & 09:53:01.4 & 0.3 & 1036.6 & 9.0 \\
132 & 2.5 & 2.5 &  & 37 12 37.99 & -104 29 57.23 & 1862.0 & 09:51:44.4 & 1.2 & 09:52:16.2 & 1.4 & 744.8 & 62.5 \\
133 & 4.0 & 4.0 &  & 33 37 29.20 & -111 43 38.31 & 520.0 & 09:52:02.5 & 0.2 & 09:52:46.6 & 0.2 & 1036.6 & 9.0 \\
134 & 8.0 &  &  & 37 13 24.7116 & -02 32 44.8800 & 2173.21 &  &  &  &  &  &  \\
135 & 15.0 &  &  & 36 49 17.2200 & 30 20 07.9836 & 2538.72 &  &  &  &  &  &  \\
136 & 20.0 & 21.9 &  & 42 3 5.95 & 00 43 46.74 & 1564.0 & 23:47:00.9 & 4.8 & 23:47:48.3 & 5.2 & 1166.4 & 239.6 \\
137 &  &  &  & 41 01 49.0476 & 29 02 33.3996 & 110.0 &  &  &  &  &  &  \\
138 &  &  &  & 41 00 42.2964 & 28 57 56.5848 & 60.0 &  &  &  &  &  &  \\
139 &  &  &  & 41 22 03.8172 & 36 12 05.6736 & 150.0 &  &  &  &  &  &  \\
140 & 5.0 & 6.2 &  & 49 54 38.016 & 14 47 01.10 & 528.0 & 23:46:18.1 & 1.2 & 23:46:54.5 & 1.5 & 896.8 & 71.8 \\
141 & 2.0 & 2.0 &  & 47 55 04.96 & 19 53 39.48 & 932.0 & 23:46:00.5 & 0.9 & 23:46:41.9 & 0.6 & 1015.2 & 34.3 \\
142 & 10.0 & 11.5 &  & 37 03 50.88 & -03 23 4.92 & 2930.0 & 23:47:24.5 & 2.3 & 23:47:58.0 & 1.6 & 825.5 & 99.2 \\
143 & 5.0 &  &  & 50 38 17.9556 & 13 50 48.3000 & 277.94 &  &  &  &  &  &  \\
144 & 15.0 & 15.4 &  & 40 02 30.48 & -01 00 58.68 & 1957.0 & 23:47:07.0 & 7.7 & 23:47:56.1 & 5.1 & 1196.8 & 361.7 \\
145 & 5.0 & 5.5 &  & 43 26 43 & 01 49 03 & 247.0 & 23:46:56.6 & 0.8 & 23:47:42.7 & 1.2 & 1133.5 & 49.5 \\
146 &  &  &  & 39 38 11.8752 & 32 48 14.9652 & 950.0 &  &  &  &  &  &  \\
147 & 5.0 & 7.0 &  & 25 16 56.92 & 55 27 42.95 & 50.0 & 23:02:54.9 & 1.5 & 23:03:36.6 & 0.5 & 1044.4 & 48.4 \\
148 &  &  &  & 24 13 10.5780 & 54 55 12.6912 & 90.0 &  &  &  &  &  &  \\
149 & 4.0 &  &  & 30 35 48.5862 & 34 45 44.1366 & 862.27 &  &  &  &  &  &  \\
150 & 4.0 &  &  & 30 35 48.5862 & 34 45 44.1366 & 862.27 &  &  &  &  &  &  \\
151 & 4.0 &  &  & 30 35 48.5862 & 34 45 44.1366 & 862.27 &  &  &  &  &  &  \\
152 &  &  &  & 29 56 02.4000 & 31 49 37.2000 & 476.0 &  &  &  &  &  &  \\
153 & 0.125 &  &  & 41 10 15.8999 & -73 19 39.2999 & 88.0 &  &  &  &  &  &  \\
154 & 2.0 &  &  & 41 31 32.6064 & -73 25 33.8160 & 76.87 &  &  &  &  &  &  \\
155 & 2.0 & 2.0 &  & 39 25 51.12 & -103 15 59.27 & 1559.0 &  &  &  &  &  &  \\
156 & 0.533 & 0.533 & -0.284 & 41 44 48.79 & -77 19 0.39 & 436.0 & 02:38:41.0 & 1.2 & 02:38:58.5 & 1.5 & 220.7 & 31.7 \\
157 & 0.533 & 0.533 & -0.284 & 41 44 48.79 & -77 19 0.39 & 436.0 & 02:39:07.9 & 0.5 & 02:40:00.4 & 0.5 & 656.6 & 12.1 \\
158 & 0.1 & 0.1 & 13.6 & 32 46 49.80 & -105 49 14.08 & 2791.0 & 02:41:29.7 & 0.5 & 02:42:49.3 & 0.5 & 1002.6 & 12.6 \\
159 & 0.1 & 0.1 & 13.6 & 32 46 49.80 & -105 49 14.08 & 2791.0 & 02:42:00.3 & 20.0 & 02:43:33.8 & 0.2 & 1177.7 & 254.4 \\
160 & 3.0 &  &  & -31 16 22.5696 & 149 04 14.9160 & 1118.33 &  &  &  &  &  &  \\
161 & 2.56 & 2.56 & -1.3 & -34 57 31.31 & 148 59 56.00 & 595.0 & 13:41:16.6 & 0.5 & 13:42:59.1 & 0.9 & 1130.3 & 14.8 \\
162 & 5.0 & 5.0 &  & -34 51 50.89 & 148 58 35.06 & 536.0 & 13:41:15.1 & 0.7 & 13:42:55.7 & 0.7 & 1108.5 & 16.2 \\
\enddata
\end{deluxetable}
\end{longrotatetable}

\clearpage

\bibliography{sample631}{}

\begin{thebibliography}{}
\expandafter\ifx\csname natexlab\endcsname\relax\def\natexlab#1{#1}\fi
\providecommand{\url}[1]{\href{#1}{#1}}
\providecommand{\dodoi}[1]{doi:~\href{http://doi.org/#1}{\nolinkurl{#1}}}
\providecommand{\doeprint}[1]{\href{http://ascl.net/#1}{\nolinkurl{http://ascl.net/#1}}}
\providecommand{\doarXiv}[1]{\href{https://arxiv.org/abs/#1}{\nolinkurl{https://arxiv.org/abs/#1}}}

\bibitem[{{Arimatsu} {et~al.}(2019){Arimatsu}, {Ohsawa}, {Hashimoto}, {Urakawa}, {Takahashi}, {Tozuka}, {Itoh}, {Yamashita}, {Usui}, {Aoki}, {Arima}, {Doi}, {Ichiki}, {Ikeda}, {Ita}, {Kasuga}, {Kobayashi}, {Kokubo}, {Konishi}, {Maehara}, {Matsunaga}, {Miyata}, {Morii}, {Morokuma}, {Motohara}, {Nakada}, {Okumura}, {Sako}, {Sarugaku}, {Sato}, {Shigeyama}, {Soyano}, {Takahashi}, {Tarusawa}, {Tominaga}, {Watanabe}, {Yamashita}, \& {Yoshikawa}}]{Arimatsu_2019}
{Arimatsu}, K., {Ohsawa}, R., {Hashimoto}, G.~L., {et~al.} 2019, \aj, 158, 236, \dodoi{10.3847/1538-3881/ab5058}

\bibitem[{{Assafin}(2023{\natexlab{a}})}]{Assafin2023b}
{Assafin}, M. 2023{\natexlab{a}}, \planss, 238, 105801, \dodoi{10.1016/j.pss.2023.105801}

\bibitem[{{Assafin}(2023{\natexlab{b}})}]{Assafin_2023}
---. 2023{\natexlab{b}}, \planss, 239, 105816, \dodoi{10.1016/j.pss.2023.105816}

\bibitem[{{Astropy Collaboration} {et~al.}(2022){Astropy Collaboration}, {Price-Whelan}, {Lim}, {Earl}, {Starkman}, {Bradley}, {Shupe}, {Patil}, {Corrales}, {Brasseur}, {N{\"o}the}, {Donath}, {Tollerud}, {Morris}, {Ginsburg}, {Vaher}, {Weaver}, {Tocknell}, {Jamieson}, {van Kerkwijk}, {Robitaille}, {Merry}, {Bachetti}, {G{\"u}nther}, {Aldcroft}, {Alvarado-Montes}, {Archibald}, {B{\'o}di}, {Bapat}, {Barentsen}, {Baz{\'a}n}, {Biswas}, {Boquien}, {Burke}, {Cara}, {Cara}, {Conroy}, {Conseil}, {Craig}, {Cross}, {Cruz}, {D'Eugenio}, {Dencheva}, {Devillepoix}, {Dietrich}, {Eigenbrot}, {Erben}, {Ferreira}, {Foreman-Mackey}, {Fox}, {Freij}, {Garg}, {Geda}, {Glattly}, {Gondhalekar}, {Gordon}, {Grant}, {Greenfield}, {Groener}, {Guest}, {Gurovich}, {Handberg}, {Hart}, {Hatfield-Dodds}, {Homeier}, {Hosseinzadeh}, {Jenness}, {Jones}, {Joseph}, {Kalmbach}, {Karamehmetoglu}, {Ka{\l}uszy{\'n}ski}, {Kelley}, {Kern}, {Kerzendorf}, {Koch}, {Kulumani}, {Lee}, {Ly}, {Ma}, {MacBride}, {Maljaars}, {Muna}, {Murphy}, {Norman},
  {O'Steen}, {Oman}, {Pacifici}, {Pascual}, {Pascual-Granado}, {Patil}, {Perren}, {Pickering}, {Rastogi}, {Roulston}, {Ryan}, {Rykoff}, {Sabater}, {Sakurikar}, {Salgado}, {Sanghi}, {Saunders}, {Savchenko}, {Schwardt}, {Seifert-Eckert}, {Shih}, {Jain}, {Shukla}, {Sick}, {Simpson}, {Singanamalla}, {Singer}, {Singhal}, {Sinha}, {Sip{\H{o}}cz}, {Spitler}, {Stansby}, {Streicher}, {{\v{S}}umak}, {Swinbank}, {Taranu}, {Tewary}, {Tremblay}, {de Val-Borro}, {Van Kooten}, {Vasovi{\'c}}, {Verma}, {de Miranda Cardoso}, {Williams}, {Wilson}, {Winkel}, {Wood-Vasey}, {Xue}, {Yoachim}, {Zhang}, {Zonca}, \& {Astropy Project Contributors}}]{Astropy_2022}
{Astropy Collaboration}, {Price-Whelan}, A.~M., {Lim}, P.~L., {et~al.} 2022, \apj, 935, 167, \dodoi{10.3847/1538-4357/ac7c74}

\bibitem[{{Barry} {et~al.}(2015){Barry}, {Gault}, {Bolt}, {McEwan}, {Filipovi{\'c}}, \& {White}}]{Barry_2015}
{Barry}, M.~A.~T., {Gault}, D., {Bolt}, G., {et~al.} 2015, \pasa, 32, e014, \dodoi{10.1017/pasa.2015.15}

\bibitem[{{Bernardes} {et~al.}(2025){Bernardes}, {Junior}, {Rodrigues}, {Rodrigues}, {Fraga}, {Martioli}, {Gneiding}, {Alves}, {Rom{\~a}o}, {Andrade}, {Almeida}, {Mattiuci}, {Ribeiro}, {Schlindwein}, {Santos}, {Jablonski}, {Campagnolo}, \& {Laporte}}]{Bernardes_2025}
{Bernardes}, D., {Junior}, O.~V., {Rodrigues}, F., {et~al.} 2025, \pasp, 137, 035003, \dodoi{10.1088/1538-3873/ada187}

\bibitem[{Braga-Ribas {et~al.}(2025)Braga-Ribas, Vachier, Desmars, Margoti, \& Sicardy}]{Ribas_2025}
Braga-Ribas, F., Vachier, F., Desmars, J., Margoti, G., \& Sicardy, B. 2025, Philosophical Transactions A, 383, 20240200

\bibitem[{{Braga-Ribas} {et~al.}(2013){Braga-Ribas}, {Sicardy}, {Ortiz}, {Lellouch}, {Tancredi}, {Lecacheux}, {Vieira-Martins}, {Camargo}, {Assafin}, {Behrend}, {Vachier}, {Colas}, {Morales}, {Maury}, {Emilio}, {Amorim}, {Unda-Sanzana}, {Roland}, {Bruzzone}, {Almeida}, {Rodrigues}, {Jacques}, {Gil-Hutton}, {Vanzi}, {Milone}, {Schoenell}, {Salvo}, {Almenares}, {Jehin}, {Manfroid}, {Sposetti}, {Tanga}, {Klotz}, {Frappa}, {Cacella}, {Colque}, {Neves}, {Alvarez}, {Gillon}, {Pimentel}, {Giacchini}, {Roques}, {Widemann}, {Magalh{\~a}es}, {Thirouin}, {Duffard}, {Leiva}, {Toledo}, {Capeche}, {Beisker}, {Pollock}, {Cede{\~n}o Monta{\~n}a}, {Ivarsen}, {Reichart}, {Haislip}, \& {Lacluyze}}]{Ribas_2013}
{Braga-Ribas}, F., {Sicardy}, B., {Ortiz}, J.~L., {et~al.} 2013, \apj, 773, 26, \dodoi{10.1088/0004-637X/773/1/26}

\bibitem[{{Brown} \& {Suer}(2007)}]{Brown_2007}
{Brown}, M.~E., \& {Suer}, T.~A. 2007, \iaucirc, 8812, 1

\bibitem[{{Desmars} {et~al.}(2015){Desmars}, {Camargo}, {Braga-Ribas}, {Vieira-Martins}, {Assafin}, {Vachier}, {Colas}, {Ortiz}, {Duffard}, {Morales}, {Sicardy}, {Gomes-J{\'u}nior}, \& {Benedetti-Rossi}}]{Desmars_2015}
{Desmars}, J., {Camargo}, J.~I.~B., {Braga-Ribas}, F., {et~al.} 2015, \aap, 584, A96, \dodoi{10.1051/0004-6361/201526498}

\bibitem[{{Dhillon} {et~al.}(2021){Dhillon}, {Bezawada}, {Black}, {Dixon}, {Gamble}, {Gao}, {Henry}, {Kerry}, {Littlefair}, {Lunney}, {Marsh}, {Miller}, {Parsons}, {Ashley}, {Breedt}, {Brown}, {Dyer}, {Green}, {Pelisoli}, {Sahman}, {Wild}, {Ives}, {Mehrgan}, {Stegmeier}, {Dubbeldam}, {Morris}, {Osborn}, {Wilson}, {Casares}, {Mu{\~n}oz-Darias}, {Pall{\'e}}, {Rodr{\'\i}guez-Gil}, {Shahbaz}, {Torres}, {de Ugarte Postigo}, {Cabrera-Lavers}, {Corradi}, {Dom{\'\i}nguez}, \& {Garc{\'\i}a-Alvarez}}]{Dhillon_2021}
{Dhillon}, V.~S., {Bezawada}, N., {Black}, M., {et~al.} 2021, \mnras, 507, 350, \dodoi{10.1093/mnras/stab2130}

\bibitem[{{Dias-Oliveira} {et~al.}(2015){Dias-Oliveira}, {Sicardy}, {Lellouch}, {Vieira-Martins}, {Assafin}, {Camargo}, {Braga-Ribas}, {Gomes-J{\'u}nior}, {Benedetti-Rossi}, {Colas}, {Decock}, {Doressoundiram}, {Dumas}, {Emilio}, {Fabrega Polleri}, {Gil-Hutton}, {Gillon}, {Girard}, {Hau}, {Ivanov}, {Jehin}, {Lecacheux}, {Leiva}, {Lopez-Sisterna}, {Mancini}, {Manfroid}, {Maury}, {Meza}, {Morales}, {Nagy}, {Opitom}, {Ortiz}, {Pollock}, {Roques}, {Snodgrass}, {Soulier}, {Thirouin}, {Vanzi}, {Widemann}, {Reichart}, {LaCluyze}, {Haislip}, {Ivarsen}, {Dominik}, {J{\o}rgensen}, \& {Skottfelt}}]{dias2015}
{Dias-Oliveira}, A., {Sicardy}, B., {Lellouch}, E., {et~al.} 2015, \apj, 811, 53, \dodoi{10.1088/0004-637X/811/1/53}

\bibitem[{{Dias-Oliveira} {et~al.}(2017){Dias-Oliveira}, {Sicardy}, {Ortiz}, {Braga-Ribas}, {Leiva}, {Vieira-Martins}, {Benedetti-Rossi}, {Camargo}, {Assafin}, {Gomes-J{\'u}nior}, {Baug}, {Chandrasekhar}, {Desmars}, {Duffard}, {Santos-Sanz}, {Ergang}, {Ganesh}, {Ikari}, {Irawati}, {Jain}, {Liying}, {Richichi}, {Shengbang}, {Behrend}, {Benkhaldoun}, {Brosch}, {Daassou}, {Frappa}, {Gal-Yam}, {Garcia-Lozano}, {Gillon}, {Jehin}, {Kaspi}, {Klotz}, {Lecacheux}, {Mahasena}, {Manfroid}, {Manulis}, {Maury}, {Mohan}, {Morales}, {Ofek}, {Rinner}, {Sharma}, {Sposetti}, {Tanga}, {Thirouin}, {Vachier}, {Widemann}, {Asai}, {Hayato}, {Hiroyuki}, {Owada}, {Yamamura}, {Hayamizu}, {Bradshaw}, {Kerr}, {Tomioka}, {Andersson}, {Dangl}, {Haymes}, {Naves}, \& {Wortmann}}]{dias_2017}
{Dias-Oliveira}, A., {Sicardy}, B., {Ortiz}, J.~L., {et~al.} 2017, \aj, 154, 22, \dodoi{10.3847/1538-3881/aa74e9}

\bibitem[{Fortin {et~al.}(2012)Fortin, De~Rainville, Gardner, Parizeau, \& Gagn{\'e}}]{deap_2012}
Fortin, F.-A., De~Rainville, F.-M., Gardner, M.-A.~G., Parizeau, M., \& Gagn{\'e}, C. 2012, The Journal of Machine Learning Research, 13, 2171

\bibitem[{{Fraser} {et~al.}(2013){Fraser}, {Batygin}, {Brown}, \& {Bouchez}}]{Fraser_2013}
{Fraser}, W.~C., {Batygin}, K., {Brown}, M.~E., \& {Bouchez}, A. 2013, \icarus, 222, 357, \dodoi{10.1016/j.icarus.2012.11.004}

\bibitem[{{French} \& {Gierasch}(1976)}]{French1976}
{French}, R.~G., \& {Gierasch}, P.~J. 1976, \aj, 81, 445, \dodoi{10.1086/111905}

\bibitem[{{Gaia Collaboration} {et~al.}(2018){Gaia Collaboration}, {Brown}, {Vallenari}, {Prusti}, {de Bruijne}, {Babusiaux}, {Bailer-Jones}, {Biermann}, {Evans}, {Eyer}, {Jansen}, {Jordi}, {Klioner}, {Lammers}, {Lindegren}, {Luri}, {Mignard}, {Panem}, {Pourbaix}, {Randich}, {Sartoretti}, {Siddiqui}, {Soubiran}, {van Leeuwen}, {Walton}, {Arenou}, {Bastian}, {Cropper}, {Drimmel}, {Katz}, {Lattanzi}, {Bakker}, {Cacciari}, {Casta{\~n}eda}, {Chaoul}, {Cheek}, {De Angeli}, {Fabricius}, {Guerra}, {Holl}, {Masana}, {Messineo}, {Mowlavi}, {Nienartowicz}, {Panuzzo}, {Portell}, {Riello}, {Seabroke}, {Tanga}, {Th{\'e}venin}, {Gracia-Abril}, {Comoretto}, {Garcia-Reinaldos}, {Teyssier}, {Altmann}, {Andrae}, {Audard}, {Bellas-Velidis}, {Benson}, {Berthier}, {Blomme}, {Burgess}, {Busso}, {Carry}, {Cellino}, {Clementini}, {Clotet}, {Creevey}, {Davidson}, {De Ridder}, {Delchambre}, {Dell'Oro}, {Ducourant}, {Fern{\'a}ndez-Hern{\'a}ndez}, {Fouesneau}, {Fr{\'e}mat}, {Galluccio}, {Garc{\'\i}a-Torres},
  {Gonz{\'a}lez-N{\'u}{\~n}ez}, {Gonz{\'a}lez-Vidal}, {Gosset}, {Guy}, {Halbwachs}, {Hambly}, {Harrison}, {Hern{\'a}ndez}, {Hestroffer}, {Hodgkin}, {Hutton}, {Jasniewicz}, {Jean-Antoine-Piccolo}, {Jordan}, {Korn}, {Krone-Martins}, {Lanzafame}, {Lebzelter}, {L{\"o}ffler}, {Manteiga}, {Marrese}, {Mart{\'\i}n-Fleitas}, {Moitinho}, {Mora}, {Muinonen}, {Osinde}, {Pancino}, {Pauwels}, {Petit}, {Recio-Blanco}, {Richards}, {Rimoldini}, {Robin}, {Sarro}, {Siopis}, {Smith}, {Sozzetti}, {S{\"u}veges}, {Torra}, {van Reeven}, {Abbas}, {Abreu Aramburu}, {Accart}, {Aerts}, {Altavilla}, {{\'A}lvarez}, {Alvarez}, {Alves}, {Anderson}, {Andrei}, {Anglada Varela}, {Antiche}, {Antoja}, {Arcay}, {Astraatmadja}, {Bach}, {Baker}, {Balaguer-N{\'u}{\~n}ez}, {Balm}, {Barache}, {Barata}, {Barbato}, {Barblan}, {Barklem}, {Barrado}, {Barros}, {Barstow}, {Bartholom{\'e} Mu{\~n}oz}, {Bassilana}, {Becciani}, {Bellazzini}, {Berihuete}, {Bertone}, {Bianchi}, {Bienaym{\'e}}, {Blanco-Cuaresma}, {Boch}, {Boeche}, {Bombrun}, {Borrachero},
  {Bossini}, {Bouquillon}, {Bourda}, {Bragaglia}, {Bramante}, {Breddels}, {Bressan}, {Brouillet}, {Br{\"u}semeister}, {Brugaletta}, {Bucciarelli}, {Burlacu}, {Busonero}, {Butkevich}, {Buzzi}, {Caffau}, {Cancelliere}, {Cannizzaro}, {Cantat-Gaudin}, {Carballo}, {Carlucci}, {Carrasco}, {Casamiquela}, {Castellani}, {Castro-Ginard}, {Charlot}, {Chemin}, {Chiavassa}, {Cocozza}, {Costigan}, {Cowell}, {Crifo}, {Crosta}, {Crowley}, {Cuypers}, {Dafonte}, {Damerdji}, {Dapergolas}, {David}, {David}, {de Laverny}, {De Luise}, {De March}, {de Martino}, {de Souza}, {de Torres}, {Debosscher}, {del Pozo}, {Delbo}, {Delgado}, {Delgado}, {Di Matteo}, {Diakite}, {Diener}, {Distefano}, {Dolding}, {Drazinos}, {Dur{\'a}n}, {Edvardsson}, {Enke}, {Eriksson}, {Esquej}, {Eynard Bontemps}, {Fabre}, {Fabrizio}, {Faigler}, {Falc{\~a}o}, {Farr{\`a}s Casas}, {Federici}, {Fedorets}, {Fernique}, {Figueras}, {Filippi}, {Findeisen}, {Fonti}, {Fraile}, {Fraser}, {Fr{\'e}zouls}, {Gai}, {Galleti}, {Garabato}, {Garc{\'\i}a-Sedano}, {Garofalo},
  {Garralda}, {Gavel}, {Gavras}, {Gerssen}, {Geyer}, {Giacobbe}, {Gilmore}, {Girona}, {Giuffrida}, {Glass}, {Gomes}, {Granvik}, {Gueguen}, {Guerrier}, {Guiraud}, {Guti{\'e}rrez-S{\'a}nchez}, {Haigron}, {Hatzidimitriou}, {Hauser}, {Haywood}, {Heiter}, {Helmi}, {Heu}, {Hilger}, {Hobbs}, {Hofmann}, {Holland}, {Huckle}, {Hypki}, {Icardi}, {Jan{\ss}en}, {Jevardat de Fombelle}, {Jonker}, {Juh{\'a}sz}, {Julbe}, {Karampelas}, {Kewley}, {Klar}, {Kochoska}, {Kohley}, {Kolenberg}, {Kontizas}, {Kontizas}, {Koposov}, {Kordopatis}, {Kostrzewa-Rutkowska}, {Koubsky}, {Lambert}, {Lanza}, {Lasne}, {Lavigne}, {Le Fustec}, {Le Poncin-Lafitte}, {Lebreton}, {Leccia}, {Leclerc}, {Lecoeur-Taibi}, {Lenhardt}, {Leroux}, {Liao}, {Licata}, {Lindstr{\o}m}, {Lister}, {Livanou}, {Lobel}, {L{\'o}pez}, {Managau}, {Mann}, {Mantelet}, {Marchal}, {Marchant}, {Marconi}, {Marinoni}, {Marschalk{\'o}}, {Marshall}, {Martino}, {Marton}, {Mary}, {Massari}, {Matijevi{\v{c}}}, {Mazeh}, {McMillan}, {Messina}, {Michalik}, {Millar}, {Molina}, {Molinaro},
  {Moln{\'a}r}, {Montegriffo}, {Mor}, {Morbidelli}, {Morel}, {Morris}, {Mulone}, {Muraveva}, {Musella}, {Nelemans}, {Nicastro}, {Noval}, {O'Mullane}, {Ord{\'e}novic}, {Ord{\'o}{\~n}ez-Blanco}, {Osborne}, {Pagani}, {Pagano}, {Pailler}, {Palacin}, {Palaversa}, {Panahi}, {Pawlak}, {Piersimoni}, {Pineau}, {Plachy}, {Plum}, {Poggio}, {Poujoulet}, {Pr{\v{s}}a}, {Pulone}, {Racero}, {Ragaini}, {Rambaux}, {Ramos-Lerate}, {Regibo}, {Reyl{\'e}}, {Riclet}, {Ripepi}, {Riva}, {Rivard}, {Rixon}, {Roegiers}, {Roelens}, {Romero-G{\'o}mez}, {Rowell}, {Royer}, {Ruiz-Dern}, {Sadowski}, {Sagrist{\`a} Sell{\'e}s}, {Sahlmann}, {Salgado}, {Salguero}, {Sanna}, {Santana-Ros}, {Sarasso}, {Savietto}, {Schultheis}, {Sciacca}, {Segol}, {Segovia}, {S{\'e}gransan}, {Shih}, {Siltala}, {Silva}, {Smart}, {Smith}, {Solano}, {Solitro}, {Sordo}, {Soria Nieto}, {Souchay}, {Spagna}, {Spoto}, {Stampa}, {Steele}, {Steidelm{\"u}ller}, {Stephenson}, {Stoev}, {Suess}, {Surdej}, {Szabados}, {Szegedi-Elek}, {Tapiador}, {Taris}, {Tauran}, {Taylor},
  {Teixeira}, {Terrett}, {Teyssandier}, {Thuillot}, {Titarenko}, {Torra Clotet}, {Turon}, {Ulla}, {Utrilla}, {Uzzi}, {Vaillant}, {Valentini}, {Valette}, {van Elteren}, {Van Hemelryck}, {van Leeuwen}, {Vaschetto}, {Vecchiato}, {Veljanoski}, {Viala}, {Vicente}, {Vogt}, {von Essen}, {Voss}, {Votruba}, {Voutsinas}, {Walmsley}, {Weiler}, {Wertz}, {Wevers}, {Wyrzykowski}, {Yoldas}, {{\v{Z}}erjal}, {Ziaeepour}, {Zorec}, {Zschocke}, {Zucker}, {Zurbach}, \& {Zwitter}}]{Gaia_2018}
{Gaia Collaboration}, {Brown}, A.~G.~A., {Vallenari}, A., {et~al.} 2018, \aap, 616, A1, \dodoi{10.1051/0004-6361/201833051}

\bibitem[{{Gaia Collaboration} {et~al.}(2021){Gaia Collaboration}, {Brown}, {Vallenari}, {Prusti}, {de Bruijne}, {Babusiaux}, {Biermann}, {Creevey}, {Evans}, {Eyer}, {Hutton}, {Jansen}, {Jordi}, {Klioner}, {Lammers}, {Lindegren}, {Luri}, {Mignard}, {Panem}, {Pourbaix}, {Randich}, {Sartoretti}, {Soubiran}, {Walton}, {Arenou}, {Bailer-Jones}, {Bastian}, {Cropper}, {Drimmel}, {Katz}, {Lattanzi}, {van Leeuwen}, {Bakker}, {Cacciari}, {Casta{\~n}eda}, {De Angeli}, {Ducourant}, {Fabricius}, {Fouesneau}, {Fr{\'e}mat}, {Guerra}, {Guerrier}, {Guiraud}, {Jean-Antoine Piccolo}, {Masana}, {Messineo}, {Mowlavi}, {Nicolas}, {Nienartowicz}, {Pailler}, {Panuzzo}, {Riclet}, {Roux}, {Seabroke}, {Sordo}, {Tanga}, {Th{\'e}venin}, {Gracia-Abril}, {Portell}, {Teyssier}, {Altmann}, {Andrae}, {Bellas-Velidis}, {Benson}, {Berthier}, {Blomme}, {Brugaletta}, {Burgess}, {Busso}, {Carry}, {Cellino}, {Cheek}, {Clementini}, {Damerdji}, {Davidson}, {Delchambre}, {Dell'Oro}, {Fern{\'a}ndez-Hern{\'a}ndez}, {Galluccio}, {Garc{\'\i}a-Lario},
  {Garcia-Reinaldos}, {Gonz{\'a}lez-N{\'u}{\~n}ez}, {Gosset}, {Haigron}, {Halbwachs}, {Hambly}, {Harrison}, {Hatzidimitriou}, {Heiter}, {Hern{\'a}ndez}, {Hestroffer}, {Hodgkin}, {Holl}, {Jan{\ss}en}, {Jevardat de Fombelle}, {Jordan}, {Krone-Martins}, {Lanzafame}, {L{\"o}ffler}, {Lorca}, {Manteiga}, {Marchal}, {Marrese}, {Moitinho}, {Mora}, {Muinonen}, {Osborne}, {Pancino}, {Pauwels}, {Petit}, {Recio-Blanco}, {Richards}, {Riello}, {Rimoldini}, {Robin}, {Roegiers}, {Rybizki}, {Sarro}, {Siopis}, {Smith}, {Sozzetti}, {Ulla}, {Utrilla}, {van Leeuwen}, {van Reeven}, {Abbas}, {Abreu Aramburu}, {Accart}, {Aerts}, {Aguado}, {Ajaj}, {Altavilla}, {{\'A}lvarez}, {{\'A}lvarez Cid-Fuentes}, {Alves}, {Anderson}, {Anglada Varela}, {Antoja}, {Audard}, {Baines}, {Baker}, {Balaguer-N{\'u}{\~n}ez}, {Balbinot}, {Balog}, {Barache}, {Barbato}, {Barros}, {Barstow}, {Bartolom{\'e}}, {Bassilana}, {Bauchet}, {Baudesson-Stella}, {Becciani}, {Bellazzini}, {Bernet}, {Bertone}, {Bianchi}, {Blanco-Cuaresma}, {Boch}, {Bombrun}, {Bossini},
  {Bouquillon}, {Bragaglia}, {Bramante}, {Breedt}, {Bressan}, {Brouillet}, {Bucciarelli}, {Burlacu}, {Busonero}, {Butkevich}, {Buzzi}, {Caffau}, {Cancelliere}, {C{\'a}novas}, {Cantat-Gaudin}, {Carballo}, {Carlucci}, {Carnerero}, {Carrasco}, {Casamiquela}, {Castellani}, {Castro-Ginard}, {Castro Sampol}, {Chaoul}, {Charlot}, {Chemin}, {Chiavassa}, {Cioni}, {Comoretto}, {Cooper}, {Cornez}, {Cowell}, {Crifo}, {Crosta}, {Crowley}, {Dafonte}, {Dapergolas}, {David}, {David}, {de Laverny}, {De Luise}, {De March}, {De Ridder}, {de Souza}, {de Teodoro}, {de Torres}, {del Peloso}, {del Pozo}, {Delbo}, {Delgado}, {Delgado}, {Delisle}, {Di Matteo}, {Diakite}, {Diener}, {Distefano}, {Dolding}, {Eappachen}, {Edvardsson}, {Enke}, {Esquej}, {Fabre}, {Fabrizio}, {Faigler}, {Fedorets}, {Fernique}, {Fienga}, {Figueras}, {Fouron}, {Fragkoudi}, {Fraile}, {Franke}, {Gai}, {Garabato}, {Garcia-Gutierrez}, {Garc{\'\i}a-Torres}, {Garofalo}, {Gavras}, {Gerlach}, {Geyer}, {Giacobbe}, {Gilmore}, {Girona}, {Giuffrida}, {Gomel}, {Gomez},
  {Gonzalez-Santamaria}, {Gonz{\'a}lez-Vidal}, {Granvik}, {Guti{\'e}rrez-S{\'a}nchez}, {Guy}, {Hauser}, {Haywood}, {Helmi}, {Hidalgo}, {Hilger}, {H{\l}adczuk}, {Hobbs}, {Holland}, {Huckle}, {Jasniewicz}, {Jonker}, {Juaristi Campillo}, {Julbe}, {Karbevska}, {Kervella}, {Khanna}, {Kochoska}, {Kontizas}, {Kordopatis}, {Korn}, {Kostrzewa-Rutkowska}, {Kruszy{\'n}ska}, {Lambert}, {Lanza}, {Lasne}, {Le Campion}, {Le Fustec}, {Lebreton}, {Lebzelter}, {Leccia}, {Leclerc}, {Lecoeur-Taibi}, {Liao}, {Licata}, {Lindstr{\o}m}, {Lister}, {Livanou}, {Lobel}, {Madrero Pardo}, {Managau}, {Mann}, {Marchant}, {Marconi}, {Marcos Santos}, {Marinoni}, {Marocco}, {Marshall}, {Martin Polo}, {Mart{\'\i}n-Fleitas}, {Masip}, {Massari}, {Mastrobuono-Battisti}, {Mazeh}, {McMillan}, {Messina}, {Michalik}, {Millar}, {Mints}, {Molina}, {Molinaro}, {Moln{\'a}r}, {Montegriffo}, {Mor}, {Morbidelli}, {Morel}, {Morris}, {Mulone}, {Munoz}, {Muraveva}, {Murphy}, {Musella}, {Noval}, {Ord{\'e}novic}, {Orr{\`u}}, {Osinde}, {Pagani}, {Pagano},
  {Palaversa}, {Palicio}, {Panahi}, {Pawlak}, {Pe{\~n}alosa Esteller}, {Penttil{\"a}}, {Piersimoni}, {Pineau}, {Plachy}, {Plum}, {Poggio}, {Poretti}, {Poujoulet}, {Pr{\v{s}}a}, {Pulone}, {Racero}, {Ragaini}, {Rainer}, {Raiteri}, {Rambaux}, {Ramos}, {Ramos-Lerate}, {Re Fiorentin}, {Regibo}, {Reyl{\'e}}, {Ripepi}, {Riva}, {Rixon}, {Robichon}, {Robin}, {Roelens}, {Rohrbasser}, {Romero-G{\'o}mez}, {Rowell}, {Royer}, {Rybicki}, {Sadowski}, {Sagrist{\`a} Sell{\'e}s}, {Sahlmann}, {Salgado}, {Salguero}, {Samaras}, {Sanchez Gimenez}, {Sanna}, {Santove{\~n}a}, {Sarasso}, {Schultheis}, {Sciacca}, {Segol}, {Segovia}, {S{\'e}gransan}, {Semeux}, {Shahaf}, {Siddiqui}, {Siebert}, {Siltala}, {Slezak}, {Smart}, {Solano}, {Solitro}, {Souami}, {Souchay}, {Spagna}, {Spoto}, {Steele}, {Steidelm{\"u}ller}, {Stephenson}, {S{\"u}veges}, {Szabados}, {Szegedi-Elek}, {Taris}, {Tauran}, {Taylor}, {Teixeira}, {Thuillot}, {Tonello}, {Torra}, {Torra}, {Turon}, {Unger}, {Vaillant}, {van Dillen}, {Vanel}, {Vecchiato}, {Viala}, {Vicente},
  {Voutsinas}, {Weiler}, {Wevers}, {Wyrzykowski}, {Yoldas}, {Yvard}, {Zhao}, {Zorec}, {Zucker}, {Zurbach}, \& {Zwitter}}]{Gaia_2021}
---. 2021, \aap, 649, A1, \dodoi{10.1051/0004-6361/202039657}

\bibitem[{{Gazeas}(2016)}]{Gazeas2016}
{Gazeas}, K. 2016, in Revista Mexicana de Astronomia y Astrofisica Conference Series, Vol.~48, Revista Mexicana de Astronomia y Astrofisica Conference Series, 22--23

\bibitem[{{Gomes-J{\'u}nior} {et~al.}(2022){Gomes-J{\'u}nior}, {Morgado}, {Benedetti-Rossi}, {Boufleur}, {Rommel}, {Banda-Huarca}, {Kilic}, {Braga-Ribas}, \& {Sicardy}}]{Gomes_2022}
{Gomes-J{\'u}nior}, A.~R., {Morgado}, B.~E., {Benedetti-Rossi}, G., {et~al.} 2022, \mnras, 511, 1167, \dodoi{10.1093/mnras/stac032}

\bibitem[{Hanisch {et~al.}(2001)Hanisch, Farris, Greisen, Pence, Schlesinger, Teuben, Thompson, \& Warnock}]{Hanisch_2001}
Hanisch, R.~J., Farris, A., Greisen, E.~W., {et~al.} 2001, Astronomy \& Astrophysics, 376, 359

\bibitem[{Harris {et~al.}(2020)Harris, Millman, Van Der~Walt, Gommers, Virtanen, Cournapeau, Wieser, Taylor, Berg, Smith, {et~al.}}]{numpy_2020}
Harris, C.~R., Millman, K.~J., Van Der~Walt, S.~J., {et~al.} 2020, Nature, 585, 357

\bibitem[{{Jehin} {et~al.}(2011){Jehin}, {Gillon}, {Queloz}, {Magain}, {Manfroid}, {Chantry}, {Lendl}, {Hutsem{\'e}kers}, \& {Udry}}]{Jehin_2011}
{Jehin}, E., {Gillon}, M., {Queloz}, D., {et~al.} 2011, The Messenger, 145, 2

\bibitem[{{Johnson} \& {McGetchin}(1973)}]{Johnson_1973}
{Johnson}, T.~V., \& {McGetchin}, T.~R. 1973, \icarus, 18, 612, \dodoi{10.1016/0019-1035(73)90064-X}

\bibitem[{{Kilic} {et~al.}(2022){Kilic}, {Braga-Ribas}, {Kaplan}, {Erece}, {Souami}, {Dindar}, {Desmars}, {Sicardy}, {Morgado}, {Shameoni}, {Rommel}, \& {Gomes-J{\'u}nior}}]{Kilic_2022}
{Kilic}, Y., {Braga-Ribas}, F., {Kaplan}, M., {et~al.} 2022, \mnras, 515, 1346, \dodoi{10.1093/mnras/stac1595}

\bibitem[{{Kiss} {et~al.}(2024){Kiss}, {M{\"u}ller}, {Marton}, {Szak{\'a}ts}, {P{\'a}l}, {Moln{\'a}r}, {Vilenius}, {Rengel}, {Ortiz}, \& {Fern{\'a}ndez-Valenzuela}}]{Kiss_2024}
{Kiss}, C., {M{\"u}ller}, T.~G., {Marton}, G., {et~al.} 2024, \aap, 684, A50, \dodoi{10.1051/0004-6361/202348054}

\bibitem[{{Klioner}(2003)}]{Klioner_2003}
{Klioner}, S.~A. 2003, \aj, 125, 1580, \dodoi{10.1086/367593}

\bibitem[{{Leiva} {et~al.}(2017){Leiva}, {Sicardy}, {Camargo}, {Ortiz}, {Desmars}, {B{\'e}rard}, {Lellouch}, {Meza}, {Kervella}, {Snodgrass}, {Duffard}, {Morales}, {Gomes-J{\'u}nior}, {Benedetti-Rossi}, {Vieira-Martins}, {Braga-Ribas}, {Assafin}, {Morgado}, {Colas}, {De Witt}, {Sickafoose}, {Breytenbach}, {Dauvergne}, {Schoenau}, {Maquet}, {Bath}, {Bode}, {Cool}, {Lade}, {Kerr}, \& {Herald}}]{Leiva_2017}
{Leiva}, R., {Sicardy}, B., {Camargo}, J.~I.~B., {et~al.} 2017, \aj, 154, 159, \dodoi{10.3847/1538-3881/aa8956}

\bibitem[{{Lykawka} \& {Mukai}(2007)}]{Lykawka_2007}
{Lykawka}, P.~S., \& {Mukai}, T. 2007, \icarus, 189, 213, \dodoi{10.1016/j.icarus.2007.01.001}

\bibitem[{Margoti(2024)}]{margoti2024determinaccao}
Margoti, G. 2024, Master's thesis, Universidade Tecnol{\'o}gica Federal do Paran{\'a}

\bibitem[{{Morgado} {et~al.}(2021){Morgado}, {Sicardy}, {Braga-Ribas}, {Desmars}, {Gomes-J{\'u}nior}, {B{\'e}rard}, {Leiva}, {Ortiz}, {Vieira-Martins}, {Benedetti-Rossi}, {Santos-Sanz}, {Camargo}, {Duffard}, {Rommel}, {Assafin}, {Boufleur}, {Colas}, {Kretlow}, {Beisker}, {Sfair}, {Snodgrass}, {Morales}, {Fern{\'a}ndez-Valenzuela}, {Amaral}, {Amarante}, {Artola}, {Backes}, {Bath}, {Bouley}, {Buie}, {Cacella}, {Colazo}, {Colque}, {Dauvergne}, {Dominik}, {Emilio}, {Erickson}, {Evans}, {Fabrega-Polleri}, {Garcia-Lambas}, {Giacchini}, {Hanna}, {Herald}, {Hesler}, {Hinse}, {Jacques}, {Jehin}, {J{\o}rgensen}, {Kerr}, {Kouprianov}, {Levine}, {Linder}, {Maley}, {Machado}, {Maquet}, {Maury}, {Melia}, {Meza}, {Mondon}, {Moura}, {Newman}, {Payet}, {Pereira}, {Pollock}, {Poltronieri}, {Quispe-Huaynasi}, {Reichart}, {de Santana}, {Schneiter}, {Sieyra}, {Skottfelt}, {Soulier}, {Starck}, {Thierry}, {Torres}, {Trabuco}, {Unda-Sanzana}, {Yamashita}, {Winter}, {Zapata}, \& {Zuluaga}}]{Morgado_2021}
{Morgado}, B.~E., {Sicardy}, B., {Braga-Ribas}, F., {et~al.} 2021, \aap, 652, A141, \dodoi{10.1051/0004-6361/202141543}

\bibitem[{{Morgado} {et~al.}(2022){Morgado}, {Bruno}, {Gomes-J{\'u}nior}, {Pagano}, {Sicardy}, {Fortier}, {Desmars}, {Maxted}, {Braga-Ribas}, {Queloz}, {Sousa}, {Ortiz}, {Brandeker}, {Collier Cameron}, {Pereira}, {Flor{\'e}n}, {Hara}, {Souami}, {Isaak}, {Olofsson}, {Santos-Sanz}, {Wilson}, {Broughton}, {Alibert}, {Alonso}, {Anglada}, {B{\'a}rczy}, {Barrado}, {Barros}, {Baumjohann}, {Beck}, {Beck}, {Benz}, {Billot}, {Bonfils}, {Broeg}, {Cabrera}, {Charnoz}, {Csizmadia}, {Davies}, {Deleuil}, {Delrez}, {Demangeon}, {Demory}, {Ehrenreich}, {Erikson}, {Fossati}, {Fridlund}, {Gandolfi}, {Gillon}, {G{\"u}del}, {Heng}, {Hoyer}, {Kiss}, {Laskar}, {Lecavelier des Etangs}, {Lendl}, {Lovis}, {Magrin}, {Marafatto}, {Nascimbeni}, {Ottensamer}, {Pall{\'e}}, {Peter}, {Piazza}, {Piotto}, {Pollacco}, {Ragazzoni}, {Rando}, {Ratti}, {Rauer}, {Reimers}, {Ribas}, {Santos}, {Scandariato}, {S{\'e}gransan}, {Simon}, {Smith}, {Steller}, {Szab{\'o}}, {Thomas}, {Udry}, {Van Grootel}, {Walton}, \& {Westerdorff}}]{morgado2022}
{Morgado}, B.~E., {Bruno}, G., {Gomes-J{\'u}nior}, A.~R., {et~al.} 2022, \aap, 664, L15, \dodoi{10.1051/0004-6361/202244221}

\bibitem[{{Morgado} {et~al.}(2023){Morgado}, {Sicardy}, {Braga-Ribas}, {Ortiz}, {Salo}, {Vachier}, {Desmars}, {Pereira}, {Santos-Sanz}, {Sfair}, {de Santana}, {Assafin}, {Vieira-Martins}, {Gomes-J{\'u}nior}, {Margoti}, {Dhillon}, {Fern{\'a}ndez-Valenzuela}, {Broughton}, {Bradshaw}, {Langersek}, {Benedetti-Rossi}, {Souami}, {Holler}, {Kretlow}, {Boufleur}, {Camargo}, {Duffard}, {Beisker}, {Morales}, {Lecacheux}, {Rommel}, {Herald}, {Benz}, {Jehin}, {Jankowsky}, {Marsh}, {Littlefair}, {Bruno}, {Pagano}, {Brandeker}, {Collier-Cameron}, {Flor{\'e}n}, {Hara}, {Olofsson}, {Wilson}, {Benkhaldoun}, {Busuttil}, {Burdanov}, {Ferrais}, {Gault}, {Gillon}, {Hanna}, {Kerr}, {Kolb}, {Nosworthy}, {Sebastian}, {Snodgrass}, {Teng}, \& {de Wit}}]{Morgado_2023}
{Morgado}, B.~E., {Sicardy}, B., {Braga-Ribas}, F., {et~al.} 2023, \nat, 614, 239, \dodoi{10.1038/s41586-022-05629-6}

\bibitem[{{Nolthenius} {et~al.}(2025){Nolthenius}, {Bender}, {Cotton}, {Proudfoot}, \& {Irwin}}]{Nolthenius_2025}
{Nolthenius}, R., {Bender}, K., {Cotton}, D.~V., {Proudfoot}, B. C.~N., \& {Irwin}, J. 2025, Research Notes of the American Astronomical Society, 9, 226, \dodoi{10.3847/2515-5172/adfeda}

\bibitem[{{Ortiz} {et~al.}(2003){Ortiz}, {Guti{\'e}rrez}, {Sota}, {Casanova}, \& {Teixeira}}]{Ortiz_2003}
{Ortiz}, J.~L., {Guti{\'e}rrez}, P.~J., {Sota}, A., {Casanova}, V., \& {Teixeira}, V.~R. 2003, \aap, 409, L13, \dodoi{10.1051/0004-6361:20031253}

\bibitem[{{Ortiz} {et~al.}(2012){Ortiz}, {Sicardy}, {Braga-Ribas}, {Alvarez-Candal}, {Lellouch}, {Duffard}, {Pinilla-Alonso}, {Ivanov}, {Littlefair}, {Camargo}, {Assafin}, {Unda-Sanzana}, {Jehin}, {Morales}, {Tancredi}, {Gil-Hutton}, {de La Cueva}, {Colque}, {da Silva Neto}, {Manfroid}, {Thirouin}, {Guti{\'e}rrez}, {Lecacheux}, {Gillon}, {Maury}, {Colas}, {Licandro}, {Mueller}, {Jacques}, {Weaver}, {Milone}, {Salvo}, {Bruzzone}, {Organero}, {Behrend}, {Roland}, {Vieira-Martins}, {Widemann}, {Roques}, {Santos-Sanz}, {Hestroffer}, {Dhillon}, {Marsh}, {Harlingten}, {Campo Bagatin}, {Alonso}, {Ortiz}, {Colazo}, {Lima}, {Oliveira}, {Kerber}, {Smiljanic}, {Pimentel}, {Giacchini}, {Cacella}, \& {Emilio}}]{Ortiz_2012}
{Ortiz}, J.~L., {Sicardy}, B., {Braga-Ribas}, F., {et~al.} 2012, \nat, 491, 566, \dodoi{10.1038/nature11597}

\bibitem[{{Pavlov} {et~al.}(2020){Pavlov}, {Anderson}, {Barry}, {Gault}, {Glover}, {Meister}, \& {Schweizer}}]{Pavlov_2020}
{Pavlov}, H., {Anderson}, R., {Barry}, T., {et~al.} 2020, Journal for Occultation Astronomy, 10, 8

\bibitem[{{Pedregosa} {et~al.}(2011){Pedregosa}, {Varoquaux}, {Gramfort}, {Michel}, {Thirion}, {Grisel}, {Blondel}, {M{\"u}ller}, {Nothman}, {Louppe}, {Prettenhofer}, {Weiss}, {Dubourg}, {Vanderplas}, {Passos}, {Cournapeau}, {Brucher}, {Perrot}, \& {Duchesnay}}]{Scikitlearn_2011}
{Pedregosa}, F., {Varoquaux}, G., {Gramfort}, A., {et~al.} 2011, Journal of Machine Learning Research, 12, 2825, \dodoi{10.48550/arXiv.1201.0490}

\bibitem[{{Pereira} {et~al.}(2023){Pereira}, {Sicardy}, {Morgado}, {Braga-Ribas}, {Fern{\'a}ndez-Valenzuela}, {Souami}, {Holler}, {Boufleur}, {Margoti}, {Assafin}, {Ortiz}, {Santos-Sanz}, {Epinat}, {Kervella}, {Desmars}, {Vieira-Martins}, {Kilic}, {Gomes J{\'u}nior}, {Camargo}, {Emilio}, {Vara-Lubiano}, {Kretlow}, {Albert}, {Alcock}, {Ball}, {Bender}, {Buie}, {Butterfield}, {Camarca}, {Castro-Chac{\'o}n}, {Dunford}, {Fisher}, {Gamble}, {Geary}, {Gnilka}, {Green}, {Hartman}, {Huang}, {Januszewski}, {Johnston}, {Kagitani}, {Kamin}, {Kavelaars}, {Keller}, {de Kleer}, {Lehner}, {Luken}, {Marchis}, {Marlin}, {McGregor}, {Nikitin}, {Nolthenius}, {Patrick}, {Redfield}, {Rengstorf}, {Reyes-Ruiz}, {Seccull}, {Skrutskie}, {Smith}, {Sproul}, {Stephens}, {Szentgyorgyi}, {S{\'a}nchez-Sanju{\'a}n}, {Tatsumi}, {Verbiscer}, {Wang}, {Yoshida}, {Young}, \& {Zhang}}]{Pereira_2023}
{Pereira}, C.~L., {Sicardy}, B., {Morgado}, B.~E., {et~al.} 2023, \aap, 673, L4, \dodoi{10.1051/0004-6361/202346365}

\bibitem[{{Person} {et~al.}(2011){Person}, {Elliot}, {Bosh}, {Zangari}, {Zuluaga}, {Brothers}, {Sallum}, {Levine}, {Bright}, {Sheppard}, \& {Tilleman}}]{Person_2011}
{Person}, M.~J., {Elliot}, J.~L., {Bosh}, A.~S., {et~al.} 2011, in American Astronomical Society Meeting Abstracts, Vol. 218, American Astronomical Society Meeting Abstracts \#218, 224.12

\bibitem[{{Poiani} {et~al.}(2026){Poiani}, {Gomes-Júnior}, {Camargo}, {Morgado}, {Assafin}, \& {Vieira-Martins}}]{Poiani_2026}
{Poiani}, M., {Gomes-Júnior}, A.~R., {Camargo}, J.~I.~B., {et~al.} 2026

\bibitem[{{Proudfoot} {et~al.}(2025){Proudfoot}, {Holler}, {Arimatsu}, {Rommel}, {Collyer}, \& {Fern{\'a}ndez-Valenzuela}}]{proudfoot2025}
{Proudfoot}, B., {Holler}, B.~J., {Arimatsu}, K., {et~al.} 2025, \psj, 6, 146, \dodoi{10.3847/PSJ/addd02}

\bibitem[{{Rabinowitz} {et~al.}(2007){Rabinowitz}, {Schaefer}, \& {Tourtellotte}}]{Rabinowitz_2007}
{Rabinowitz}, D.~L., {Schaefer}, B.~E., \& {Tourtellotte}, S.~W. 2007, \aj, 133, 26, \dodoi{10.1086/508931}

\bibitem[{{Rizos} {et~al.}(2025){Rizos}, {Ortiz}, {Rommel}, {Sicardy}, {Morales}, {Santos-Sanz}, {Fern{\'a}ndez-Valenzuela}, {Desmars}, {Souami}, {Kretlow}, {Alvarez-Candal}, {G{\'o}mez-Lim{\'o}n}, {Duffard}, {Kilic}, {Morales}, {Holler}, {Vara-Lubiano}, {Marciniak}, {Kashuba}, {Koshkin}, {Kashuba}, {Pal}, {Szab{\'o}}, {Derekas}, {Szigeti}, {Ellington}, {Schreurs}, {Mottola}, {Iglesias-Marzoa}, {Ma{\'\i}cas}, {Galindo-Guil}, {Organero}, {Ana}, {Getrost}, {Nikitin}, {Verbiscer}, {Skrutskie}, {Gray}, {Malacarne}, {Jacques}, {Cacella}, {Canales}, {Lafuente}, {Calavia}, {Oncins}, {Assafin}, {Braga-Ribas}, {Camargo}, {Gomes-J{\'u}nior}, {Morgado}, {Gradovski}, {Vieira-Martins}, {Colas}, {Tekes}, {Erece}, {Kaplan}, {Schweizer}, \& {Kubanek}}]{Rizos_2025}
{Rizos}, J.~L., {Ortiz}, J.~L., {Rommel}, F.~L., {et~al.} 2025, \aap, 697, A62, \dodoi{10.1051/0004-6361/202554154}

\bibitem[{{Rodr{\'\i}guez} {et~al.}(2023){Rodr{\'\i}guez}, {Morgado}, \& {Callegari}}]{Rodriguez_2023}
{Rodr{\'\i}guez}, A., {Morgado}, B.~E., \& {Callegari}, Jr., N. 2023, \mnras, 525, 3376, \dodoi{10.1093/mnras/stad2413}

\bibitem[{{Rommel} {et~al.}(2020){Rommel}, {Braga-Ribas}, {Desmars}, {Camargo}, {Ortiz}, {Sicardy}, {Vieira-Martins}, {Assafin}, {Santos-Sanz}, {Duffard}, {Fern{\'a}ndez-Valenzuela}, {Lecacheux}, {Morgado}, {Benedetti-Rossi}, {Gomes-J{\'u}nior}, {Pereira}, {Herald}, {Hanna}, {Bradshaw}, {Morales}, {Brimacombe}, {Burtovoi}, {Carruthers}, {de Barros}, {Fiori}, {Gilmore}, {Hooper}, {Hornoch}, {Jacques}, {Janik}, {Kerr}, {Kilmartin}, {Winkel}, {Naletto}, {Nardiello}, {Nascimbeni}, {Newman}, {Ossola}, {P{\'a}l}, {Pimentel}, {Pravec}, {Sposetti}, {Stechina}, {Szak{\'a}ts}, {Ueno}, {Zampieri}, {Broughton}, {Dunham}, {Dunham}, {Gault}, {Hayamizu}, {Hosoi}, {Jehin}, {Jones}, {Kitazaki}, {Kom{\v{z}}{\'\i}k}, {Marciniak}, {Maury}, {Miku{\v{z}}}, {Nosworthy}, {F{\'a}brega Polleri}, {Rahvar}, {Sfair}, {Siqueira}, {Snodgrass}, {Sogorb}, {Tomioka}, {Tregloan-Reed}, \& {Winter}}]{Rommel_2020}
{Rommel}, F.~L., {Braga-Ribas}, F., {Desmars}, J., {et~al.} 2020, \aap, 644, A40, \dodoi{10.1051/0004-6361/202039054}

\bibitem[{{Rommel} {et~al.}(2025){Rommel}, {Fern{\'a}ndez-Valenzuela}, {Proudfoot}, {Ortiz}, {Morgado}, {Sicardy}, {Morales}, {Braga-Ribas}, {Desmars}, {Vieira-Martins}, {Holler}, {Kilic}, {Grundy}, {Rizos}, {Camargo}, {Benedetti-Rossi}, {Gomes-J{\'u}nior}, {Assafin}, {Santos-Sanz}, {Kretlow}, {Vara-Lubiano}, {Leiva}, {Ragozzine}, {Duffard}, {Ku{\v{c}}{\'a}kov{\'a}}, {Hornoch}, {Nikitin}, {Santana-Ros}, {Canales-Moreno}, {Lafuente-Aznar}, {Calavia-Belloc}, {Perell{\'o}}, {Selva}, {Organero}, {Hernandez}, {Cueva}, {Yuste-Moreno}, {Garc{\'\i}a-Navarro}, {Donate-Lucas}, {Izquierdo-Carri{\'o}n}, {Iglesias-Marzoa}, {Lacruz}, {Gon{\c{c}}alves}, {Staels}, {Goossens}, {Henden}, {Walker}, {Reyes}, {Pastor}, {Kaspi}, {Skrutskie}, {Verbiscer}, {Martinez}, {Andr{\'e}}, {Maestre}, {Aceituno}, {Bacci}, {Maestripieri}, {Grazia}, {Castro-Tirado}, {P{\'e}rez-Garcia}, {Fern{\'a}ndez Garc{\'\i}a}, {Fern{\'a}ndez}, {Messner}, {Scarfi}, {Miku{\v{z}}}, {Prat}, {Martorell}, {Nardiello}, {Nascimbeni}, {Sfair}, {Siqueira}, {Lattari},
  {Liberato}, {Pinheiro}, {de Santana}, {Pereira}, {Alava-Amat}, {Ciabattari}, {Gonz{\'a}lez-Rodriguez}, \& {Schnabel}}]{Rommel_2025}
{Rommel}, F.~L., {Fern{\'a}ndez-Valenzuela}, E., {Proudfoot}, B.~C.~N., {et~al.} 2025, \psj, 6, 48, \dodoi{10.3847/PSJ/adabc1}

\bibitem[{{Rossi} {et~al.}(1999){Rossi}, {Marzari}, \& {Farinella}}]{Rossi_1999}
{Rossi}, A., {Marzari}, F., \& {Farinella}, P. 1999, Earth, Planets and Space, 51, 1173, \dodoi{10.1186/BF03351592}

\bibitem[{{Sicardy}(2023)}]{Sicardy_2023}
{Sicardy}, B. 2023, Comptes Rendus Physique, 23, 213, \dodoi{10.5802/crphys.109}

\bibitem[{{Sicardy} {et~al.}(2024){Sicardy}, {Braga-Ribas}, {Buie}, {Ortiz}, \& {Roques}}]{Sicardy_2024}
{Sicardy}, B., {Braga-Ribas}, F., {Buie}, M.~W., {Ortiz}, J.~L., \& {Roques}, F. 2024, \aapr, 32, 6, \dodoi{10.1007/s00159-024-00156-x}

\bibitem[{{Sicardy} \& {Dettwiller}(2025)}]{sicardy2025}
{Sicardy}, B., \& {Dettwiller}, L. 2025, \aap, submitted

\bibitem[{{Tancredi} \& {Favre}(2008)}]{Tancredi_2008}
{Tancredi}, G., \& {Favre}, S. 2008, \icarus, 195, 851, \dodoi{10.1016/j.icarus.2007.12.020}

\bibitem[{{Vachier} {et~al.}(2012){Vachier}, {Berthier}, \& {Marchis}}]{Vachier_2012}
{Vachier}, F., {Berthier}, J., \& {Marchis}, F. 2012, \aap, 543, A68, \dodoi{10.1051/0004-6361/201118408}

\bibitem[{Van~Belle(1999)}]{vanbelle_1999}
Van~Belle, G.~T. 1999, Publications of the Astronomical Society of the Pacific, 111, 1515

\bibitem[{{Young} {et~al.}(2020){Young}, {Braga-Ribas}, \& {Johnson}}]{Young_2022}
{Young}, L.~A., {Braga-Ribas}, F., \& {Johnson}, R.~E. 2020, in The Trans-Neptunian Solar System, ed. D.~{Prialnik}, M.~A. {Barucci}, \& L.~{Young}, 127--151, \dodoi{10.1016/B978-0-12-816490-7.00006-0}

\end{thebibliography}
\bibliographystyle{aasjournal}

%% This command is needed to show the entire author+affiliation list when
%% the collaboration and author truncation commands are used.  It has to
%% go at the end of the manuscript.
%\allauthors

%% Include this line if you are using the \added, \replaced, \deleted
%% commands to see a summary list of all changes at the end of the article.
%\listofchanges
\end{document}